\documentclass[pdflatex,sn-mathphys,Numbered,iicol]{sn-jnl}

\usepackage{nicefrac}
\usepackage{multirow}

\usepackage{amssymb,amsbsy,amsmath,amsfonts}
\usepackage{color,graphicx}
\usepackage{multicol}
\usepackage{xcolor}
\usepackage{epsf,epsfig,float,latexsym,amsthm,fancyhdr,rotating}
\usepackage{graphics,psfrag}
\usepackage{slashed}
\usepackage{comment}
\usepackage{lipsum}

\usepackage[labelfont=bf,font+=smaller,justification=centerlast]{caption}
\usepackage[labelfont=normal,font=normal,font=small,labelformat=simple]{subcaption}
\usepackage{upgreek}
\hypersetup{colorlinks,citecolor=blue,linktoc=all,linkcolor=cyan,urlcolor=blue}

\newcommand{\cG}{{\cal G}}
\newcommand{\cF}{{\cal F}}

 \usepackage{setspace}
 
 \usepackage[normalem]{ulem}

\def\beq{\begin{equation}}
\def\eeq{\end{equation}}
\def\bea{\begin{eqnarray}}
\def\eea{\end{eqnarray}}
\def\beqa{\begin{equation}\begin{array}{l}}
\def\eeqa{\end{array}\end{equation}}
\def\eqlab#1{\label{eq:#1}}

\def\seclab#1{\label{sec:#1}}

\def\Eqref#1{Eq.~\ref{eq:#1}}

\def\Figref#1{Fig.~\ref{fig:#1}}

\def\secref#1{Sec.~\ref{sec:#1}}

\def\barr{\left(\begin{array}{c}}
\def\earr{\end{array}\right)}
\def\bmat{\left(\begin{array}{cc}}
\def\emat{\end{array}\right)}
\def\al{\alpha}

  \def\eps{\epsilon}

\def\nn{\nonumber}
\def\dd{\mathrm{d}}

\def\AMP{{\mathcal A}}

\DeclareMathOperator\im{Im}
\DeclareMathOperator\re{Re}\def\3d{3-D}

\definecolor{darkred}{rgb}{0.9, 0.0, 0.0}

\def\TCop{\textcopyright}

\newcommand{\mycomment}[1]{}
\newcommand{\ep}{\varepsilon}
\newcommand{\ds}{\displaystyle}
\newcommand{\GeV}{\rm GeV}

\newcommand\RotText[1]{\rotatebox{90}{\parbox{3cm}{#1}}}

\makeatletter
\renewcommand\addressfont{\reset@font\fontsize{10bp}{13.5bp}\selectfont\titraggedcenter}
\makeatother

\def\preprint{LA-UR-22-32680, IPPP/23/28}

\title[Radiative Corrections]%
{Radiative Corrections:\\ From Medium to High Energy Experiments}

\makeatletter
\let\@titleT\@title
\renewcommand\@title{%
\vskip-4cm%
\hfill {\normalsize\preprint}\\%
\vskip0cm%
\centering\textbf\@titleT\\[5mm]%
\hrule}
\makeatother

\author*[1]{\fnm{Andrei}~\sur{Afanasev}}\email{afanas@gwu.edu}
\author*[2,3]{\fnm{Jan C.}~\sur{Bernauer}}\email{jan.bernauer@stonybrook.edu}
\author[4]{\fnm{Peter}~\sur{Blunden}}
\author*[5]{\fnm{Johannes}~\sur{Bl\"umlein}}\email{johannes.bluemlein@desy.de}
\author*[2]{\fnm{Ethan W.}~\sur{Cline}}\email{ethan.cline@stonybrook.edu}
\author[6,7]{\fnm{Jan M.}~\sur{Friedrich}}
\author*[8,9,10]{\fnm{Franziska}~\sur{Hagelstein}}\email{hagelste@uni-mainz.de}
\author*[11,12]{\fnm{Tom\'a\v{s}}~\sur{Husek}}\email{husek@ipnp.mff.cuni.cz}
\author[13]{\fnm{Michael}~\sur{Kohl}}
\author[14]{\fnm{Fred}~\sur{Myhrer}}
\author[15]{\fnm{Gil}~\sur{Paz}}
\author*[16]{\fnm{Susan}~\sur{Schadmand}}\email{s.schadmand@gsi.de}
\author[17]{\fnm{Axel}~\sur{Schmidt}}
\author[10,18]{\fnm{Vladyslava}~\sur{Sharkovska}}
\author*[10,18]{\fnm{Adrian}~\sur{Signer}}\email{adrian.signer@psi.ch}
\author[19]{\fnm{Oleksandr}~\sur{Tomalak}}
\author[20]{\fnm{Egle}~\sur{Tomasi-Gustafsson}}
\author*[21]{\fnm{Yannick}~\sur{Ulrich}}\email{yannick.ulrich@durham.ac.uk}
\author[8,9]{\fnm{Marc}~\sur{Vanderhaeghen}}

\affil[1]{\orgdiv{Department of Physics}, \orgname{The George Washington University}, \orgaddress{\city{Washington, DC}, \postcode{20052}, \country{USA}}}
\affil[2]{\orgdiv{Center for Frontiers in Nuclear Science}, \orgname{Stony Brook University}, \orgaddress{\city{Stony Brook}, \postcode{11794}, \state{NY}, \country{USA}}}
\affil[3]{\orgdiv{RIKEN BNL Research Center}, \orgname{Brookhaven National Lab}, \orgaddress{\city{Upton}, \postcode{11973}, \state{NY}, \country{USA}}}
\affil[4]{\orgdiv{Department of Physics and Astronomy}, \orgname{University of Manitoba}, \orgaddress{\city{Winnipeg},  \state{MB}, \postcode{R3T 2N2}, \country{Canada}}}
\affil[5]{\orgdiv{Deutsches Elektronen-Synchrotron} \orgname{DESY}, \orgaddress{\city{Zeuthen}, \postcode{15738}, \country{Germany}}}
\affil[6]{\orgdiv{Physics Department}, \orgname{Technical University of Munich},\orgaddress{\city{Garching}, \postcode{85748}, \country{Germany}}}
\affil[7]{\orgdiv{Excellence Cluster ``Origins"}, \orgaddress{\city{Garching}, \postcode{85748}, \country{Germany}}}
\affil[8]{\orgdiv{Institute of Nuclear Physics}, \orgname{Johannes Gutenberg-Universit\"at}, \orgaddress{\city{Mainz}, \postcode{55099}, \country{Germany}}}
\affil[9]{\orgdiv{PRISMA+ Cluster of Excellence}, \orgname{Johannes Gutenberg-Universit\"at}, \orgaddress{\city{Mainz}, \postcode{55099}, \country{Germany}}}
\affil[10]{\orgdiv{Laboratory for Particle Physics}, \orgname{Paul Scherrer Institute}, \orgaddress{\city{Villigen}, \postcode{5232}, \country{Switzerland}}}
\affil[11]{\orgdiv{Department of Astronomy and Theoretical Physics}, \orgname{Lund University}, \orgaddress{\city{Lund}, \postcode{223-62}, \country{Sweden}}}
\affil[12]{\orgdiv{Institute of Particle and Nuclear Physics}, \orgname{Charles University}, \orgaddress{\city{Prague}, \postcode{180 00}, \country{Czech Republic}}}
\affil[13]{\orgdiv{Department of Physics}, \orgname{Hampton University}, \orgaddress{\city{Hampton}, \postcode{23668}, \country{USA}}}
\affil[14]{\orgdiv{Department of Physics and Astronomy}, \orgname{University of South Carolina}, \orgaddress{\city{Columbia}, \postcode{29208}, \country{USA}}}
\affil[15]{\orgdiv{Department of Physics and Astronomy}, \orgname{Wayne State University}, \orgaddress{\city{Detroit}, \postcode{48201}, \country{USA}}}
\affil[16]{\orgdiv{Helmholtzzentrum für Schwerionenforschung GmbH}, \orgname{GSI}, \orgaddress{\city{Darmstadt}, \postcode{64291}, \country{Germany}}}
\affil[17]{\orgdiv{Department of Physics}, \orgname{The George Washinmgton University}, \orgaddress{\city{Washington, D.C.}, \postcode{20052}, \country{USA}}}
\affil[18]{\orgdiv{Department of Physics}, \orgname{University of Zurich}, \orgaddress{\city{Zurich}, \postcode{CH-8057}, \country{Switzerland}}}
\affil[19]{\orgdiv{Theoretical Division}, \orgname{Los Alamos National Laboratory}, \orgaddress{\city{Los Alamos}, \postcode{87545}, \country{USA}}}
\affil[20]{\orgdiv{Département de Physique Nucléaire, IRFU}, \orgname{Commissariat à l'énergie atomique et Université Paris Saclay}, \orgaddress{\city{Paris}, \postcode{91190}, \country{France}}}
\affil[21]{\orgdiv{Institute for Particle Physics Phenomenology}, \orgname{University of Durham}, \orgaddress{\city{Durham}, \postcode{DH1 3LE}, \country{United Kingdom}}}

\abstract{
Radiative corrections are crucial for modern high-precision physics experiments, and are an area of active research in the experimental and theoretical community. Here we provide an overview of the state of the field of radiative corrections with a focus on several topics: lepton-proton scattering, QED corrections in deep-inelastic scattering, and in radiative light-hadron decays. Particular emphasis is placed on the two-photon exchange, believed to be responsible for the proton form-factor discrepancy, and associated Monte-Carlo codes. We encourage the community to continue developing theoretical techniques to treat radiative corrections, and perform experimental tests of these corrections.
}


\begin{document}

\maketitle
\clearpage
\setcounter{tocdepth}{3}
\tableofcontents

\clearpage

\section{Introduction}

Radiative corrections are of central importance for modern high-precision, sub-atomic physics experiments. In order to extract physical observables from experiments with percent precision, radiative corrections on the experimental observables need to be known to sub-percent precision. Modern tools to calculate the size, and to estimate the uncertainty, of these corrections are a subject of active research in the field. Frameworks for determining these effects can be developed and must be checked experimentally.

In modern experiments the radiative corrections for QCD and QED interactions are of particular interest. QCD systems (e.g., nucleons or light nuclei) are often probed in scattering experiments, which allow for precision extractions of form factors and structure functions. Typical probes are leptons or photons, since they have no internal structure themselves. Traditionally, the electron would be used as the lepton in scattering experiments as it is stable, but its low mass allows for significant external radiation. This becomes a challenge as experiments move to ever higher energies in the deeply inelastic scattering regime. In modern experiments, the muon serves as an attractive alternative to the electron, as its higher mass typically leads to smaller radiative corrections. In the case of QCD systems being studied via meson decay channels, hadronic and QED corrections can be of similar and significant size. In $e^+e^-$ interactions, special care must be taken to not neglect the mass of the electron in order to achieve the requisite experimental and theoretical precision. 

The topic of radiative corrections is quite broad, and as such in this review article we focus on a few selected topics. In Secs.~\ref{LPScatt} to \ref{sec:Implementations}, we focus on the process of lepton-proton ($\ell p$) scattering where $\ell=e,\mu$ can be an electron or a muon. In Sec.~\ref{LPScatt}, we start with the theoretical base and the classification of radiative corrections. In Sec.~\ref{sec:TPE}, we study an important class of radiative corrections --- the two-photon exchange (TPE) contributions, which for hadron targets require knowledge of hadronic structure. In Sec.~\ref{sec:Implementations}, event generators for scattering experiments are discussed. In Sec.~\ref{sec:ho}, we consider higher-order corrections to the purely QED process of $e^+e^-$ annihilation in the deeply inelastic regime. In Sec.~\ref{sec:mesons}, we discuss the corrections for radiative decays of light mesons. Finally, we summarize the state of the field, and give recommendations to those involved for a path forward to push the global understanding of radiative corrections to higher precision than exists today in Sec.~\ref{sec:Summary}.

\section{Lepton-proton scattering}\label{LPScatt}

\begin{figure}[b]
\begin{center}
\includegraphics[scale=0.45,angle=0]{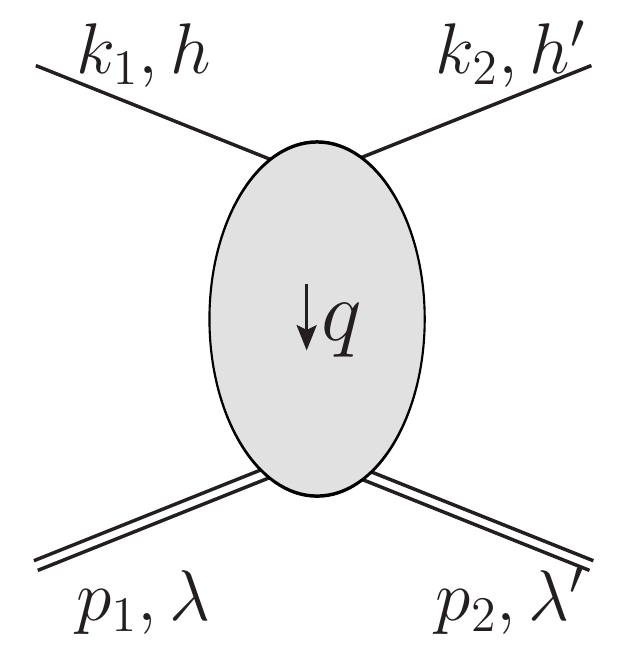}
\qquad
\includegraphics[scale=0.45,angle=0]{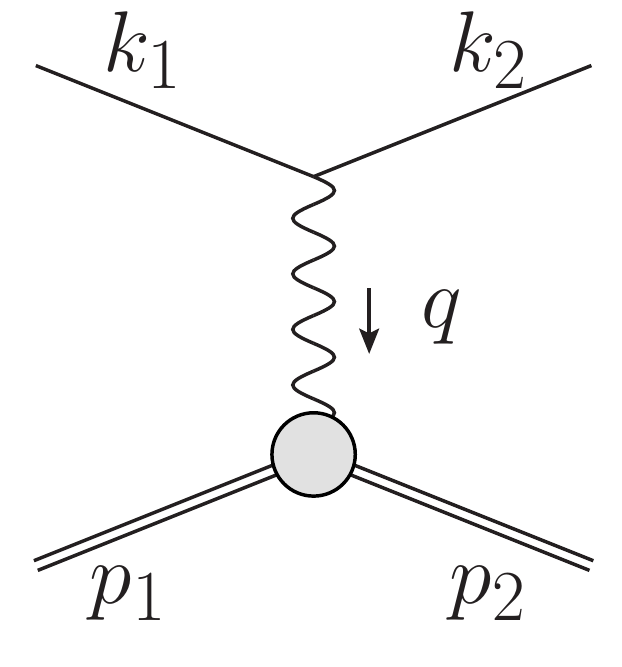}
\end{center}
\caption{
\label{fig:rc:elLP}
Generic $\ell p$ scattering (left panel), tree-level Born diagram (right panel).}
\end{figure}

Scattering experiments allow experimenters to probe the structure of the target. From a measurement of the unpolarized $\ell p$ scattering cross section, one can for instance determine the proton electric and magnetic Sachs form factors (FFs), $G_E(Q^2)$ and $G_M(Q^2)$, functions of the momentum transfer $t=-Q^2$. 
The Born cross section of elastic $\ell p$ scattering, given by the tree-level one-photon exchange (OPE) diagram in Fig.~\ref{fig:rc:elLP} (right panel), can be written in a compact form as

\beq
\eqlab{BornXS}
\frac{\mathrm{d} \sigma^{(0)}}{\mathrm{d} \Omega} = \frac{\sigma_\text{Mott} \,\tau}{\varepsilon(1+\tau)}\,\sigma_R,
\eeq
with the Mott cross section for scattering off a point-like particle
\beq \label{SigmaRed}
\sigma_\text{Mott}=\frac{\alpha^2 \epsilon_2 \cos^2{\theta/2}}{4 \epsilon_1^3 \sin^4{\theta/2}},
\eeq
the reduced Born cross section
\beq
\sigma_R =G_M^2(Q^2)+\frac{\varepsilon}{\tau}\, G_E^2(Q^2),\eqlab{reducedCS}
\eeq
and the 
photon polarization parameter 
\beq 
\label{eq:epsilon_def}
\varepsilon = \frac{\nu^2 - M^4 \tau ( 1 + \tau )}{ \nu^2 + M^4 \tau ( 1 + \tau )  ( 1 - 2\varepsilon_{\ell})}.
\eeq
 Here, we introduced the 
 dimensionless quantities $\tau=Q^2/4M^2$, $\varepsilon_\ell = 1/(2\tau_\ell)$ and $\tau_\ell = Q^2 / (4 m_\ell^2)$, and the crossing-symmetric variable $ \nu \equiv k_1 \cdot (p_1 + p_2)/2 = (s - u)/4$, with $M$ ($m_\ell)$ the proton (lepton) mass, and where $s$ and $u$ denote the usual Mandelstam variables of the elastic scattering process of Fig.~\ref{fig:rc:elLP}. Furthermore, $\alpha \simeq 1/137$ is the usual fine structure constant, $\epsilon_1$ ($\epsilon_2$) is the energy of the incoming (outgoing) lepton, and $\theta$ is the scattering angle in the laboratory frame, see \Eqref{eq2}.

The extraction of the FFs, cf.\ \Eqref{reducedCS}, is done via Rosenbluth separation  \cite{Rosenbluth:1950yq}, which  requires measurements at the same momentum transfer $Q^2$ but for different energies and angles. To recover the Born cross section, the experimentally measured cross section needs to be corrected for radiative corrections, cf. Refs.~\cite{Maximon:2000hm,Mo:1968cg,Ent:2001hm,Gramolin:2014pva,Gerasimov:2015aoa,Bystritskiy:2007hw}. These change not only the absolute value of the cross section but also its dependence on the relevant kinematical variables. In general, the  $\mathcal{O}(\al^2)$ Born cross section has to be corrected by additional virtual photons or inelastic scattering with emission of real bremsstrahlung.

\subsection{Radiative corrections}\label{Sec:RadCorrAdrian}

In this subsection, we discuss the calculation of cross sections in a
perturbative expansion of the electromagnetic coupling $\alpha$. While
many remarks are valid for arbitrary processes, we will illustrate the
general procedure for elastic $\ell p$ scattering \beq
\ell( k_1 , h ) + p( p_1, \lambda ) \to \ell( k_2, h^\prime) + p(p_2, \lambda^\prime) \label{eq:eq1}
\eeq
where $k_1(k_2)$ and $p_1(p_2)$ are the incoming (outgoing) lepton and proton momenta, and $h(h^\prime)$ and $\lambda(\lambda^\prime)$ the respective helicities.

\begin{figure}[ht]
\begin{center}
\includegraphics[scale=0.45,angle=0]{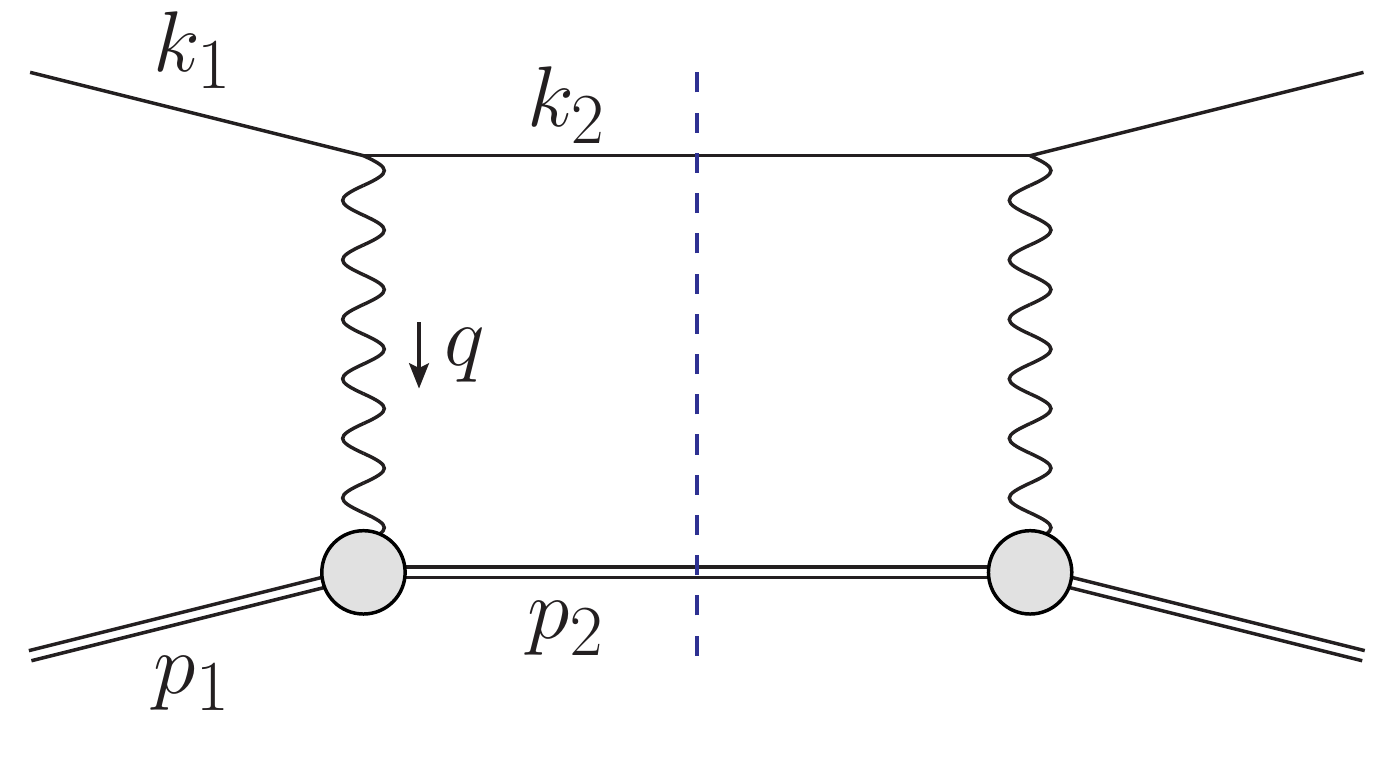}
\end{center}
\caption[]{
\label{fig:rc:LO}
Schematic illustration of the squared LO amplitude
$\AMP^{(0)}\AMP^{(0)\,*}$ for lepton-proton scattering. Leptons
(protons) are depicted as single (double) lines and the grey blobs
represent the form factors.  }
\end{figure}

\begin{figure}[ht]
\begin{center}
\includegraphics[scale=0.45,angle=0]{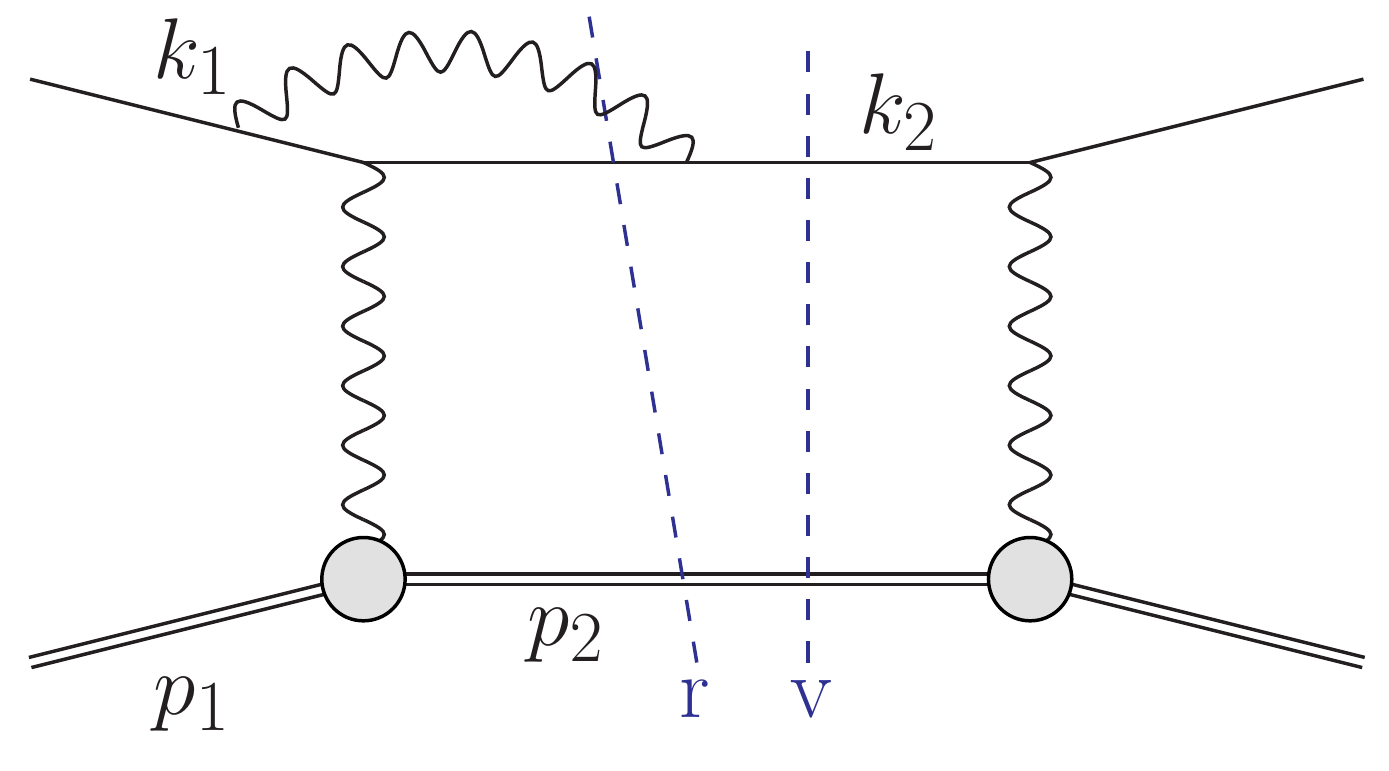}
\end{center}
\caption[]{
\label{fig:rc:lNLO}
An example contribution $\mathcal{O}(\alpha^3\,Z^2)$ to the squared NLO
amplitude with photon emission restricted to the lepton line. The
virtual (v) cut represents $\AMP^{(1)}\,\AMP^{(0)\,*}$ whereas the
real (r) cut represents $|\AMP_\gamma^{(0)}|^2$. }
\end{figure}

\begin{figure}[ht]
\begin{center}
\includegraphics[scale=0.45,angle=0]{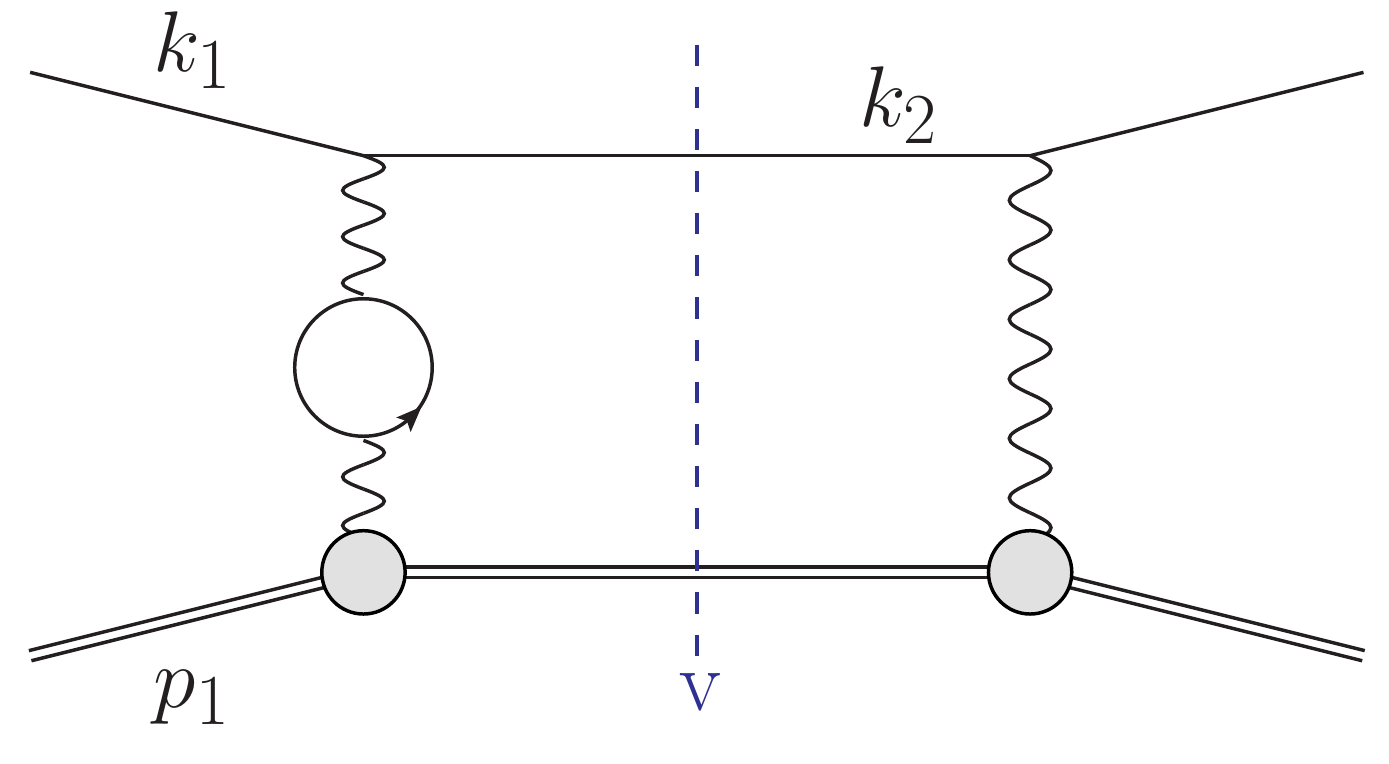}
\end{center}
\caption[]{
\label{fig:rc:vpNLO}
An NLO correction due to VP contributions. The virtual (v) cut
represents $\AMP^{(1)}\,\AMP^{(0)\,*}$. }
\end{figure}

\begin{figure}[ht]
\begin{center}
\includegraphics[scale=0.45,angle=0]{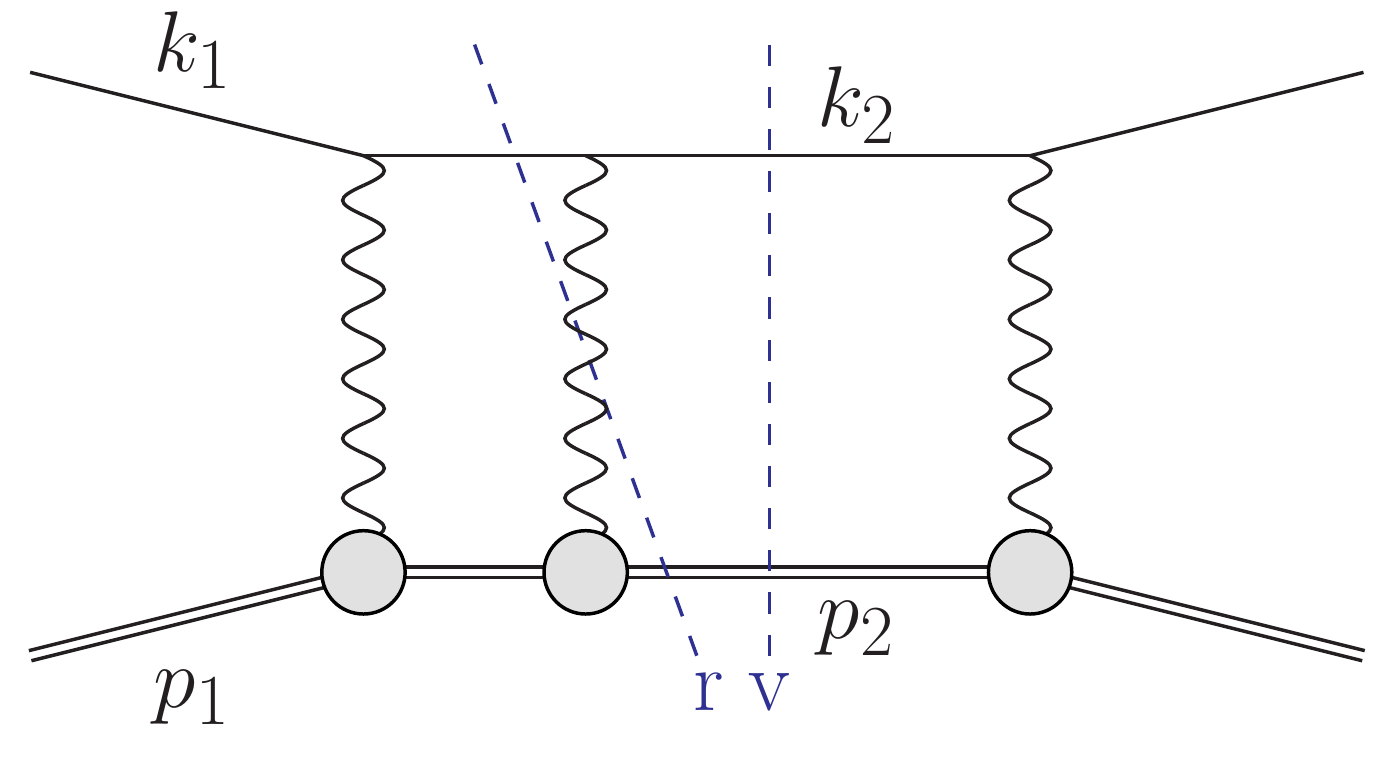}
\end{center}
\caption[]{
\label{fig:rc:pNLO}
An example contribution to the squared NLO amplitude
$\mathcal{O}(\alpha^3 Z^3)$ with elastic TPE. The
virtual (v) cut represents $\AMP^{(1)}\,\AMP^{(0)\,*}$ whereas the
real (r) cut represents $|\AMP_\gamma^{(0)}|^2$. }
\end{figure}

\begin{figure}[ht]
\begin{center}
\includegraphics[scale=0.45,angle=0]{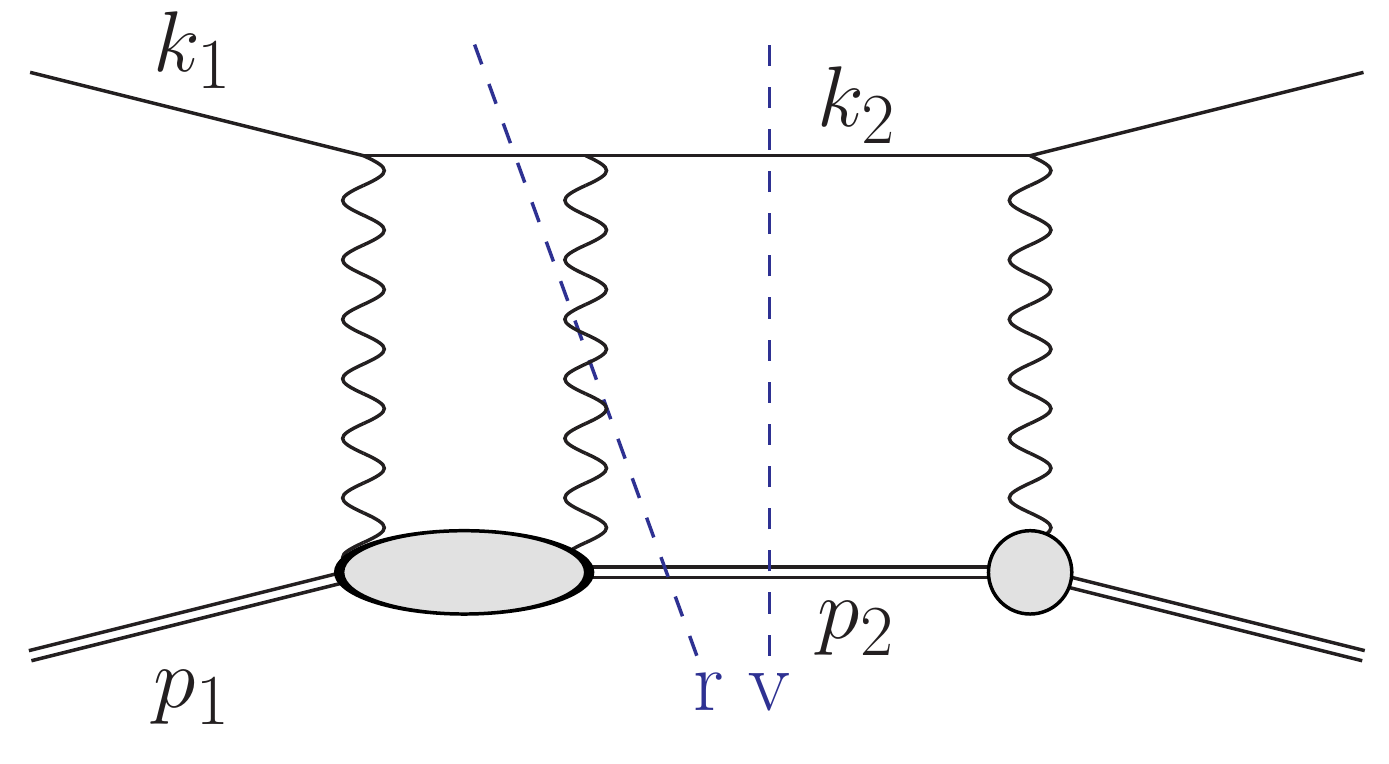}
\end{center}
\caption[]{
\label{fig:rc:messNLO}
An example contribution to the squared NLO amplitude with inelastic
TPE. The virtual cut (v) represents
$\AMP^{(1)}\,\AMP^{(0)\,*}$ whereas the real cut (r) represents
$|\AMP_\gamma^{(0)}|^2$. }
\end{figure}

\begin{figure}[ht]
\begin{center}
\includegraphics[scale=0.45,angle=0]{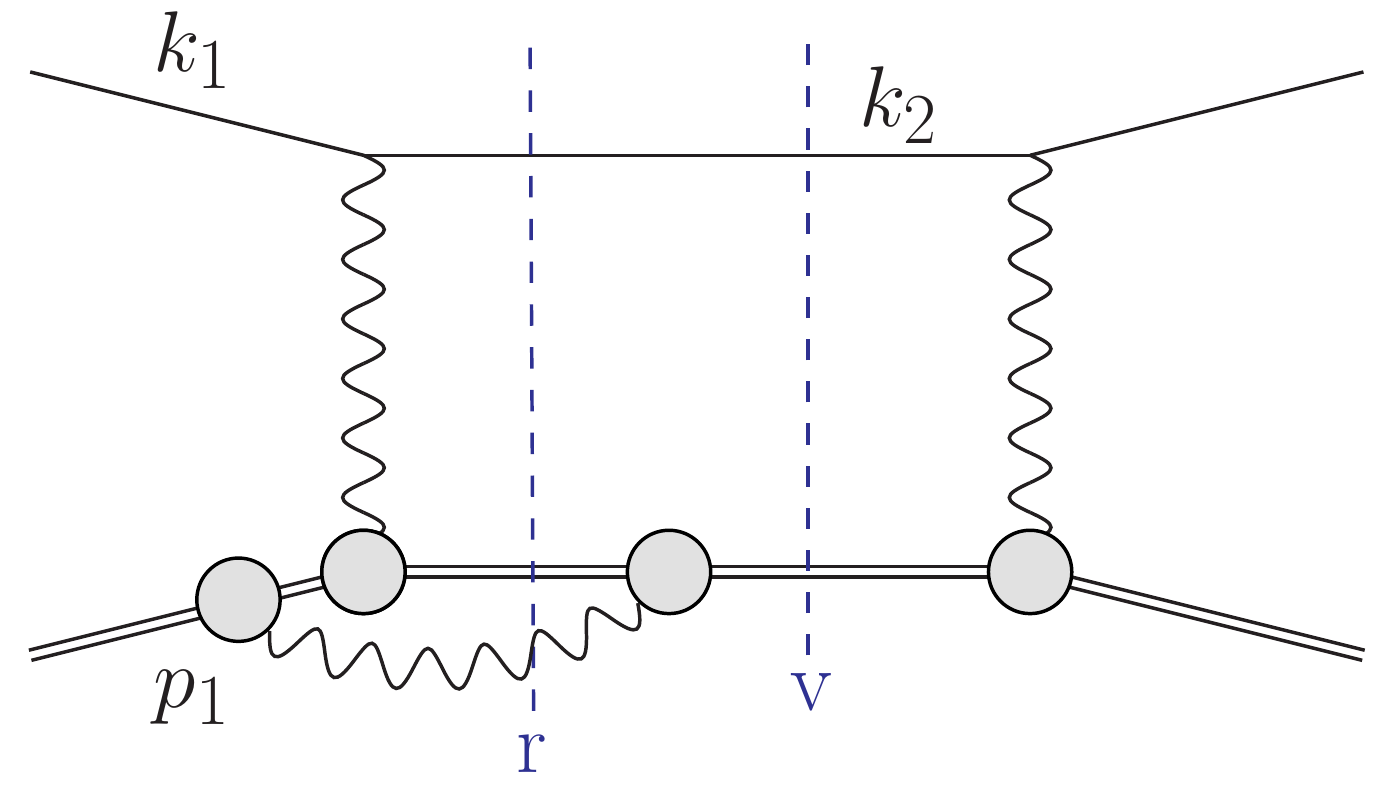}
\end{center}
\caption{
\label{fig:rc:FF-NLO}
An example contribution $\mathcal{O}(\alpha^3\,Z^4)$ to the squared NLO
amplitude with photon emission restricted to the proton line. The
virtual (v) cut represents $\AMP^{(1)}\,\AMP^{(0)\,*}$ whereas the
real (r) cut represents $|\AMP_\gamma^{(0)}|^2$. }
\end{figure}

The Born or leading order (LO) cross section for elastic $\ell$p scattering, given in \Eqref{BornXS}, is of
$\mathcal{O}(\alpha^2 Z^2)$ and is a function of $G_E(Q^2)$ and $G_M(Q^2)$.  These form factors are depicted as grey blobs
in~\Figref{rc:LO} and, for each appearance, a factor $Z$, the total charge, is included in
the counting of the couplings. To improve the theory description,
next-to-leading order (NLO) corrections to \Eqref{BornXS} have been
computed in Refs.~\cite{Maximon:2000hm, Mo:1968cg, Ent:2001hm,
  Gramolin:2014pva, Gerasimov:2015aoa, Bystritskiy:2007hw}. The NLO
corrections can be decomposed into several gauge-invariant parts. The one-photon exchange (OPE) contributions are of the form
$\mathcal{O}(\alpha^3 Z^2)$ and also include vacuum polarisation (VP)
corrections. They are illustrated in Figs.~\ref{fig:rc:lNLO} and
\ref{fig:rc:vpNLO}, respectively. The TPE
contributions, depicted in
Figs.~\ref{fig:rc:pNLO} and \ref{fig:rc:messNLO}, are $\mathcal{O}(\alpha^3 Z^3)$. Finally, corrections to the
proton line are of $\mathcal{O}(\alpha^3 Z^4)$, with a sample
contribution shown in \Figref{rc:FF-NLO}. These corrections change
not only the absolute value of the observables but also their
dependence on the relevant kinematical variables, e.g., the square of
the transferred momentum $Q^2$ and the photon polarization
parameter $\varepsilon$, defined in~\Eqref{epsilon_def}.

If the radiative corrections are small, an NLO calculation might be
sufficient. In particularly simple situations, these corrections can be
included as a multiplicative factor. The
error associated to this procedure is at the percent level. Radiative corrections can lead to
contributions that are enhanced by large logarithms of the form
$L=\ln{(Q^2/m_\ell^2)}$.  As we will discuss below, these are associated with hard
collinear radiation. Another source of large logarithms are stringent
experimental cuts that restrict the phase space for soft emission. As a consequence, modern high-precision experiments are required to introduce higher order corrections
and a more sophisticated treatment of real radiation.


This leads us to consider fully differential computations of cross
sections which allows one to obtain precise predictions for fiducial cross
sections. To keep full generality, the phase-space integration needs
to be adaptable to the experimental setting. As a result, it is
necessary to revert to numerical integrations, which are most
efficiently carried out using Monte Carlo (MC) techniques.

For the LO cross section $\dd \sigma^{(0)}\sim\alpha^2\,Z^2$, the
tree-level matrix element squared $|\AMP^{(0)}|^2$, depicted in
\Figref{rc:LO}, is to be integrated over the two-body phase space
$\Phi_2$. Thus,
\begin{equation}\label{mcLO}
\dd\sigma^{(0)} = \int \dd\Phi_2\, |\AMP^{(0)}|^2\, S(k_2,p_2)\,,
\end{equation}
where the so-called measurement function $S(k_2,p_2)$ defines the
quantity to be computed in terms of the momenta of the final state
particles. For simplicity, trivial elements such as the flux 
factor are also included in $S$. If a cut is to be
applied, $S$ is set to 0 if the event does not pass the cut. 

Going beyond LO, we encounter ultraviolet (UV) and infrared (IR)
divergences. After renormalization, in a QED calculation where all
fermion masses are kept at their finite values, only soft IR
divergences remain. These soft singularities cancel between the
virtual and real corrections. At NLO this is illustrated in
\Figref{rc:lNLO}, where some corrections due to photon emission from
the lepton line are shown. The virtual corrections are obtained as
\begin{equation}\label{mcVNLO}
\dd\sigma^{(1)}_{\rm v} = \int \dd\Phi_2\, 2\re(\AMP^{(1)}\AMP^{(0)\,*})\,
S(k_2,p_2)\, ,
\end{equation}
whereas the real corrections are given by 
\begin{equation}\label{eq:mcRNLO}
\dd\sigma^{(1)}_{\rm r} = \int \dd\Phi_3\, |\AMP_\gamma^{(0)}|^2\,
S(k_2,p_2,k_\gamma)\, .
\end{equation}
Both terms are $\mathcal{O}(\alpha^3\,Z^2)$. Here $\AMP^{(1)}$ is the one-loop amplitude of the process \Eqref{eq1} and  $\AMP_\gamma^{(0)}$
is the tree-level amplitude with an additional photon of momentum
$k_\gamma$ in the final state. The integration in \Eqref{mcRNLO} has
to be carried out over the three-particle final state $\Phi_3$. Also,
the definition of $q$ is more subtle now, since $k_1-k_2\neq p_2-p_1$.

To give mathematical meaning individually to $\dd \sigma^{(1)}_{\rm v}$
and $\dd \sigma^{(1)}_{\rm r}$, the IR singularities need to be
regularized. Many calculations in QED still use a (small) fictitious
photon mass to do so. Then, the IR singularities manifest themselves as
logarithms of this mass. Another approach that is more in line with
the tremendous progress in computational techniques that has been made
for QCD calculations is to use dimensional regularization (i.e. work in
$4-2\epsilon$ dimensions) also for IR singularities. In this case,
IR singularites show up as poles $1/\epsilon$. See \secref{IRsubtraction} for a detailed discussion. Independent of the
chosen regularization, in the combination of $\dd \sigma^{(1)}_{\rm v}$
and $\dd \sigma^{(1)}_{\rm r}$ the remnants of the IR singularities
(i.e. the photon mass logarithms or the $1/\epsilon$ poles)
cancel. This is ensured by the well-known limiting behavior of the
real matrix element in the soft limit ($\mathcal{S}@$)~\cite{Yennie:1961ad}, where it
can be written as an eikonal factor $\mathcal{E}$ times the Born
matrix element
\begin{equation}\label{eikonal}
\mathcal{S}@ |\AMP_\gamma^{(0)}|^2 = \mathcal{E}\,|\AMP^{(0)}|^2 . 
\end{equation}
Hence, in the complete NLO corrections to Eq.~\ref{mcLO},
\begin{equation}\label{mcNLO}
\dd \sigma^{(1)} = \dd \sigma^{(1)}_{\rm v} + \dd \sigma^{(1)}_{\rm r} ,
\end{equation}
the regularization can be removed (i.e. the photon mass
or $\epsilon$ set to zero). For this cancellation to occur, the
observable has to be IR safe. This means that the observable must not
change whether or not an arbitrarily soft photon is emitted. At a
technical level, we must require
\begin{equation}\label{irsafe}
  \lim_{k_\gamma \to 0} S(k_2,p_2,k_\gamma) = S(k_2,p_2) .
\end{equation}  
From a mathematical point of view, it is sufficient if Eq.~\ref{irsafe}
holds in the strict limit. However, if experimental cuts (directly or
indirectly affecting real radiation) induce a difference between
$S(k_2,p_2,k_\gamma)$ and $S(k_2,p_2)$ for finite, but small photon
energies $\Delta{E_\gamma}$, there can be a large logarithm
$\ln\left(\Delta{E_\gamma^2}/Q^2\right)$ as a remnant. Such (soft) logarithms can lead
to enhanced QED corrections.

Another source of large corrections is collinear emission. While
collinear singularites are regulated by the fermion masses, they still
can also lead to enhanced logarithmic terms $\alpha L$ after
combination of real and virtual corrections. These logarithms often
form the dominant part of the higher-order corrections. This also
entails that in fully differential QED calculations -- contrary to QCD
calculations -- it is not possible to set the lepton masses to
zero. Furthermore, the neglect of hard collinear emission potentially leads
to a loss of accuracy. Thus, using the soft approximation for the real
matrix elements can have severe implications on the accuracy of the
results. 

The extraction of the IR poles from \Eqref{mcRNLO} for an arbitrary
IR-safe observable, i.e. an arbitrary $S(k_2,p_2,k_\gamma)$ satisfying
Eq.~\ref{irsafe}, is by now a standard procedure at NLO and the focus
turns towards next-to-next-to leading order (NNLO) applications as discussed in Ref.~\cite{TorresBobadilla:2020ekr}. Again,
at NLO there are two widely used options, the slicing method and the
subtraction method.

In the slicing method, the phase space is split into two parts,
depending on the photon energy
\begin{align}\label{slicing}
  \dd\sigma^{(1)}_{\rm r}\big|_\text{sl}
  &=
  \int\displaylimits_{E_\gamma<\delta} \hspace*{-5pt}
  \dd\Phi_3\, |\AMP_\gamma^{(0)}|^2\,S(k_2,p_2)
\nonumber \\
 &+ \int\displaylimits_{E_\gamma\ge\delta} \hspace*{-5pt}
 \dd\Phi_3\, |\AMP_\gamma^{(0)}|^2\,S(k_2,p_2,k_\gamma)
\, .
\end{align}
In the part, where the photon energy is larger than a chosen resolution
parameter $\delta$, the integration can be carried out for a massless
photon without encountering singularities. In the part, where the
photon energy is smaller than the resolution parameter, the soft
(eikonal) approximation Eq.~\ref{eikonal} is used for the matrix
element. The integration over the photon phase space then simplifies
and can be carried out analytically. In this term, the photon mass
logarithms that cancel those of the virtual
corrections are generated. This method relies on a suitable choice for $\delta$. It should not be too large, to ensure
that the soft approximation is good enough. It should also not be too
small, otherwise numerical problems (large cancellations between the
resolved and unresolved parts) occur. For more details, see Sec.~\ref{sec:Implementations}.

In the subtraction method a term (the soft limit of the integrand) is
subtracted and added back
\begin{align}\label{subtraction}
  &\dd\sigma^{(1)}_{\rm r}\big|_\text{su}  =  \int \dd\Phi_3\ \mathcal{S}@ |\AMP_\gamma^{(0)}|^2  \,S(k_2,p_2) \ +\nonumber \\
   &\int \dd\Phi_3 \bigg(|\AMP_\gamma^{(0)}|^2S(k_2,p_2,k_\gamma) -\mathcal{S}@ |\AMP_\gamma^{(0)}|^2S(k_2,p_2)\bigg).
\end{align}
The first term on the r.h.s. of Eq.~\ref{subtraction} is very similar to
the corresponding term in Eq.~\ref{slicing}. The only differences are
that the photon is treated as massless and the integration is done
over the full phase space in Eq.~\ref{subtraction}. This results in
IR $1/\epsilon$ poles that cancel against the IR poles of the virtual
contributions. The second term on the r.h.s. of Eq.~\ref{subtraction} is
finite and can be integrated in four dimensions. An efficient numerical
evaluation of this term including the subtraction requires an adapted
phase-space parameterization. Within the subtraction method, there is
no motivation to split the real corrections into hard and soft
parts. It is actually simpler to always include the full real matrix
element and make no approximation whatsoever.

The OPE corrections due to emission from the lepton line
$\mathcal{O}(\alpha^3\,Z^2)$ and the VP corrections, illustrated in Figs.~\ref{fig:rc:lNLO} and
\ref{fig:rc:vpNLO}, are simple from a conceptual
point of view as they involve standard QED pointlike interactions
only. However, there are also corrections of $\mathcal{O}(\alpha^3\,Z^3)$
that involve multiple exchange of photons between the two fermion
 lines, in particular the notorious TPE
corrections. The class of corrections depicted in \Figref{rc:pNLO} is
more complicated than those of \Figref{rc:lNLO} for several
reasons. Even if the proton was pointlike, these corrections now
involve one-loop box diagrams at NLO. In addition, there is the complication that the
photon-proton vertex is more involved than for a pointlike
particle. Furthermore, for $\ell p$ scattering there are also
contributions where the intermediate state is not just a proton, as
depicted in \Figref{rc:messNLO}. These aspects will be discussed in
\secref{TPE}.

A final class of NLO corrections are those where the emission of
additional photons is restricted to the proton line. They are
$\mathcal{O}(\alpha^3\,Z^4)$ and a sample contribution is depicted in
\Figref{rc:FF-NLO}. It is tempting to include these diagrams into
the definition of the form factors. However, they contain IR
divergences that, once more, cancel between the real and virtual
parts. Thus, from a practical point of view it is more convenient to
explicitly include these diagrams into the radiative corrections and define the form factors
accordingly, i.e. exclude all QED effects from their definition. 
The numerical impact of these contributions is rather minor, as they
do not generate collinear logarithms. Hence, it is often possible to
simply neglect the $\mathcal{O}(\alpha^3\,Z^4)$ contribution.

\begin{figure}[ht]
\begin{center}
\includegraphics[scale=0.45,angle=0]{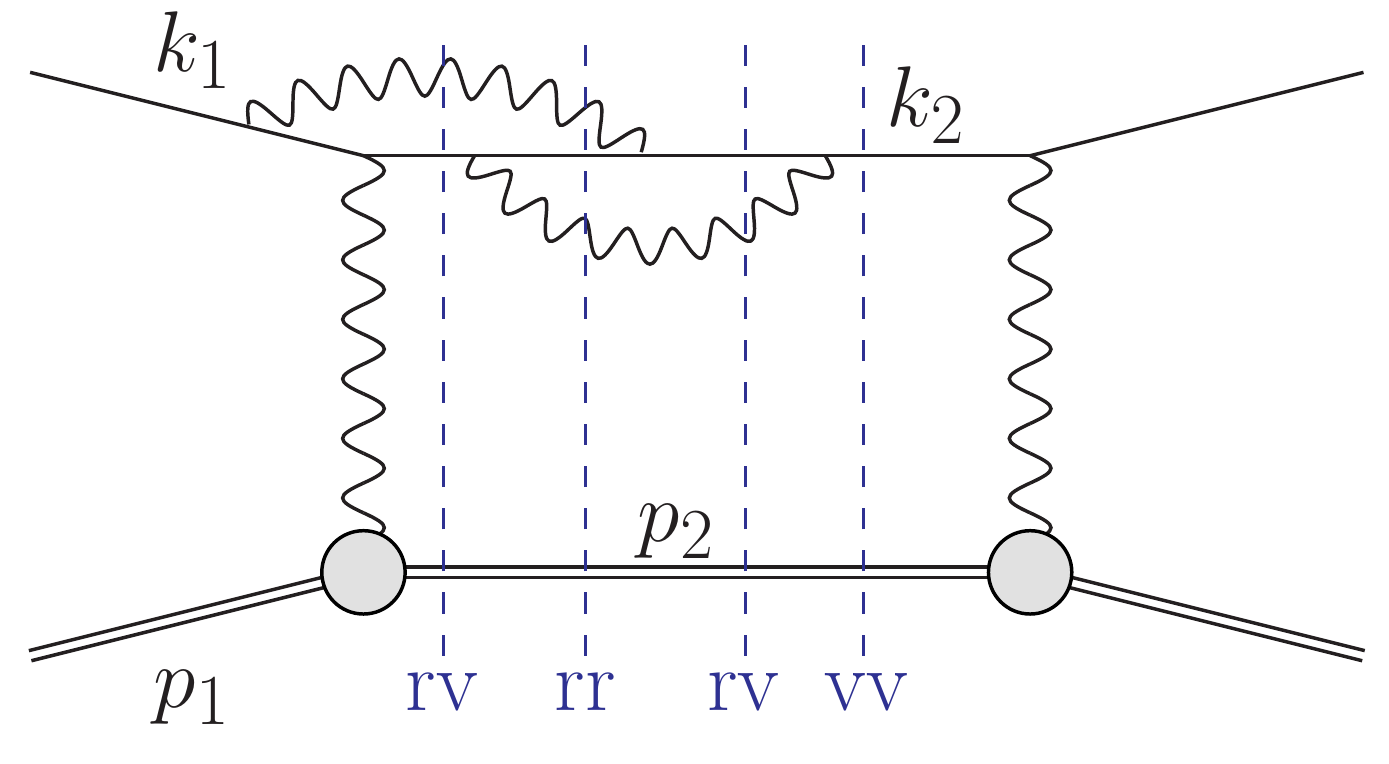}
\end{center}
\caption[]{
\label{fig:rc:NNLO}
Example of a NNLO contribution $\mathcal{O}(\alpha^4 Z^2)$ to the
squared amplitude for $\ell p$ scattering. The double-virtual
(vv) cut represents $\AMP^{(2)}\AMP^{(0)\,*}$, the real-virtual (rv)
cuts $\AMP^{(0)}_\gamma\AMP^{(1)\,*}_\gamma$ and
$\AMP^{(1)}_\gamma\AMP^{(0)\,*}_\gamma$, and the double-real (rr) cut
$|\AMP^{(0)}_{\gamma\gamma}|^2$.}
\end{figure}

As discussed above, at each order in $\alpha$ the QED corrections can
be enhanced by a collinear logarithm $L$ and possibly a soft logarithm
$\ln\left(\Delta{E_\gamma}^2/Q^2\right)$. Hence, the corrections can be considerably
larger than naive scaling with $\alpha$ implies. Thus, for very
precise predictions it is necessary to go beyond NLO. Regarding
OPE corrections restricted to emission from the lepton line
$\mathcal{O}(\alpha^4\,Z^2)$, the computations have been extended to
NNLO. This involves double-virtual, real-virtual, and double-real
corrections, as illustrated in \Figref{rc:NNLO}. Results have been
obtained using photon-mass regularization and the slicing
method in Refs.~\cite{Bucoveanu:2018soy,CarloniCalame:2020yoz} as well as using
dimensional regularization with the subtraction
method in Ref.~\cite{Banerjee:2020rww}. This class of corrections is a gauge
invariant subset of the full corrections. The virtual part involves
only the vector part of the two-loop heavy-lepton form
factor, cf. Refs.~\cite{Mastrolia:2003yz,Bonciani:2003ai,Bernreuther:2004ih}. For
lepton masses much smaller than the proton mass, these corrections are
typically also dominant. They involve up to two additional photon
emissions and a precise description how these photons are treated is
required. This amounts to a definition of
$S(k_2,p_2,k_{\gamma_1},k_{\gamma_2})$. In particular, both photons
can be collinear, leading to enhanced NNLO correction of the form
$(\alpha\,L)^2$. 

Corrections at NNLO that go beyond the OPE approximation lead to
substantially more complicated calculations. If the proton is treated
as pointlike, the NNLO results for electron-muon
scattering in Refs.~\cite{Engel:2022kde, Broggio:2022htr} can be adapted by
simple replacements $m_\mu\to{M}$ and $m_e\to{m}_\ell$. This requires the computation of the two-loop amplitude $\AMP^{(2)} $(for double-virtual) involving (crossed) double-box diagrams~\cite{Bonciani:2021okt}. For the real-virtual corrections $\AMP_\gamma^{(1)}$ is required where one-loop pentagon diagrams contribute. Efficient techniques for one-loop computations have been developed in connection with high multiplicity processes at hadron colliders, as reviewed in Ref.~\cite{Ellis:2011cr}.
As part of this topical collection, the effect of pointlike NNLO corrections were compared to present uncertainties from hadronic correction~\cite{McMule}.
It turns out that the foremer can be sizeable and need to be properly taken into account.
Beyond pointlike photon-proton interaction make the formidable task of the
two-loop amplitude computation even more
daunting. At NNLO, depending on the experimental set up, it might also
be necessary to include the process with  emission of an additional
lepton pair which also has been computed for electron-muon
scattering in Ref.~\cite{Budassi:2021twh}.

First steps to move beyond NNLO have already been taken. The form factor $\gamma^*\to\ell^+\ell^-$ is known
at three loops~\cite{Fael:2022rgm} and the subtraction method for
massive QED has been extended beyond NNLO~\cite{Engel:2019nfw}. Hence,
a next-to-next-to-next-to leading order (NNNLO) calculation of OPE contributions $\mathcal{O}(\alpha^5\,Z^2)$
seems to be feasible within a reasonable time frame. However, a full
computation of $2\to{2}$ scattering processes with massive particles at
NNNLO is currently out of reach.

As alluded to above, the corrections are often dominated by large
logarithms associated with collinear or soft emission. A convenient
method resumming all contributions of large logarithms at all orders
has been proposed in Ref.~\cite{Kuraev:1985hb}. All leading contributions
of order $(\alpha L)^n$ are taken into account.
The accuracy can be further improved calculating non-leading
contributions of order $\alpha (\alpha L)^n$ as a $K$-factor. The application to $ep$ elastic scattering can be found in
Ref.~\cite{Bystritskiy:2006ju}, and, including hard photon emission,
in Ref.~\cite{Kuraev:2013dra}. 

A more generic approach to include logarithmically enhanced
corrections is to use QED parton showers. A recent review of this
activity can be found in Ref.~\cite{Frixione:2022ofv} and will be discussed
in some more detail in Sec.~\ref{sec:Implementations}.  Parton
showers allow for the inclusion of the leading logarithmic
contributions in a process independent way. Combined with fixed-order
calculations this is a powerful tool to improve further the precision
for fully differential cross sections. Including next-to-leading
logarithms from initial-state (or final-state) collinear emission is also possible in
a generic way~\cite{Bertone:2019hks}. However, in general the
systematic inclusion of subleading logarithms is observable dependent
and some examples will be discussed in Sec.~\ref{sec:ho}. Contrary
to QCD, in QED it is typically not required to actually resum the
logarithms. The suppression by $\alpha$ is strong enough to ensure
that $\alpha\,L$ is still reasonably small. Hence, the inclusion of
the logarithms at the first few orders in $\alpha$ beyond the
fixed-order results is sufficient.

\subsection{Experimental observables}
In the following, we discuss different experimental observables and how they are affected by the LO in $\alpha$ virtual corrections, cf.\ Eq.~\ref{mcVNLO}. Of particular interest are the unpolarized cross section, the longitudinal and transverse polarization transfer observables $P_t$ and $P_l$, as well as the target and beam asymmetries $A_n$ and $B_n$, introduced in Secs.~\ref{sec:PTobservable} and \ref{sec:beam_targ_normal} respectively.

\subsubsection{Elastic lepton-proton scattering}

Let us start by presenting the tensor decomposition of elastic $\ell p$ scattering. 
Taking into account discrete symmetries, assuming parity and time reversal invariance of the electromagnetic interaction, this process can be described in terms of $6$ independent complex scalar amplitudes: $\cG_M$ and $\cF_i$ ($i=2,\dots 6)$, which are generally functions of two
independent kinematical variables, e.g., the Mandelstam variables $t=(k_1-k_2)^2$ and $s=(k_1+p_1)^2$.  The helicity amplitudes of the process can be split into a sum of helicity-conserving and helicity-flip  contributions, see, e.g.,\ Refs.~\cite{Guichon:2003qm,Gorchtein:2004ac,Gakh:2014zva},

\begin{eqnarray}
\label{eq:str_ampl}
\AMP_{h^\prime \lambda^\prime, h \lambda}  &=&  \AMP_{h^\prime \lambda^\prime, h \lambda}^{\mathrm{non-flip}}  +  \AMP_{h^\prime \lambda^\prime, h \lambda}^{\mathrm{flip}},\\
\AMP_{h^\prime \lambda^\prime, h \lambda}^{\mathrm{non-flip}} &=&  \frac{4\pi \al}{Q^2}\, \bar{u}(k_2,h^\prime) \gamma_\mu u(k_1,h)\;  \bar{N}(p_2,\lambda^\prime)\nn\\&& \times\left( \cG_M \gamma^\mu -   \cF_2 \frac{P^{\mu}}{M} +  \cF_3 \frac{\gamma \cdot K P^{\mu}}{M^2} \right)\nn\\ &&\times N(p_1,\lambda),  \label{eq:str_ampl1}  \\
 \AMP_{h^\prime \lambda^\prime, h \lambda}^{\mathrm{flip}} &=& \frac{4\pi \al}{Q^2} \frac{m_\ell}{M}\, \bar{u}(k_2,h^\prime) u(k_1,h) \; \bar{N}(p_2,\lambda^\prime)\nn\\ &&\times\left( \cF_4  +  \frac{\gamma \cdot K}{M} \cF_5  \right) N(p_1,\lambda)  \nonumber \\
 &+ &\frac{4\pi \al}{Q^2} \frac{m_\ell}{M}\, \bar{u}(k_2,h^\prime) \gamma_5 u(k_1,h) \;\nn\\ &&\times\bar{N}(p_2,\lambda^\prime) \cF_6 \gamma_5 N(p_1,\lambda), \label{eq:str_ampl2}
\end{eqnarray}

where $q=k_1-k_2=p_2-p_1$, $K=(k_1+k_2)/2$,
$P=(p_1+p_2)/2$, and $\gamma \cdot a \equiv \gamma^\mu a_\mu$.  
The $6$ complex amplitudes $\cG_M$ and $\cF_i$, sometimes called generalized FFs, fully describe the
spin structure of reaction Eq.~\ref{eq:eq1} for any
number of exchanged virtual photons. In the limit of $m_\ell \to 0$, the contribution from the helicity-flip amplitudes to observables vanishes, see Eq.~\ref{eq:str_ampl2}. Relations between the helicity amplitudes and the generalized FFs can be found f.i.\ in Ref.~\cite{Tomalak:2014dja}.

In the Born approximation, i.e., considering only the tree-level OPE diagram with proton FFs in \Figref{rc:elLP} (right panel), the number of amplitudes in \Eqref{str_ampl} reduces to $2$. To be more precise, the Born amplitudes read
\begin{eqnarray}
\cG_M^\text{Born}(Q^2, s)&=&G_{M}(Q^2), \\
\cF_2^\text{Born}(Q^2,
s)&=&F_{2}(Q^2),  \\
\cF_i^\text{Born}(Q^2, s)&=&0 \quad (i=3,\, \dots, 6),
\label{eq:eq6}
\end{eqnarray}
where $F_{2}(Q^2)$ is the Pauli
 FF of the proton. The fact that the FFs  in the space-like region are 
real functions of the virtuality of the exchanged photon is a consequence of unitarity, the spin-$1$ nature of the
virtual photon, parity conservation, and the identity of the initial
and final states, see Refs.~\cite{Halzen:1984mc,Tomasi-Gustafsson:2011sda}
for an explicit derivation.

To separate the Born contribution from effects of virtual radiative corrections to the elastic scattering, we define the following
decomposition of the scalar amplitudes:
    \begin{eqnarray}
\cG_M(Q^2, s)&=&G_M(Q^2)+\Delta\cG_M(Q^2, s), \\
 \cG_E(Q^2, s)&=&G_E(Q^2)+\Delta\cG_E(Q^2, s),\\
 \cF_i(Q^2, s)&=&\Delta\cF_i(Q^2, s) \quad (i=3,\, \dots, 6),\quad
\label{eq:eq8}
\end{eqnarray}
where $\cG_E$ has been introduced via
\beq
\cF_2(Q^2, s)=\frac{1}{1+\tau}\left[\cG_M(Q^2, s)-\cG_{E}(Q^2, s)\right].
\eeq
The order of magnitude of these quantities is given by
$ G_{M,E}\sim\alpha^0$ and 
$\Delta\cG_M(q^2,s)\sim\Delta\cG_E(q^2, s)~
\sim\Delta\cF_i(q^2, s) (i=3,\, \dots, 6)\sim\alpha  $.

\subsubsection{Laboratory frame}

The proton
and lepton four-momenta in the laboratory frame can be written as
\begin{eqnarray}\label{eq:eq2}
p_1&=&(M,\boldsymbol{0}),\\
p_2&=&(E_2, \boldsymbol{p_2}),\\
k_1&=&(\epsilon_1, \boldsymbol{k_1}),\\ k_2&=&(\epsilon_2, \boldsymbol{k_2}),
\end{eqnarray}
where three momenta are denoted by bold symbols. The scattered lepton
energy is written in terms of $\theta $ as
\begin{eqnarray}    
\epsilon_2&=& \frac{1}
{(\epsilon_1+M)^2-\boldsymbol{k_1}^{\!2}\cos^2\theta }\nonumber\\
&\times& \Bigg[(\epsilon_1+M)(M\epsilon_1 +
m_\ell^2)\nonumber\\
&&+\boldsymbol{k_1}^{\!2}\cos\theta\sqrt{ M^2-m_\ell^2\sin^2\theta}\Bigg],\qquad
\label{eq:eqe12}
\end{eqnarray}
while the Mandelstam variables $s$ and $t$ are expressed as
\begin{eqnarray}\label{eq:eqq2}
t&=&-Q^2=2m_\ell^2-2(\epsilon_1\epsilon_2-\vert\boldsymbol{k_1}\vert \vert\boldsymbol{k_2}\vert\cos\theta),\qquad\\
s&=& M^2 + m_\ell^2 + 2 M \epsilon_1.
\end{eqnarray}
The differential cross section in terms of the  matrix element squared is given by
\begin{eqnarray}
\dd\sigma&=&\frac{(2\pi )^4}{4I}\vert{\AMP}\vert^2\frac{\dd{\boldsymbol {k_2}}\,\dd{\boldsymbol{
p_2}}}{(2\pi )^64\epsilon_2E_2}\nonumber\\
&\times&\,\delta^{(4)}(k_1+p_1-k_2-p_2),
\label{eq:eq10}
\end{eqnarray}
with the invariant $I^2=(k_1\cdot p_1)^2-m_\ell^2M^2$ and the
energy of the recoil proton $E_2$.

For the case where the scattered lepton is detected in the final state, one obtains 
\begin{equation}
\frac{\dd\sigma}{\dd\Omega}=\frac{\vert \AMP \vert^2}{(4\pi)^2 \,4M}\frac{\boldsymbol{k_2}^{\!2}}{\mathfrak{D}\vert\boldsymbol{ k_1}\vert},
\label{eq:eq11}
\end{equation}
where $\mathfrak{D}=(M+\epsilon_1)\vert\boldsymbol{ k_2}\vert-\epsilon_2\vert\boldsymbol{ k_1}\vert\cos\theta $, and $\dd \Omega$ is the differential solid angle of the scattered lepton. For the case where the recoil proton
is detected in the final state, one obtains
\begin{equation}
\frac{\dd\sigma}{\dd\Omega_p}=\frac{\vert{\AMP}\vert^2}{(4\pi)^2\,4M}\frac{\boldsymbol{p_2}^{\!2}}{\bar{ \mathfrak{D}}\vert\boldsymbol{ k_1}\vert},
\label{eq:eq13}
\end{equation}
with $\bar{ \mathfrak{D}}=(M+\epsilon_1)\vert\boldsymbol{ p_2}\vert-E_2\vert\boldsymbol{ k_1}\vert\cos\theta_p $, where $\theta_p $  is the angle between the directions 
of the lepton beam and the recoil proton, and $\dd \Omega_p$ is the differential solid angle of the scattered proton. Using the relation
\beq
\dd Q^2=\vert\boldsymbol{ k_1}\vert \vert\boldsymbol{ p_2}\vert\frac{1}{\pi}\frac{E_2+M}{\epsilon_1+M}\dd\Omega_p, 
\eeq
we obtain the following expression for the differential cross section as a function of $Q^2$
\begin{equation}
\frac{\dd\sigma}{\dd Q^2}=\frac{\vert \AMP\vert^2}{64\pi M\,\boldsymbol{k_1}^{\!2}}.
\label{eq:eq14}
\end{equation}

\subsubsection{Unpolarized cross section}

The interference of the Born diagram (\Figref{rc:elLP}, right panel) with any higher-order in $\alpha$ diagram of elastic $\ell p$ scattering (\Figref{rc:elLP}, left panel), can be expressed through a multiplicative correction, 
\begin{eqnarray} \label{eq:delta_TPE_massive}
&&\delta^{(1)}_{\rm v} = \frac{2}{ G_M^2 + \frac{\varepsilon}{\tau} G_E^2}
\bigg[ G_M \re \Delta\cG_1 + \frac{\varepsilon}{\tau} G_E \re \Delta\cG_2 \nonumber \\
&&\hspace{-0.4cm}+(1-\varepsilon_T)\left(\frac{\varepsilon_\ell}{\tau}\frac{\nu}{M^2}  G_E \re \Delta\cG_4  - G_M \re \Delta\cG_3 \right) \bigg],
\end{eqnarray}
 to the $\mathcal{O}(\al^2)$ Born cross section in \Eqref{reducedCS}.  
Here, $\re$ denotes the real part of the auxiliary amplitudes
\begin{eqnarray}\eqlab{cGamps}
 \cG_1 & = & \cG_M + \frac{\nu}{M^2} \cF_3 + \frac{m_\ell^2}{M^2} \cF_5, \label{eq:G1} \\
 \cG_2 & = & \cG_M - ( 1 + \tau ) \cF_2 + \frac{\nu}{M^2} \cF_3,  \\
 \cG_3 & = & \frac{m_\ell^2}{M^2} \cF_5 + \frac{\nu}{M^2} \cF_3,   \\
 \cG_4 & = &    \cF_4 + \frac{\nu}{M^2 (1+\tau)} \cF_5,\label{eq:G4}
\end{eqnarray}
 and
$\varepsilon_T$ is
the degree of linear polarization of transverse photons
\beq
\label{eq:transverse_epsilon}
\varepsilon_T= \frac{\nu^2  - M^4 \tau ( 1 + \tau ) ( 1 + 2 \varepsilon_\ell )}{ \nu^2 + M^4 \tau ( 1 + \tau )  ( 1 - 2 \varepsilon_\ell )}.
\eeq
In this way, one can also describe the leading TPE correction, cf.\ Eq.~\ref{delta2gTPE}. 

\subsubsection{Polarization transfer observables}\seclab{PTobservable}

The longitudinal and transverse polarization transfer asymmetries, $P_l$ and $P_t$, are defined as (with, e.g., $h=+$) 
\begin{eqnarray}
\eqlab{Pl}
P_l&=&\frac{\dd \sigma(\lambda'=+)-\dd \sigma(\lambda'=-)}{\dd \sigma(\lambda'=+)+\dd \sigma(\lambda'=-)},\\
P_t&=&\frac{\dd \sigma(S'=S_\perp)-\dd \sigma(S'=-S_\perp)}{\dd \sigma(S'=S_\perp)+\dd \sigma(S'=-S_\perp)},\quad
\eqlab{Pt}
\end{eqnarray}
where $S'=\pm S_\perp$ is the spin direction of the recoil proton in the scattering plane transverse to its momentum.
In the Born approximation, their ratio is related to
the ratio of electric to magnetic 
Sachs FFs \cite{Akhiezer:1968ek,Akhiezer:1974em}
\begin{eqnarray}
\eqlab{PTOPE}
\frac{P_t}{P_l}=-\sqrt{\frac{2\varepsilon}{\tau (1+\varepsilon)}} \frac{G_E}{G_M}.
\end{eqnarray}
Equivalent information on the proton FFs can be obtained from measuring double-spin asymmetries on a polarized proton target \cite{Dombey1969}. Taking into account the leading virtual corrections to the asymmetries, the ratio modifies, up to terms of order $\mathcal{O}(\alpha^2)$, as described in Refs.~\cite{Guichon:2003qm,Guttmann:2010au,Meziane:2010xc} as
\begin{align}
\label{PtPlradcorr}
\frac{P_t}{P_l}=-\sqrt{\frac{2\varepsilon}{\tau (1+\varepsilon)}} \Bigg[&R_{EM} +Y_E - R_{EM} Y_M \nn\\&  +\Bigg (1-\frac{2 \varepsilon}{1+\varepsilon} R_{EM} \Bigg )Y_3 \Bigg],
\end{align}
with
\begin{align}
R_{EM} \equiv \frac{G_E}{G_M},
\eqlab{REM}
\end{align}
where the shorthand notations for ratio of two-photon amplitudes relative to the magnetic FF have been used: 
\begin{align}
&Y_M \equiv \frac{\re \Delta \cG_M }{G_M}, \\ 
&Y_E \equiv \frac{\re \Delta \cG_E }{G_M}, \\ 
&Y_3 \equiv \frac{\nu}{M^2}\frac{\re \cF_3}{G_M}.
\label{eq:2gaampl}
\end{align}
Furthermore for the longitudinal polarization transfer $P_l$ separately, its expression relative to the $1 \gamma$-result $P_l^\text{Born}$ 
is given in Refs.~\cite{Guichon:2003qm,Guttmann:2010au}, up to terms of order $\mathcal{O}(\alpha^2)$, by
\begin{eqnarray}
\frac{P_l}{P_l^\text{Born}}  & =    1 - 2 \varepsilon 
\left( 1 + \frac{\varepsilon}{\tau} R_{EM}^2  \right)^{-1} 
\nn \\
&\times \left\{ 
\left[ \frac{\varepsilon}{1 + \varepsilon} \bigg( 1 - \frac{R_{EM}^2}{\tau}  \right) 
+ \frac{R_{EM}}{\tau}  \right] Y_3 
\nn \\ 
&+  \frac{R_{EM}}{\tau} \left[  Y_E  - R_{EM} Y_M \right]  \bigg\},  	\label{pl_2gamma}  
\end{eqnarray}
with
\begin{align}
& P_l^\text{Born} = \sqrt{1 - \varepsilon^2} \, 
\left( 1 + \frac{\varepsilon}{\tau} R_{EM}^2 \right)^{-1}, 
\eqlab{pl_1gamma} 
\end{align}
Note that in Eqs.~\ref{PtPlradcorr} and \ref{pl_2gamma} the lepton mass has been neglected, $m_\ell=0$, which is not a valid approximation for low-energy muon scattering \cite{Tomalak:2016kyd}.

\subsubsection{Beam and target normal single-spin asymmetries}\seclab{beam_targ_normal}
\label{SSAintro}
The target (beam) normal single-spin asymmetries (SSA) $A_n$ ($B_n$) is defined for the scattering of an unpolarized beam (a beam polarized normal to the scattering plane, with $s= \pm s_n$ the direction of incoming lepton,) off a target with normal to the scattering plane polarization $S= \pm S_n$ (an unpolarized target) 

  \begin{eqnarray}
     A_n&=&\frac{\text{d} \sigma (S=S_n)-\text{d} \sigma (S=-S_n)}{\text{d} \sigma (S=S_n)+\text{d} \sigma (S=-S_n)},\\
     B_n&=&\frac{\text{d} \sigma (s=s_n)-\text{d} \sigma (s=-s_n)}{\text{d} \sigma (s=s_n)+\text{d} \sigma (s=-s_n)}.
 \end{eqnarray}
 Both these asymmetries are zero in the  Born  approximation. Taking into account leading virtual corrections, the  asymmetries can be expressed as~\cite{Gorchtein:2004ac,Chen:2004tw,Afanasev:2005prd}
 \begin{eqnarray}
A_n&=&\sqrt{\frac{2(\varepsilon -\varepsilon_\ell)(1+\varepsilon-2\varepsilon\varepsilon_\ell)}{\tau (1-\varepsilon_\ell)^2}}
\left(1 +\frac{\varepsilon}{\tau} R_{EM}^2 \right)^{-1} \nn\\
&\times& \frac{1}{G_M} \Bigg \{- \im \bigg (\Delta \cG_2+\frac{(1+\tau)m_\ell^2}{\nu}  \cG_4\bigg ) \nn\\
&&+ R_{EM} \im \bigg(\Delta \cG_1-\frac{\tau (1+\tau) M^2}{\nu} \cF_3 \bigg ) \Bigg \},\\
B_n&=& -\frac{m_\ell}{M} \sqrt{ \frac{2 (1-\varepsilon)(\varepsilon-\varepsilon_\ell)}{ (1-\varepsilon_\ell)^2}} \frac{\sqrt{1+\tau}}{\tau} 
\left(1+\frac{\varepsilon}{\tau} R_{EM}^2 \right)^{-1} \nonumber \\
&\times& \frac{1}{G_M} \Big \{\tau \im \bigg( \cF_3+\frac{\nu}{M^2}\frac{\cF_5}{1+\tau} \bigg) + R_{EM} \im \cG_4 \Big \},\nn\\
\end{eqnarray}
 
where $\im$ denotes the imaginary part.

\section{Two-photon exchange}\label{sec:TPE}

The TPE contribution to lepton scattering, shown in \Figref{TPE_diagram}, is of particular importance for two main reasons: Firstly,  hadronic corrections, cf.\ the blob in \Figref{TPE_diagram} (bottom panel), are notoriously difficult to calculate and often have a large relative uncertainty. Secondly, as will be explained in Sec.~\ref{sec:RoPTcomp}, the OPE may not be a good approximation for the extraction of FFs from the unpolarized cross section at large $Q^2$. Therefore, the leading TPE correction 
\begin{equation}
\label{delta2gTPE}
\delta_{2\gamma}   \simeq   \frac{2\re \big(\AMP^{(0)*} \AMP_{2\gamma}\big)} {|\AMP^{(0)}|^2},
\end{equation}
following from the interference of the Born and TPE amplitudes, $\AMP^{(0)}$ and $\AMP_{2\gamma}$, has to be taken into account appropriately. A complete calculation should go beyond the soft-photon approximation and include the hard TPE with all possible intermediate states. In the last decade, predictions of the TPE correction have been systematically improved and model dependence has been reduced by employing dispersion relations and effective field theories.

This part of the paper is organized as follows. In Sec.~\ref{sec:RoPTcomp}, we illustrate the importance of the TPE by comparing Rosenbluth and polarization-transfer measurements of the proton FFs. In Sec.~\ref{sec:TPEtheory}, we review theoretical predictions for the leading TPE correction, distinguishing between proton and inelastic intermediate states, as well as regions from small to large momentum transfer. In Sec.~\ref{sec:EFTCalc}, we discuss effective field theory calculations. In Sec.~\ref{sec:TPEempirical}, empirical extractions of TPE corrections and amplitudes are presented and updated based on new data.  In Sec.~\ref{Sec:Experiments}, results of past experiments  aimed at extracting the TPE  are presented and compared to theoretical predictions. We finish with an outlook on future experiments and theory advances in Sec.~\ref{sec:TPEOutlook}. For further reading, we refer to the following reviews in Refs.~\cite{Carlson:2007sp, Arrington:2011dn,Afanasev:2017gsk,Borisyuk:2019gym} and the recent CFNS whitepaper, Ref.~\cite{Afanasev:2020hwg}.

\subsection{Rosenbluth vs. polarization transfer experiments}\seclab{RoPTcomp}

\begin{figure}[ht]
\includegraphics[width=\columnwidth]{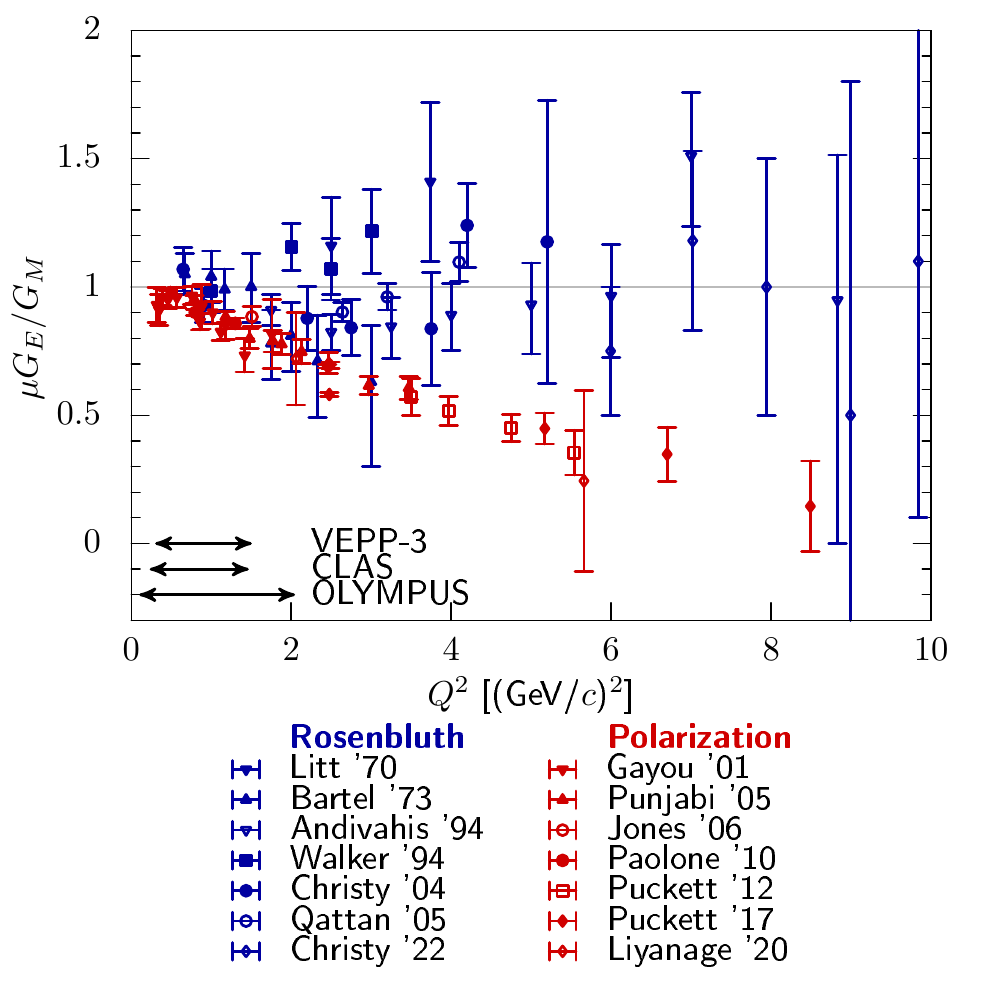}
\caption{Ratio of proton electric and magnetic FFs, $G_E \,\mu_p/G_M$, with $\mu_p$ the magnetic moment of the proton, as obtained from two types of measurements: Rosenbluth separation \cite{LITT197040,BARTEL1973429,Andivahis:1994rq,PhysRevD.49.5671,PhysRevC.70.015206,Qattan:2004ht,Christy:2021snt} and polarization transfer \cite{Gayou:2001qt,Punjabi:2005wq,ResonanceSpinStructure:2006oim,Paolone:2010qc,Puckett:2011xg,Puckett:2017flj,SANE:2018cub}. Shown are a subset of the available data to not overcrowd the figure. The Rosenbluth result only contains the TPE corrections in the form of the well-known Maximon-Tjon radiative corrections \cite{Maximon:2000hm} (no hard TPE was included), whereas for the polarized extraction no radiative corrections were taken into account. }
\label{fig.GEGM}
\end{figure}

Polarization experiments in elastic $ep$ scattering at Jefferson Lab Hall A \cite{JeffersonLabHallA:1999epl,Gayou:2001qt,Punjabi:2005wq,Puckett:2011xg}, revealed that the ratios of proton FFs, $G_E(Q^2)/G_M(Q^2)$, extracted based on the Rosenbluth \cite{Rosenbluth:1950yq} or polarization transfer methods \cite{Akhiezer:1968ek,Akhiezer:1974em} in the OPE approximation deviate with increasing $Q^2$. The FF ratio from Rosenbluth extractions is nearly constant as a function of $Q^2$, while it decreases for polarization transfer measurements, as can be seen in Fig.\ \ref{fig.GEGM}. In the limit of OPE, one defines an unpolarized reduced cross section that is linear in $\varepsilon$ and the corresponding ratio of measurements with transverse or longitudinal polarization of the recoiling proton is constant in $\varepsilon$  as given in Eqs.~\ref{eq:reducedCS} and \ref{eq:PTOPE}. What is measured in experiment, however, will always include radiative corrections. Taking these into account and under some assumptions, in presence of TPE, the expressions modify according to Eqs.~\ref{eq:delta_TPE_massive} and \ref{PtPlradcorr}, respectively.
One can see that after inclusion of the TPE correction, the interpretation of what is measured in Rosenbluth and polarization transfer experiments changes.  
Model independent considerations  \cite{Rekalo:2003xa,Rekalo:2003km,Rekalo:2004wa} show that it is still possible to recover experimentally the electric and magnetic proton FFs, even in presence of TPE, but this requires either the measurement of three time-odd or five time-even polarization observables (including triple spin observables, of the order of $\alpha$) or, alternatively, the generalization of the Akhiezer-Rekalo recoil proton polarization method  \cite{Akhiezer:1968ek,Akhiezer:1974em} with longitudinally polarized electrons and positron beams in identical kinematical conditions.

In a series of papers in the early 2000's, it has been shown that the inclusion of ``hard'' TPE may reconcile the FF extractions from polarized and unpolarized $\ell p$ scattering, within some assumptions (real-valued FFs and  $\varepsilon$ linearity of the TPE contribution). Guichon and Vanderhaeghen identified in Ref.~\cite{Guichon:2003qm} the experimental observables used to extract the $G_E/G_M$ ratio as the $\varepsilon$-slope in the reduced cross section
\begin{align}
\sigma_R(\varepsilon, Q^2)
= G_{M}^2 \left(1 + \frac{\varepsilon}{\tau}  
R_{EM}^2 \right) \left\{ 1 + \delta_{2 \gamma}\right\} .
\label{eq:sigmaR}
\end{align}
Neglecting lepton mass terms and using the shorthand notations for the TPE amplitudes of 
Eq.~\ref{eq:2gaampl}, $\delta_{2\gamma}$ 
has the form
\begin{align}
&\delta_{2 \gamma}(\varepsilon, Q^2) = \left(1 + \frac{\varepsilon}{\tau}  R_{EM}^2 \right)^{-1} 
\nn \\ &
\times \left\{ 2 \, Y_M
+ 2 \varepsilon  \frac{R_{EM}}{\tau} \, Y_E + 2 \varepsilon \left( 1 + \frac{R_{EM}}{\tau} \right) Y_3 \right\}.
\label{crossen} 
\end{align} 
On the other hand, the polarization transfer experiments yield the ratio given by Eq.~\ref{PtPlradcorr}
\begin{align}
-\sqrt{\frac{\tau (1+\varepsilon)}{2\varepsilon}}  \frac{P_t}{P_l}&= R_{EM} +Y_E - R_{EM} Y_M \nn\\&  +\Bigg (1-\frac{2 \varepsilon}{1+\varepsilon} R_{EM} \Bigg )Y_3.
\eqlab{Rratios}
\end{align}
Reference \cite{Guichon:2003qm} showed that the TPE in the Rosenbluth method effectively corrects a small number, cf.\ the 
$\varepsilon$ dependent term in Eq.~\ref{eq:sigmaR} with $R_{EM}^2(0)\sim (1/2.79)^2 \sim 0.13$ which has an additional $1/\tau$ suppression at large $Q^2$. The FF ratio extracted from $P_t/P_l$ is only mildly affected by radiative corrections as it involves asymmetries,  whereas the cross sections themselves are strongly modified by radiative corrections at moderately and large momenta.  
Around the same time, Blunden et al.\  in Ref.~\cite{Blunden:2003sp} showed with a simple hadronic model calculation, including the finite size of the proton, that the dominant TPE with a proton intermediate state allows to partially resolve the discrepancy.
Inclusion of the $\Delta(1232)$ resonance further improved the agreement, as shown by Kondratyuk et al.\ in Ref.~\cite{Kondratyuk:2005kk}. A similar conclusion was drawn by Chen et al.\ and Afanasev et al.\ in Refs.~\cite{Chen:2004tw,Afanasev:2005prd} respectively, with a partonic model calculation.

 Note that alternative explanations have been put forward, such as different calculations of radiative corrections, including also  lepton structure functions \cite{Gramolin:2016hjt,Bystritskiy:2006ju}, correlations between the Rosenbluth parameters \cite{Tomasi-Gustafsson:2006dhf},
or acceptance problems in the analysis or experiment \cite{Pacetti:2016tqi}.

\begin{figure}[ht]
\includegraphics[width=\columnwidth]{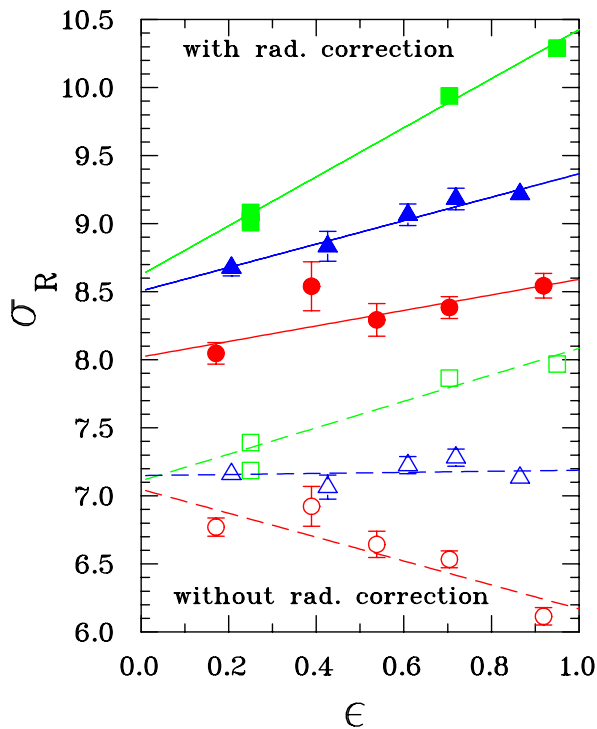}
\caption{Rosenbluth plot for the data from Ref.~\cite{Andivahis:1994rq} at momentum transfers of 1.75 (green squares), 3.25 (blue triangles) and 5 GeV$^2$ (red circles). Empty and full symbols show data before and after radiative corrections \cite{Mo:1968cg} (no hard TPE included), respectively. Figure taken from Ref.~\cite{Perdrisat:2006hj}.
}
\label{Fig:AndiRad}
\end{figure}

The electric FF is determined by the slope of the Rosenbluth plot, at fixed $Q^2$, derived from the radiatively corrected cross section. Corrections at large $Q^2$ may reach 40\% \cite {Andivahis:1994rq} and the $\varepsilon$-slope may essentially change. Even more dramatically, above 3 GeV$^2$ the slope of the measured cross section may even be negative before corrections, cf.\ Fig.~\ref{Fig:AndiRad} \cite{Tomasi-Gustafsson:2004fss,Perdrisat:2006hj} where the effect of radiative corrections \cite{Mo:1968cg}, including ``soft'' but no ``hard'' TPE, is shown. Note that $G_E$ must be real in the space-like region. This means that $G_E^2$ is essentially determined by the $\varepsilon$-dependence of the applied radiative corrections. Different calculations show a difference of few percent in the slope, already at NLO \cite{Gramolin:2014pva,Gerasimov:2015aoa}.

Concerning polarization observables, it is commonly assumed that the FF ratio given by polarization measurements is not (or less) affected by radiative corrections, giving therefore a more reliable result on this ratio than the one extracted by the Rosenbluth method. 
The polarization asymmetries $P_t$ and $P_l$ are ratios of cross sections, cf.\ Eqs.~\ref{eq:Pl} and \ref{eq:Pt}, in which the bulk of the radiative corrections that factorize (virtual corrections on lepton side and soft-photon emission corrections) drop out. Therefore, they are less sensitive to radiative corrections than the unpolarized cross section.

Already in the 1970's, a reason why the TPE correction could become important at large $Q^2$ was put forward \cite{Gunion:1972bj,Boitsov:1972if,Franco:1973uq}:
when the transferred momentum is equally shared between the two photons the scaling in $\alpha$ may be compensated by the steep decrease of the FFs with $Q^2$. The effect is therefore expected to increase with $Q^2$ and with the hadron mass.
The advent of the high duty cycle and highly polarized electron beam at the Jefferson Lab, together with the availability of large solid angle spectrometers and hadron polarimetry in the GeV region, opened the way to high-precision experiments in elastic and inelastic electron-hadron scattering. Two experiments of elastic electron-deuteron ($ed$) scattering at large $Q^2$ in Hall A \cite{JeffersonLabHallA:1998xrv} and Hall C \cite{JeffersonLabt20:1998vrh} claiming an error of 5\%, showed a discrepancy of up to 15\% in the elastic $ed$ cross section at the same $Q^2$, but different energies and angles. A possible explanation was brought up that the TPE contribution could be at the origin of these findings \cite{Rekalo:1999mt}. Finally the discrepancy was attributed to a systematic error of the Hall C spectrometer setting, as no $Q^2$ dependence was observed.

\subsection{Theoretical predictions} \seclab{TPEtheory}

\begin{figure}[ht]
\centering
\includegraphics[scale=0.45,angle=0]{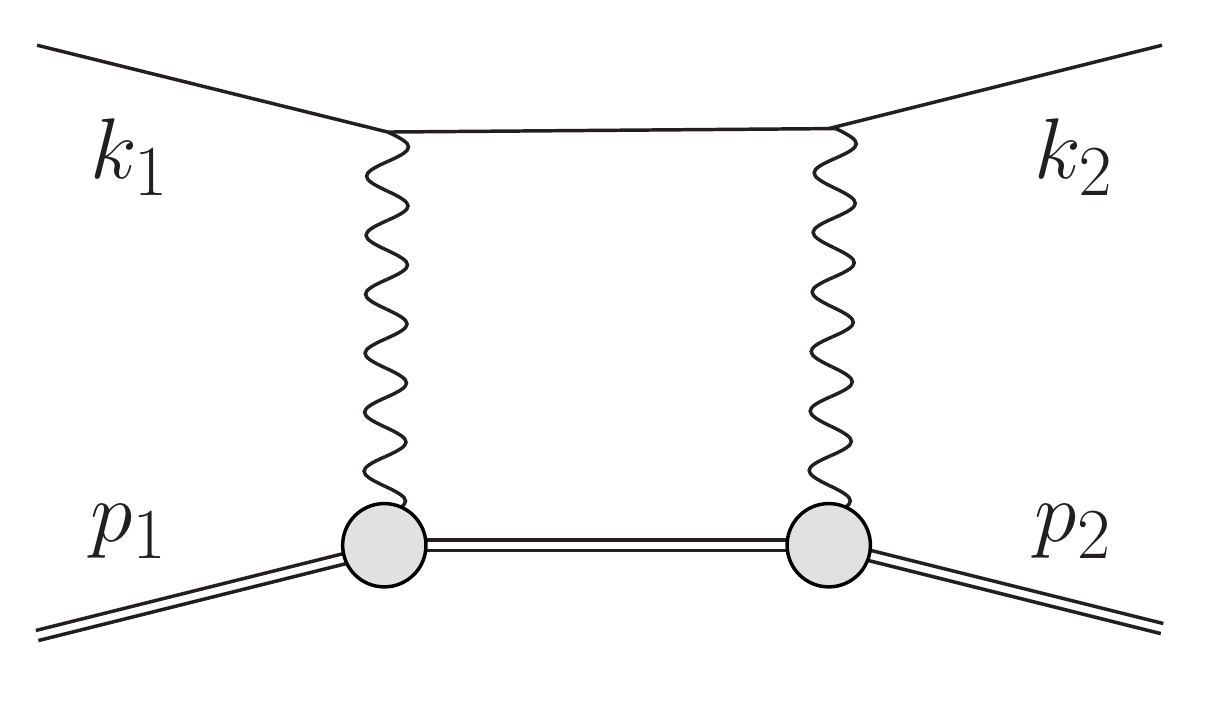}\\
\vspace{0.5cm}
\includegraphics[scale=0.45,angle=0]{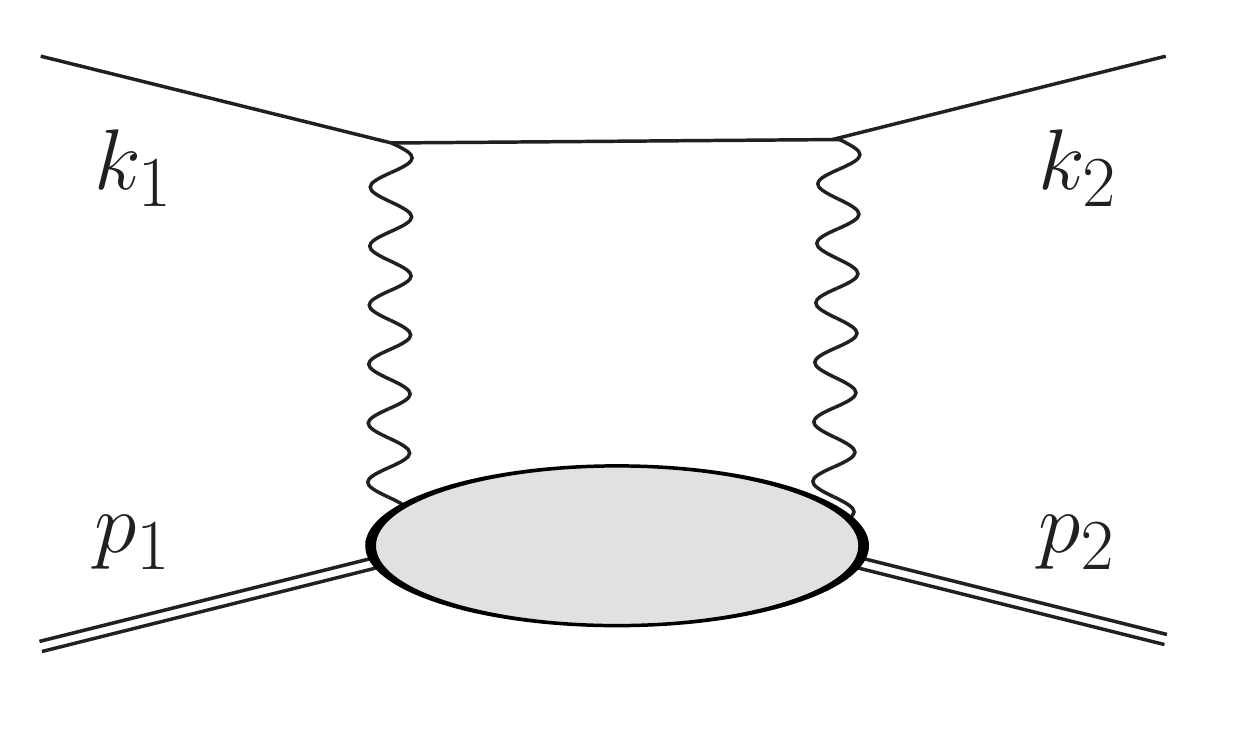}
\caption{Off-forward TPE contribution to elastic $\ell p$ scattering with proton intermediate state (top panel), where the circles represent the proton FFs, and inelastic intermediate states (bottom panel), where the blob represents all possible excitations. The horizontal lines correspond to the lepton and proton (double line). Crossed TPE  diagrams are implied in all references to this figure. \label{fig:TPE_diagram}}
\label{OffTPE}
\end{figure}

In this subsection, we review theoretical predictions for the TPE contributions to $\ell p$ scattering. We will refer to the (standard) TPE approximations by Maximon-Tjon \cite{Maximon:2000hm} (or Mo-Tsai \cite{Mo:1968cg}), conventionally included in the experimental analysis, as ``soft'' TPE. (Non-standard) refinements of the TPE, that need to be included given the accuracy of modern scattering experiments, will be referred to below as ``hard'' TPE. Starting from the so-called elastic TPE with a proton intermediate state, Fig.~\ref{OffTPE} (top panel), we subsequently discuss TPE contributions with inelastic intermediate states, Fig.~\ref{OffTPE} (bottom panel), in regions of small, medium and large momentum transfer. Furthermore, we discuss the forward TPE as the limiting factor in the theoretical description of the energy spectra of light muonic atoms.

\def\Re{{\rm Re}}
\subsubsection{Infrared subtraction schemes}\seclab{IRsubtraction}

When plotting the elastic TPE contribution to $\delta_{2\gamma}$ or $R_{2\gamma}$, defined in Eqs.~\ref{delta2gTPE} and \ref{eq:R2gammadef}, one has to somehow treat the IR divergences they contain.
Of course the only way that leads to a measurable quantity is to combine them with the corresponding real corrections as discussed in Sec.~\ref{Sec:RadCorrAdrian}.
However, this can be quite cumbersome in practise and makes comparing different TPE scenarios and parametrisations needlessly complicated.
Instead, one defines an IR subtraction scheme that removes the IR divergence (and potentially some finite parts as well).
The exact form of this subtraction term is somewhat arbitrary and solely determined by conventions since the result is no longer a physical quantity.

A common way of doing this in the high-energy community is the definition of the Catani $I$ operator~\cite{Catani:1998bh} that simply removes the $1/\epsilon$ poles in dimensional regularisation.

In the QED community it is not uncommon to add the first terms of Eqs.~\ref{slicing} or \ref{subtraction}
\begin{eqnarray}
    \delta_{2\gamma}^{\rm eik} &= \delta_{2\gamma} +&
          \int_{E_\gamma < \delta} \dd\Phi_\gamma\ \mathcal{E} \nn\\
    & = \delta_{2\gamma} +&
          \underbrace{
              \int \dd\Phi_\gamma\Big(\frac{p_1\cdot k_1}{p_1\cdot k\ k_1\cdot k}+\frac{p_2\cdot k_2}{p_2\cdot k\ k_2\cdot k}}_{\hat{\mathcal{E}}(s)}\nn\\
              &&
         \hspace*{1cm} - (s\rightarrow u)\Big).
\end{eqnarray}
This scheme, sometimes called eikonal subtraction~\cite{Engel:2019nfw}, has the added advantage of giving the remnant a physical interpretation as long as the parameter $\delta$ is chosen small enough -- it just approximately includes real corrections up to the cut-off $\delta$.
The calculation of this integral is somewhat involved but the result is well known~\cite{Engel:2019nfw,Frederix:2009yq,Ulrich:2020phd}.
Since this physical interpretation exists, it also means that the result needs to be independent of how the TPE calculation was regularised, making comparisons between different methods easier.
The TPE community usually follows either Mo-Tsai (MoT) \cite{Mo:1968cg} or Maximon-Tjon (MTj) \cite{Maximon:2000hm}.
Note that in this paper, we use the MTj convention for our Figs.~\ref{fig.Dsig}-\ref{fig.NearForw}. The latter defines the IR-subtracted TPE contribution as
\begin{align}    &\delta_{2\gamma}^{\rm MTj} = \delta_{2\gamma} -
        \frac{2 \alpha}{\pi} q^2 \,\Re\Bigg[p_1\cdot k_1 \nn\\
        &\times\int  \frac{\dd \ell}{
            \ell^2
            (\ell-q)^2
            \Big[(\ell-k_1)^2-m^2\Big]
            \Big[(\ell+p_1)^2-M^2\Big]
        } \nn\\&\qquad\quad - (s\rightarrow u)\Bigg].
\end{align}
This integral can be evaluated both using photon-mass regularisation or dimensional regularisation
    \begin{eqnarray}
    \delta_{2\gamma}^{\rm MTj} &=& \delta_{2\gamma} +
    \frac{2 \alpha}{\pi}\,\Re\Bigg[\nn\\
        \frac{m^2+M^2-s}{\sqrt{\lambda_s}}
        & \times&\ln\frac{m^2+M^2-s + \sqrt{\lambda_s}}{2mM}
    \begin{cases} \ln\frac{m_\gamma^2}{Q^2} \\ \frac1{\epsilon}\Big(\frac{\mu^2}{Q^2}\Big)^\epsilon \end{cases}\nn\\
    &
    &- (s\rightarrow u)\Bigg],
\end{eqnarray}
where we have defined $\lambda_s = m^4 + (M^2 - s)^2 - 2 m^2 (M^2 + s)$.
In the limit of $m^2\ll s,M^2,Q^2$, the expression for $e^- p$ scattering simplifies to
    \begin{equation}
    \delta_{2\gamma}^{\rm MTj} = \delta_{2\gamma} +
    \frac{2 \alpha}{\pi}
        \ln\Big|\frac{s - M^2}{u - M^2}\Big|
    \begin{cases} \ln\frac{m_\gamma^2}{Q^2} \\ \frac1{\epsilon}\Big(\frac{\mu^2}{Q^2}\Big)^\epsilon \end{cases}\!\!\!\!\!\!
    .
\end{equation}
To derive the MoT subtraction prescription, we instead start by replacing either the $\ell^2$ propagator or the $(\ell-q)^2$ propagator with $1/q^2$.
The idea is to mimic the soft behaviour of the integral as either the $\ell^2$ propagator goes to zero or the $(\ell-q)^2$ propagator.
This procedure results in two identical triangle functions
    \begin{align}
    &\delta_{2\gamma}^{\rm MoT'} = \delta_{2\gamma} -
        \frac{2\alpha}{\pi} q^2 \Re\Bigg[p_1\cdot k_1 \nn\\
      &  \hspace{-0.5cm}\times\int  \frac{\dd \ell}{
            \ell^2
            q^2
            \Big[(\ell-k_1)^2-m^2\Big]
            \Big[(\ell+p_1)^2-M^2\Big]
        } \nn\\&\qquad\quad - (s\rightarrow u)\Bigg].
        \label{eq:motorgi}
\end{align}
Note that this is not yet the correct MoT prescription, hence {\rm MoT$'$}.
The loop integral is slightly more involved than the previous one but can still be trivially written in terms of the well-known Ellis-Zanderighi functions~\cite{Ellis:2007qk}.
In this case, we need the integral $I_6^{\rm fin}(s;m^2,M^2)$ commonly referred to as Triangle 6,
\begin{eqnarray}
    \delta_{2\gamma}^{\rm MoT'} &=& \delta_{2\gamma} +
        \frac{2 \alpha}{\pi} \Bigg\{
            (m^2 + M^2 - s) \Re\Bigg[
                I_6^{\rm fin}(s;m^2,M^2) \nn\\
                &+& \frac1{\sqrt{\lambda_s}}\ln\frac{m^2+M^2-s + \sqrt{\lambda_s}}{2mM} 
            \begin{cases} \ln\frac{m_\gamma^2}{mM} \\ \frac1{\epsilon}\Big(\frac{\mu^2}{mM}\Big)^\epsilon \end{cases}\!\!\!\!\!\!\Bigg]\nn\\
        &-& (s\rightarrow u)\Bigg\}.
\end{eqnarray}
If one chooses, $I_6^{\rm fin}$ could be expressed in terms of logarithms and dilogarithms.
The resulting expression is not overly complicated but reproducing it here serves no practical purpose.
However, in the limit of $m^2\ll s,M^2,Q^2$ the expression is fairly compact
\begin{eqnarray}
    I_6^{\rm fin}(s;m^2,M^2) &=& -\frac{y}{2M^2}\Bigg(
        -\frac13\big(i\pi+\log(-y^3)\big)^2\nn\\ &+ &2\ln(-y) \ln\frac y{1+y} - 2{\rm Li}_2(-y) \nn\\&-& \ln(-y)\ln\frac{m^2}{M^2}
    \Bigg) + \mathcal{O}(m^2)\,,
\end{eqnarray}
where we have defined $y=M^2/(s-M^2)$ and used the usual definitions of the $\zeta$ function and dilogarithm.

To arrive at the correct MoT prescription~\cite{Mo:1968cg} (cf. also~\cite{Arrington:2011dn}) we need to flip $k_1\to-k_1$ in the first term of~\eqref{eq:motorgi}, resulting in
\begin{align}
    &\delta_{2\gamma}^{\rm MoT} = \delta_{2\gamma} -
        \frac{2\alpha}{\pi} q^2 \Re\Bigg[\nn\\
      &  \hspace{-0.5cm}\phantom{+}\int \dd \ell \frac{-p_1\cdot k_1 }{
            \ell^2
            q^2
            \Big[(\ell+k_1)^2-m^2\Big]
            \Big[(\ell+p_1)^2-M^2\Big]
        } \nn\\
      &  \hspace{-0.5cm}+\int \dd \ell \frac{-p_1\cdot k_2 }{
            \ell^2
            q^2
            \Big[(\ell+k_2)^2-m^2\Big]
            \Big[(\ell+p_1)^2-M^2\Big]
        } \Bigg].
\end{align}
This operation is equivalent to setting $s\to s'=2 m^2 + 2 M^2 - s$, which spoils the crossing symmetry but eliminates the $(i\pi)^2$ from the analytic continuation of the logarithm.
The resulting expression can still be written in terms of $I_6^{\rm fin}$ but now in terms of $s'$ rather than $s$
\begin{align}
    \delta_{2\gamma}^{\rm MoT} &= \delta_{2\gamma} +
        \frac{2 \alpha}{\pi} \mathrm{Re}\Bigg [
              (s - m^2 - M^2) I_6^{\rm fin}(s';m^2,M^2) \nn\\
            &\qquad+ (u - m^2 - M^2) I_6^{\rm fin}(u;m^2,M^2)\nn\\
            &+\Bigg[
                  \frac{s-m^2-M^2}{\sqrt{\lambda_{s}}}\ln\frac{-m^2-M^2+s + \sqrt{\lambda_{s}}}{2mM}\nn\\
               &+\frac{u-m^2-M^2}{\sqrt{\lambda_{u}}}\ln\frac{ m^2+M^2-u + \sqrt{\lambda_{u}}}{2mM}
            \Bigg]\nn\\
            &\begin{cases} \ln\frac{m_\gamma^2}{mM} \\ \frac1{\epsilon}\Big(\frac{\mu^2}{mM}\Big)^\epsilon \end{cases}\!\!\!\!\!\!\Bigg\} \Bigg].
\end{align}
The symmetry of the $\lambda$ functions means that $\lambda_{s'}=\lambda_s$, which helps to simplify the result.
The difference between $\delta_{2\gamma}^{\rm MoT}$ and $\delta_{2\gamma}^{\rm MoT'}$ is
\begin{eqnarray}
    &\delta_{2\gamma}^{\rm MoT'}-\delta_{2\gamma}^{\rm MoT} = \dfrac{2\alpha}\pi (m^2+M^2-s)\Re\Bigg[\nn\\
            &+I_6^{\rm fin}(s';m^2,M^2) + I_6^{\rm fin}(s;m^2,M^2)\Bigg].
\end{eqnarray}
Crucially, the poles agree after taking the real part so that the IR cancellation in $\delta_{2\gamma}$ is unaffected.

\subsubsection{Proton intermediate state} \label{subsec:proton_state}

The first terms in the low-$Q$ expansion of the TPE with proton intermediate state can be described model independently. The leading term is given by the well-known Feshbach correction
 \cite{McKinley:1948zz}
\begin{equation} \eqlab{Feshbach}
    \delta_{\text{F}}=Z \alpha \pi \frac{\sin \frac{\theta}{2}-\sin^2 \frac{\theta}{2}}{\cos^2 \frac{\theta}{2}},
\end{equation}
which stems from the interaction between a massless lepton and a structureless proton.  Inelastic intermediate states start to contribute in the subleading $Q^2 \ln Q^2$ term only \cite{Brown:1970te}, see discussion in Sec.~\ref{SmallMT}. It follows that TPE could become larger in lepton scattering on heavy ions, inducing  a large charge asymmetry even at small angles \cite{Kuraev:2009hj}.

During the last two decades,  predictions of the TPE contribution improved considerably. The simplest approach to evaluate the elastic TPE is with a hadronic model calculation of the box (and crossed-box) diagrams in Fig.~\ref{OffTPE} (top panel), assuming on-shell proton FFs, see for instance Refs.~\cite{Blunden:2003sp} and \cite{Tomalak:2014dja} for electron and muon scattering, respectively. A way to avoid model dependence is to instead use a dispersive approach to describe the scalar amplitudes $\cG_i$ and $\cF_i$, introduced in Eqs.~\ref{eq:str_ampl} and \ref{eq:cGamps}. 
Dispersion relations (DRs) allow one to express the real part of the amplitudes via their imaginary part, and the latter, using unitarity, can be related to physical observables. The $s-$channel cut in the elastic TPE diagram then permits the use of on-shell proton FFs. 
First dispersive evaluations of the elastic TPE in $ep$ scattering, neglecting the electron mass, used unsubtraced DRs for the scalar amplitudes \cite{Borisyuk:2006fh,Borisyuk:2008es}. Considering the case of $\mu p$ scattering, in which the lepton mass cannot be neglected, $\cF_4$ will require a once-subtracted DR \cite{Tomalak:2018jak}.

The contribution from TPE with inelastic intermediate states, discussed in the following, is becoming important with increasing momentum transfer. It has been  suggested that further subtractions, e.g., in the DR of the $\cF_3$ amplitude, can be introduced in order to minimize model-dependence from higher intermediate states \cite{Tomalak:2014sva}. Sum rules, which are exact in QED and approximated in QCD, give indications on the contribution of intermediate states \cite{Kuraev:2006ym}. 

\subsubsection{Zero momentum transfer}\seclab{VVCS}
In the limit of zero momentum transfer, $Q^2=0$, the TPE is not important as a radiative correction to the scattering process. $\delta_{2\gamma}$, as defined by the interference of OPE and TPE amplitudes in Eq.~\ref{delta2gTPE}, is vanishing in the forward limit ($Q^2\rightarrow 0$) at fixed $\nu$, see discussion in Ref.~\cite{Tomalak:2018jak}. At the next order in $\alpha$, the contribution from $\vert \AMP_{2\gamma}\vert^2$ to the cross section at zero momentum transfer is not vanishing, however, numerically suppressed.
On the contrary, the TPE in forward kinematics is important in the description of the spectra of light muonic atoms, where its effect is enhanced as compared to ordinary atoms  due to the heavier muon mass. Let us focus on (muonic) hydrogen. The forward TPE corresponds to a $\delta(\vec{r}\,)$ potential that gives an $\mathcal{O}(\alpha^5)$ contribution to the energy levels. For comparison, the leading contribution from the Coulomb potential is of $\mathcal{O}(\alpha^2)$ and the proton finite-size starts to contribute at $\mathcal{O}(\alpha^4)$ through an OPE diagram with proton FFs. The off-forward TPE starts to contribute at $\mathcal{O}(\alpha^6\ln \alpha)$ through the so-called Coulomb distortion effect. See Ref.~\cite{Pachucki:2022tgl} for a comprehensive theory of the Lamb shift in light muonic atoms. Here, we want to focus on the dispersive formalism used to evaluate the $\mathcal{O}(\alpha^5)$ effect of the forward TPE, since a similar approach is used to approximate the small momentum transfer corrections to the scattering process, cf.\ Sec.~\ref{SmallMT}.
 
The forward TPE between a lepton and a proton can be expressed in terms of forward doubly-virtual Compton scattering (VVCS) off the proton. The TPE-induced shift of the $nS$-level in a hydrogen-like atom is given by the unpolarized VVCS amplitudes $T_1$ and $T_2$  \cite{Carlson:2011zd}
\begin{eqnarray}
&&\Delta E^{2\gamma}(nS)= 8\pi \alpha m \,\phi_n^2\,
\frac{1}{i}\int_{-\infty}^\infty \!\frac{\mathrm{d}\bar \nu}{2\pi} \int \!\!\frac{\mathrm{d} \boldsymbol{q}}{(2\pi)^3} \qquad\\
&&\times\frac{\left(\bar Q^2-2\bar \nu^2\right)T_1(\bar \nu,\bar Q^2)-(\bar Q^2+\bar \nu^2)\,T_2(\bar \nu,\bar Q^2)}{\bar Q^4(\bar Q^4-4m^2\bar \nu^2)},\nonumber
\end{eqnarray}
where $\phi_n^2=1/(\pi n^3 a^3)$ is the wavefunction at the origin, $a=(Z \alpha m_r)^{-1}$ is the Bohr radius (in the following $Z=1$ for hydrogen), $m_r$ is the reduced mass of the lepton-proton system and $m$ is the lepton mass ($m_e$ or $m_\mu$, respectively, for hydrogen and muonic hydrogen). Furthermore, $\bar \nu=q_0$ and $\bar Q^2 = \boldsymbol{q}^2 -q_0^2$ are the energy and virtuality of the VVCS photons inside the TPE loop diagram. Similarly, the TPE contribution to the hyperfine splitting can be expressed through the spin-dependent VVCS amplitudes $S_1$ and $S_2$. The TPE effect can then be evaluated with a data-driven dispersive approach \cite{Pachucki:1999zza,Martynenko:2005rc,Carlson:2011zd,Birse:2012eb,Gorchtein:2013yga,Hill:2016bjv,Tomalak:2018uhr,Faustov:2006ve,Carlson:2008ke,Carlson:2011af}, or predicted from variants of chiral perturbation theory \cite{Alarcon:2013cba,Hagelstein:2015lph,Lensky:2017bwi,Hagelstein:2018bdi,Peset:2014yha,Peset:2014jxa,Peset:2016wjq}, and in the future, from lattice QCD \cite{Fu:2022fgh,Hagelstein:2020awq}. For recent reviews, discussing low-energy proton structure in the context of muonic-hydrogen spectroscopy and lepton scattering, including TPE and the proton radius, we refer to Refs.~\cite{Carlson:2007sp,Carlson:2015jba,Gao:2021sml,Peset:2021iul,Antognini:2022xoo,Hagelstein:2015egb}.

Data-driven evaluations of the forward TPE make use of dispersion relations for the VVCS amplitudes, expressing them through proton structure functions measured in electron scattering
\begin{align}
\eqlab{T1DR}
T_1 (\bar  \nu, \bar Q^2) &=T_1(0,\bar Q^2)  \nn\\
&+\frac{32\pi \alpha M\bar \nu^2}{\bar Q^4}\int_{0}^1 
\,\dd x \, 
\frac{x F_1 (x, \bar Q^2)}{1 - x^2 (\bar \nu/\bar \nu_{\mathrm{el}})^2},\\
\eqlab{T2DR}
T_2 (\bar  \nu, \bar Q^2)
&=\frac{16\pi \alpha M}{\bar Q^2} \int_{0}^1 
\!\dd x\, 
\frac{F_2 (x, \bar Q^2)}{1 - x^2 (\bar \nu/\bar \nu_{\mathrm{el}})^2 },
\end{align}
where $F_1(x,\bar Q^2)$ and $F_2(x,\bar Q^2)$ are the unpolarized structure functions, with $x=\bar Q^2/2M\bar \nu$ the Bjorken variable, and $\bar \nu_\mathrm{el}=\bar Q^2/2M$. Similarly, the spin-dependent VVCS amplitudes $S_1$ and $S_2$ follow from dispersion integrals over the spin structure functions $g_1(x,\bar Q^2)$ and $g_2(x,\bar Q^2)$.
Unfortunately, the $T_1$ amplitude in \Eqref{T1DR} requires a once-subtracted dispersion relation. Since the subtraction
function $T_1(0,\bar Q^2)$ cannot be fully constrained from experiment, the data-driven approach suffers from a model dependence.

The TPE contributions to the energy levels in muonic atoms are usually split into a 
so-called ``elastic'' contribution, corresponding to the TPE with proton intermediate state shown in \Figref{TPE_diagram} (top panel), and a so-called ``polarizability'' contribution, corresponding to the diagrams in \Figref{TPE_diagram} (bottom panel). The elastic part of the VVCS amplitudes is well-known and can be expressed through proton FFs. The uncertainty of the elastic TPE can therefore be reduced with improved input for the proton FFs. This has recently been done based on the FF descriptions from Refs.~\cite{Borah:2020gte,Lin:2021xrc}, which are both consistent with the small proton charge radius extracted from the muonic-hydrogen Lamb shift, see Refs.~\cite{Antognini:2022xqf} and \cite{Pachucki:2022tgl} for the elastic TPE in the muonic hydrogen Lamb shift and hyperfine splitting, respectively. Therefore, the dispersive approach is only used to reconstruct the inelastic part of the VVCS amplitudes.  To this end, the dispersion integral is restricted to the inelastic structure functions, i.e., the region of $x \in [0,x_0]$ with the inelastic threshold $x_0$, e.g., the pion production threshold. In the following section, we will discuss the extension of this formalism to the TPE correction to $\ell p$ scattering in the region of small momentum transfer.

\subsubsection{Small momentum transfer}\label{SmallMT}

In the near-forward kinematics, i.e., in the small momentum transfer region, the TPE amplitude can be approximated with a modification of the dispersive approach presented above, see Eqs.~\ref{eq:T1DR} and \ref{eq:T2DR}, that reconstructs the VVCS amplitudes based on empirical input for the inelastic proton structure functions. In the following, we review the leading-$Q^2$ behaviour of the inelastic TPE corrections to the unpolarized cross section and the beam normal spin asymmetry as functions of the transverse cross section for photoabsorption off the proton. Recall that the leading terms in the $Q^2$ expansion of the TPE stem from the proton intermediate state \cite{McKinley:1948zz,Brown:1970te}, see discussion in Sec.~\ref{subsec:proton_state}.

\paragraph{Unpolarized cross section}

The leading-$Q^2$ behaviour of the inelastic TPE correction to the unpolarized cross section reads \cite{Brown:1970te,Gorchtein:2014hla}
\begin{eqnarray}
\delta_{2\gamma}&=&\frac{Q^2}{2\pi^3}\int_{\bar\nu_\pi}^\infty \frac{\dd \bar \nu}{\bar \nu}\sigma_T (\bar \nu) \ln \left(\frac{4 M^2\,\bar \nu^2}{Q^2\,W^2}\right)\eqlab{delta2galowt} \nonumber\\
&\times&\bigg[\left(1+\frac{\bar \nu^2}{2\epsilon_1^2}\right)\ln \left\vert \frac{\epsilon_1+\bar \nu}{\epsilon_1-\bar \nu} \right\vert\nonumber\\&&+\frac{\bar \nu}{\epsilon_1}\ln \left\vert 1-\frac{\epsilon_1^2}{\bar \nu^2} \right\vert-\frac{\bar \nu}{\epsilon_1}\bigg],
\end{eqnarray}
where $W^2$ is the invariant mass of the hadronic state in the internal Compton scattering process, and $\bar\nu_\pi$ is the pion production threshold. This formula has been first published by Brown \cite{Brown:1970te} in the early 1970's. Later, \Eqref{delta2galowt} has been re-evaluated by Gorchtein \cite{Gorchtein:2014hla} based on the phenomenological fit of the total photoabsorption cross section in Ref.~\cite{Gorchtein:2011xx}. It is important to point out that the coefficient of the order $Q^2 \ln Q^2$ term in $\delta_{2\gamma}$ is a model-independent result.

Tomalak and Vanderhaeghen have extended this approach beyond the leading $Q^2 \ln Q^2$ term and approximate the $\delta_{2\gamma}$ correction for $ep$ scattering in the region up to $Q^2=0.25\,\text{GeV}^2$ in Ref.~\cite{Tomalak:2015aoa} and for 
$\mu p$ scattering in the region up to $Q^2=0.08\,\text{GeV}^2$ in Ref.~\cite{Tomalak:2015hva}, see Fig.~\ref{fig.NearForw}. As can be seen from Eqs.~\ref{eq:T1DR} and \ref{eq:T2DR}, in general, this requires input for the unpolarized proton structure functions, $F_1(x,\bar Q^2)$ and $F_2(x,\bar Q^2)$ (or equivalently, the transverse and longitudinal cross sections, $\sigma_T(\bar \nu,\bar Q^2)$ and $\sigma_L(\bar \nu,\bar Q^2)$), as well as the subtraction function $T_1(0,\bar Q^2)$, where contrary to \Eqref{delta2galowt} the $\bar Q^2$ dependence of the input is kept. Besides the limited applicability range of the near-forward approximation, the $T_1(0,\bar Q^2)$ subtraction function introduces an additional model dependence for the scattering of massive leptons, while its contribution in the elastic $ep$ scattering is suppressed by the electron mass and negligible~\cite{Tomalak:2015aoa}.

\paragraph{Beam normal spin asymmetry}

The leading-$Q^2$ behaviour of the inelastic TPE contribution to the beam normal single spin asymmetry in the diffractive limit (high-energy and forward scattering) reads  \cite{Afanasev:2004pu,Gorchtein:2005za} 
\begin{eqnarray}
\eqlab{BnlowQ}
B_n&=&\frac{m \,Q \,\sigma_T}{8\pi^2}\frac{G_E(Q^2)}{\tau \,G_M^2(Q^2)+\varepsilon \,G_E^2(Q^2)}\nn \\
&\times&\left[2-\ln \frac{Q^2}{m^2}\right],
\end{eqnarray}
where the total photoabsorption cross section is assumed to be independent of the photon virtuality and roughly constant as a function of the invariant mass of the photon-proton system. In the nuclear resonance region, where $\sigma_T$ strongly varies with energy, the cross section enters through an energy-weighted integral. Note that this result is not fully model independent, as it makes use of the Callan-Gross relation \cite{Callan:1969uq} between the longitudinal and the transverse cross section in the DIS region.
The expression \Eqref{BnlowQ} has different dependence on momentum transfer $Q$ and the beam energy $\epsilon_1$ compared to contributions from the proton intermediate state. First, it does not fall off with the beam energy as $m/\epsilon_1$; second, it has linear dependence on $\theta$ at a fixed energy $\epsilon_1$ in the near-forward kinematics, whereas for elastic intermediate states (as well as for a celebrated Mott formula) the asymmetry falls as the third power, $\propto\theta^3$. As a result, the contribution of inelastic intermediate states becomes dominant for energies above GeV \cite{pasquini2004resonance, Afanasev:2004pu, Gorchtein:2005za}.

In Ref.~\cite{Gorchtein:2006mq}, subleading terms in the $Q^2$ expansion of $B_n$ were derived and the energy dependence of the photoabsorption cross section was taken into account. A comparison to \Eqref{BnlowQ} showed that the leading-term in the $Q^2$ expansion is indeed a good approximation. Note that while $B_n$ also has a double-logarithmic enhancements due to hard collinear quasi-real photons \cite{Afanasev:2004pu,Afanasev:2004hp,Gorchtein:2005yz}, it is highly suppressed for small scattering angles as it stems only from the helicity-flip Compton amplitude. However, double-logarithmic enhancement takes place at large scattering angles from a region of quasi-real Compton scattering \cite{pasquini2004resonance}, where both exchanged photons have low virtualities, while the mass of the excited hadronic state is the largest allowed by the incoming beam energy. 

Below the threshold of two-pion production, the absorptive part of proton Compton amplitude that governs the beam asymmetry can be described, as required by unitarity, in terms of quadratic combinations of single-pion production amplitudes. This program was realized in Ref.~\cite{pasquini2004resonance} where, using MAID amplitudes for electroproduction of pions, the authors reached good agreement for the beam asymmetries measured at MAMI for a broad range of scattering angles.

\subsubsection{Medium momentum transfer}\label{MediumMT}

The dispersion relation approach is an appropriate tool to improve our knowledge of inelastic contributions to the most uncertain TPE corrections for arbitrary scattering angles in elastic $ep$ scattering at GeV energies and below. The total TPE correction is given as a sum over all possible intermediate states, proton, pion-nucleon, resonances, etc. As was mentioned in Sec.~\ref{subsec:proton_state}, the elastic (proton) intermediate state was included within unsubtracted dispersion relations in Ref.~\cite{Borisyuk:2006fh} and within subtracted dispersion relations in Refs.~\cite{Tomalak:2014sva,Tomalak:2018jak}. The first inelastic pion-nucleon intermediate state was evaluated from data-driven MAID parameterization for the pion electroproduction amplitudes~\cite{Drechsel:1998hk,Drechsel:2007if} at low momentum transfers ($Q^2 \lesssim 0.064~\mathrm{GeV}^2$) in Ref.~\cite{Tomalak:2016vbf}. This straightforward, but numerically involved calculation, consistently accounts for the contribution of all resonances and non-resonant background. Going to higher momentum transfers requires analytical continuation of $ep$ scattering amplitudes to the unphysical region of kinematics. Such a novel method of analytical continuation for multiparticle intermediate states was developed and tested in Ref.~\cite{Tomalak:2017shs}. It allowed authors to evaluate pion-nucleon contributions to TPE corrections in elastic $ep$ scattering at more than an order of magnitude larger momentum transfers ($Q^2 \lesssim 1~\mathrm{GeV}^2$) compared to their previous work~\cite{Tomalak:2016vbf}.

For higher $Q^2$, the dispersive approach for resonant intermediate states was developed in Ref.~\cite{Blunden:2017nby}, and implemented for spin-1/2 and spin-3/2 resonances with mass below 1.8~GeV in Ref.~\cite{Ahmed:2020uso}. They included a Breit-Wigner shape with a finite width for each resonance, and for the resonance electrocouplings at the hadronic vertices they used helicity amplitudes extracted from the analysis of CLAS electroproduction data~\cite{HillerBlin:2019jgp}.

\begin{figure}[ht]
\includegraphics[width=0.45\textwidth]{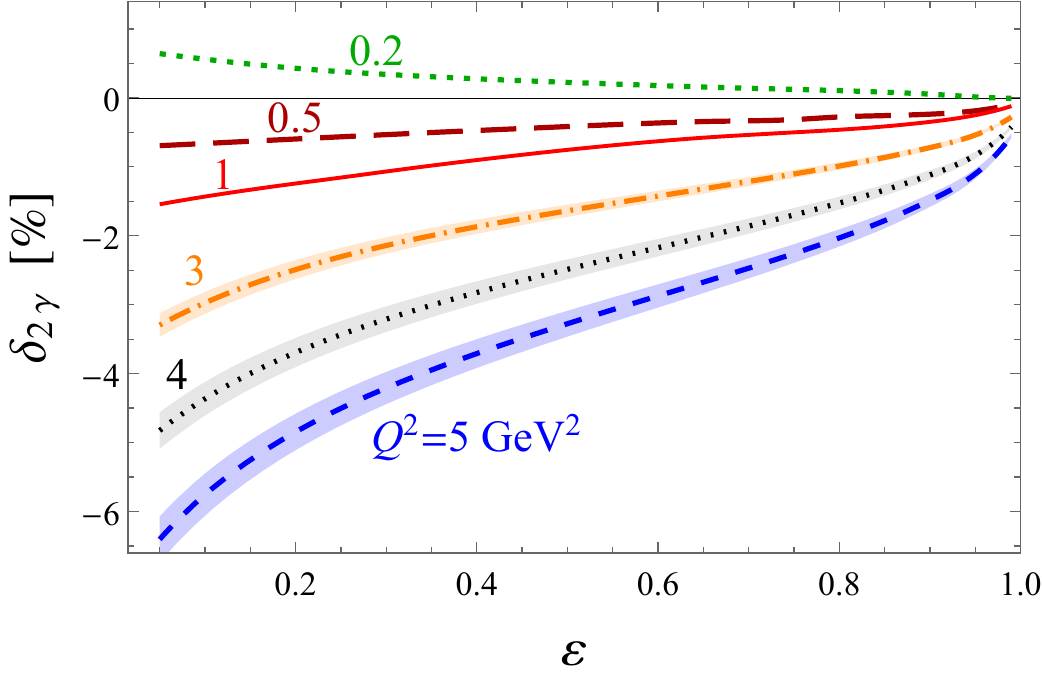}
\caption{The TPE correction $\delta_{2\gamma}$ (in \%) versus $\varepsilon$ for nucleon plus all
spin-parity $1/2^\pm$ and $3/2^\pm$ states \cite{Ahmed:2020uso} at
    $Q^2 = 0.2$~GeV$^2$ (green dashed line),
    0.5~GeV$^2$ (dark red long-dashed),
    1~GeV$^2$ (red solid),
    3~GeV$^2$ (orange dot-dashed),
    4~GeV$^2$ (black dotted), and
    5~GeV$^2$ (blue dashed). The shaded bands correspond to
    the uncertainty propagated from the input electrocouplings.}
\label{fig.Dsig}
\end{figure}

The combined effect on $\delta_{2\gamma}$ from the nucleon plus all the spin-parity $1/2^\pm$ and $3/2^\pm$ resonances is illustrated in Fig.~\ref{fig.Dsig} as a function of $\varepsilon$ for a range of fixed $Q^2$ values between 0.2 and 5~GeV$^2$.
At low $Q^2\lesssim 3$~GeV$^2$ the net excited state resonance contributions are small, and the total correction is dominated by the nucleon elastic intermediate state.
The net effect of the higher mass resonances is to increase the magnitude of the TPE correction at $Q^2 \gtrsim 3$~GeV$^2$, due primarily to the growth of the (negative) odd-parity $N(1520)~\!3/2^-$ and $N(1535)~\!1/2^-$ resonances, which overcompensates the (positive) contributions from the $\Delta(1232)~\!3/2^+$. 
At the highest $Q^2=5$~GeV$^2$ value, the total TPE correction $\delta_{\rm tot}$ reaches $\approx 6\% - 7\% $ at low $\varepsilon$.

An estimate of the theoretical uncertainties on the TPE contributions can be made by propagating the uncertainties on the fitted values of the transition electrocouplings~\cite{HillerBlin:2019jgp}, which are dominated by the $\Delta(1232)~3/2^+$ and $N(1520)~3/2^-$ intermediate states.
At low~$Q^2$, $Q^2 \lesssim 0.5$~GeV$^2$, the uncertainties are insignificant, but become more visible at higher $Q^2$ values, as illustrated by the shaded bands in Fig.~\ref{fig.Dsig}, for $Q^2 = 1-5~ \rm{GeV}^2$.

\subsubsection{Large momentum transfer}

For TPE calculations at large momentum transfers, quark degrees of freedom have to be included. In an approach based on generalized parton distributions and a handbag mechanism (with two photons coupling to the same quark) \cite{Chen:2004tw,Afanasev:2005prd}, TPE effects for the unpolarized cross section and for polarization asymmetries were calculated. Perturbative QCD (pQCD) approaches were considered in Refs.~\cite{Borisyuk:2008db, Guttmann:2010au}. In particular, in Ref.~\cite{Guttmann:2010au}, special consideration was given to the coupling of two photons to different quarks, noticing that a large momentum may be transferred with one less gluon compared to the case of pQCD description of elastic proton FFs.

\subsection{Effective field theory calculations}\seclab{EFTCalc}

For elastic $\ell p$ scattering at lepton energies well below the proton mass, we can use low-energy effective field theories (EFTs). At such energies, we do not need the full knowledge of the non-perturbative properties of the proton. For OPE, we only need a few effective couplings, e.g., the proton charge, magnetic moment, and charge radius. Similar simplifications apply to TPE. In the following, we discuss two such effective theories: heavy baryon chiral perturbation theory (HBChPT) and QED-NRQED, and review the current status of their predictions for TPE.
Before a detailed discussion, it is worth mentioning that radiative corrections in the elastic electron-proton scattering when $Q^2 >> m^2_e$ formulated recently in the Soft-Collinear Effective Theory (SCET) approach within QED~\cite{Hill:2016gdf}, which allows us also for a systematic resummation of large logarithms. Moreover, the two-photon exchange corrections in the hard momentum region were also evaluated in the SCET framework within QCD in Ref.~\cite{Kivel:2012vs}.

\subsubsection{Heavy baryon chiral perturbation theory}

Within this topical collection, an  exact analytical evaluation of the TPE contribution to elastic $\ell p$ scattering in low-energy EFT, namely, HBChPT  at NLO, is published \cite{choudhary2023analytical}.
The LO HBChPT
Lagrangian contains one derivative only. The NLO Lagrangian  will have two derivatives and terms of order $1/M$, where $M$ is the proton mass. 
In this framework, the proton being the heavy degree of freedom is treated non-relativistically. The leptons, the electron and muon, along with the photons constituting the light degrees of freedom, are treated in standard covariant QED. Throughout, the masses of the relativistic leptons are kept non-zero. 

Previously, the NLO HBChPT prediction had been calculated using the soft-photon approximation (SPA) \cite{Talukdar:2019dko,Talukdar:2020aui}. As before, Ref.~\cite{choudhary2023analytical} assumes 
that at low energies the dominant photon loop contributions arise only from the elastic proton intermediate state.  The inelastic proton intermediate states are beyond the intended accuracy of the evaluation. Notably, since the proton FFs, which parametrize the hadronic structure effects, enter in HBChPT at NNLO, the proton is point-like at NLO. All hard- and soft-photon exchanges are now included in the two-photon loop corrections which are evaluated exactly. The methodology relies on the successive use of partial fractions and integration by parts facilitating the decomposition of the rather intricate TPE four-point loop functions into a system of two- and three-point scalar master integrals, which are straightforward to evaluate analytically. 
 
To LO accuracy, our low-energy TPE result approximates the well-known McKinley-Feshbach two-photon correction in potential scattering theory, cf.\ \Eqref{Feshbach}. At NLO the finite analytical real parts of the TPE loop contributions to the elastic differential cross section are presented. The exact analytical evaluation of the NLO bremsstrahlung process is not included. This means that the systematic cancellations of the IR-singular terms that arise from the TPE loops are not discussed. While the details of the TPE evaluation are going to be presented in Ref.~\cite{choudhary2023analytical}, 
a rigorous NLO evaluation of the bremsstrahlung subtracted IR-free TPE contribution to the radiative corrections shall be a subject matter of a later publication.

\subsubsection{QED-NRQED}

An EFT related to HBChPT, is QED-NRQED, suggested in \cite{Caswell:1985ui,Pineda:1997bj,Pineda:1998kj,Hill:2012rh} and further developed in \cite{Dye:2016uep,Dye:2018rgg,Peset:2014jxa}. In this EFT the proton is treated non-relativistically, using NRQED, and the leptons and photons are treated using QED. Unlike HBChPT, the EFT does not include other hadronic degrees of freedom, such as pions. Such effects are encoded in the QED-NRQED coupling constants. Thus QED-NRQED is most useful for elastic $\ell p$ scattering.  

Up to dimension six, the independent NRQED Wilson coefficients are $c_F$ and $c_D$ corresponding to the proton magnetic moment and charge radius, respectively. There are also two dimension-six contact interaction terms: a spin-independent term and a spin-dependent term \cite{Hill:2012rh}, 
\begin{equation}
{\cal L}_{\ell\psi}^{d=6}=\dfrac{b_1}{M^2}\psi^\dagger\psi\,\bar \ell\gamma^0\ell+\dfrac{b_2}{M^2}\psi^\dagger\sigma^i\psi\,\bar \ell\gamma^i\gamma^5\ell\,,
\end{equation}
where $\psi$ ($\ell$) is the proton (lepton) field and $M$ is the proton mass, the cutoff scale of the EFT. 

In Ref.~\cite{Dye:2016uep}, it was shown that QED-NRQED $\ell p$ scattering at ${\cal O}(Z\alpha)$ and power $1/M^2$  reproduces the known Rosenbluth scattering formula expanded to $1/M^2$. It requires just the Dirac Lagrangian and the NRQED Lagrangian up to $1/M^2$, implying that $b_1$ and $b_2$ are zero at ${\cal O}(Z\alpha)$. QED-NRQED $\ell p$ scattering at ${\cal O}(Z^2\alpha^2)$ and leading power in $1/M$ reproduces the ${\cal O}(Z^2\alpha^2)$ terms in the scattering of a lepton off a static $1/r$ potential \cite{Dalitz:1951ah}. Here $Z$ denotes the proton charge in units of $|e|$ and is used to denote the appropriate class of radiative corrections.

Since the Wilson coefficients $b_1$ and $b_2$ are zero at ${\cal O}(Z\alpha)$, they are sensitive to TPE at scales above the proton mass. In Ref.~\cite{Dye:2018rgg}, $b_1$ and $b_2$ were calculated at ${\cal O}(Z^2\alpha^2)$. They were extracted by calculating the $\ell p\to\ell p$ off-shell  forward scattering amplitude at ${\cal O}(Z^2\alpha^2)$ and power $1/M^2$ in the effective and full theory  in both Feynman and Coulomb gauges. Two cases were considered:  a toy example of a non-relativistic point particle (p.p.), and the real proton described by a hadronic tensor, namely, the forward VVCS off the proton discussed in \secref{VVCS}.

For the toy example, Ref.~\cite{Dye:2018rgg} found $b_1^{\text{ p.p.}}=0$ and $b_2^\text{ p.p.}=Q_\ell^2Z^2\alpha^2\left[{16}/{3}+\ln\left({M}/{2\Lambda}\right)\right]$, where $\Lambda$ is the UV cutoff of QED-NRQED, and $Q_\ell$ is the lepton charge in units of $|e|$. Surprisingly, $b_1^{\text{ p.p.}}=0$ at ${\cal O}(Z^2\alpha^2)$. For the case of the real proton, Ref.~\cite{Dye:2018rgg} derived implicit expressions for $b_1$ and $b_2$ in terms of the components of the hadronic tensor. Considering only the contribution to the Wilson coefficients of $F_1(0)$, $F_2(0)$ and $M^2 F_1^\prime(0)$, related to the proton charge, magnetic moment and charge radius respectively, Ref.~\cite{Dye:2018rgg} finds: 
\begin{align}
b_1/(\alpha^2Q_\ell^2)&=0+\cdots\,,\\
\nn\\
b_2/(\alpha^2Q_\ell^2)&=F_1(0)^2\left[\frac{16}{3}+\ln\left(\frac{M}{2\Lambda}\right)\right] \nn\\&+\frac{16}{3}F_1(0)F_2(0)\nn\\&+\frac{F_2(0)^2}{2}\left[\frac{17}{12}-\ln\left(\frac{M}{2\Lambda}\right)+3\ln\left(\frac{Q}{M}\right)\right]\nn\\&+\cdots\,. 
\end{align}
The ellipsis denotes  $\mbox{non } F_1(0), F_2(0), M^2 F_1^\prime(0)$ terms.  Surprisingly, again there is no contribution to $b_1$. It does not follow from a symmetry of the EFT and it might be a one-loop ``accident". 

This implies that low-energy elastic $\ell p$ scattering is much less sensitive to spin-independent TPE effects above the proton mass scale compared to spin-dependent ones. On the other hand, the proton charge radius extraction will be more robust. 

With the calculation of the QED-NRQED dimension six couplings: $c_F,c_D, b_1, b_2$, at hand, the next logical step is the calculation of the differential cross section for elastic $\ell p$ scattering. Such a calculation was not performed in the literature yet. Once performed, it would be interesting to compare its result to HBChPT.

\subsection{Empirical extractions of TPE}\seclab{TPEempirical}

Besides theoretical predictions, it is possible to extract TPE amplitudes, as well as the TPE contributions to the unpolarized cross section $\delta_{2\gamma}$,  from empirical cross section and polarization transfer measurements.

\subsubsection{Phenomenological fits of \texorpdfstring{$\delta_{2\gamma}$}{TEXT}}

The size of the TPE contribution becomes more significant at larger $Q^2$. A phenomenological fit of the $Q^2$-dependent part, as an addition to the Feshbach correction, \Eqref{Feshbach}, is \cite{A1:2013fsc} 
\begin{equation}
    \delta_{2 \gamma} =-(1-\varepsilon) \, a \ln (b\, Q^2+1),
\end{equation}
where $a=0.069, \,b=0.394\, \text{GeV}^{-2}$. This fit includes the world data set on unpolarized cross section measurements (updated to a common standard on radiative corrections) and measurements of the FF ratio using polarization and assumes that the whole discrepancy is driven by hard TPE.

Similarly, Schmidt in Ref.~\cite{Schmidt:2019vpr} uses several existing parametrizations for FFs from Rosenbluth-type experiments and polarization measurements of the ratio to determine the hard TPE contribution, under the assumption that TPE preserves the linearity of the reduced cross section in $\varepsilon$.

\subsubsection{Updated extraction of TPE amplitudes}

A measurement of the three observables for elastic $ep$ scattering, $\sigma_R, P_l,$ and $P_t$,  as a function of $\varepsilon$ for a fixed value of $Q^2$ allows one to extract in a model-independent fashion the three TPE amplitudes, which conserve the lepton helicity, as can be seen from Eqs.~\ref{crossen}, \ref{PtPlradcorr} and \ref{pl_2gamma}. Such an analysis has been performed in \cite{Guttmann:2010au} at $Q^2 = 2.5$~GeV$^2$, based on 
available cross section data as well as the $\varepsilon$-dependence of  
  $P_t/P_l$ and $P_l/P_l^{Born}$ 
  as measured by the JLab/Hall C experiment~\cite{GEp2gamma:2010gvp}. 
    The combination of both unpolarized and polarization experiments at a same $Q^2$ value 
provided the necessary three observables to extract the 
  $\varepsilon$-dependence of the three TPE amplitudes $Y_M$, $Y_E$, and $Y_3$, introduced in \Eqref{2gaampl}. A concise update of this analysis in view of more recent data is given here. 

\begin{figure}[ht]
\includegraphics[width=0.45\textwidth]{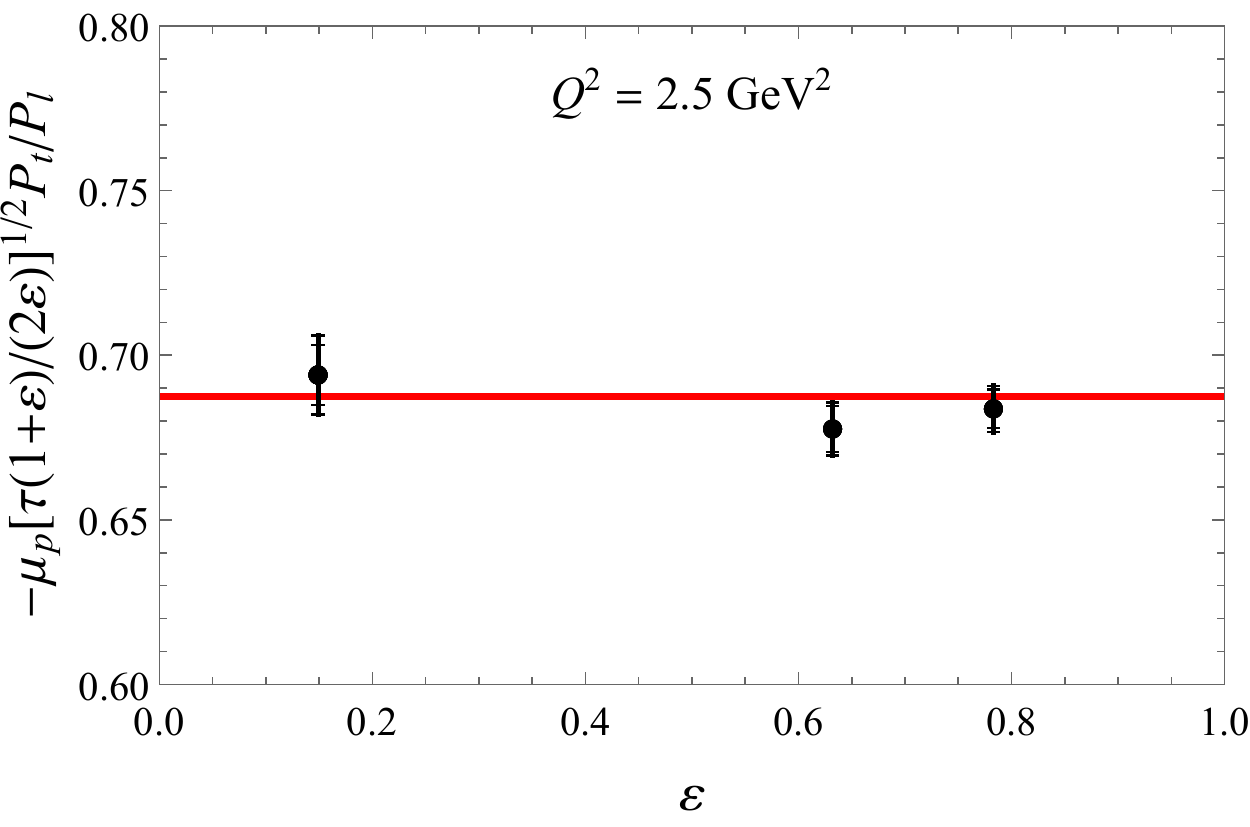}
\includegraphics[width=0.45
\textwidth]{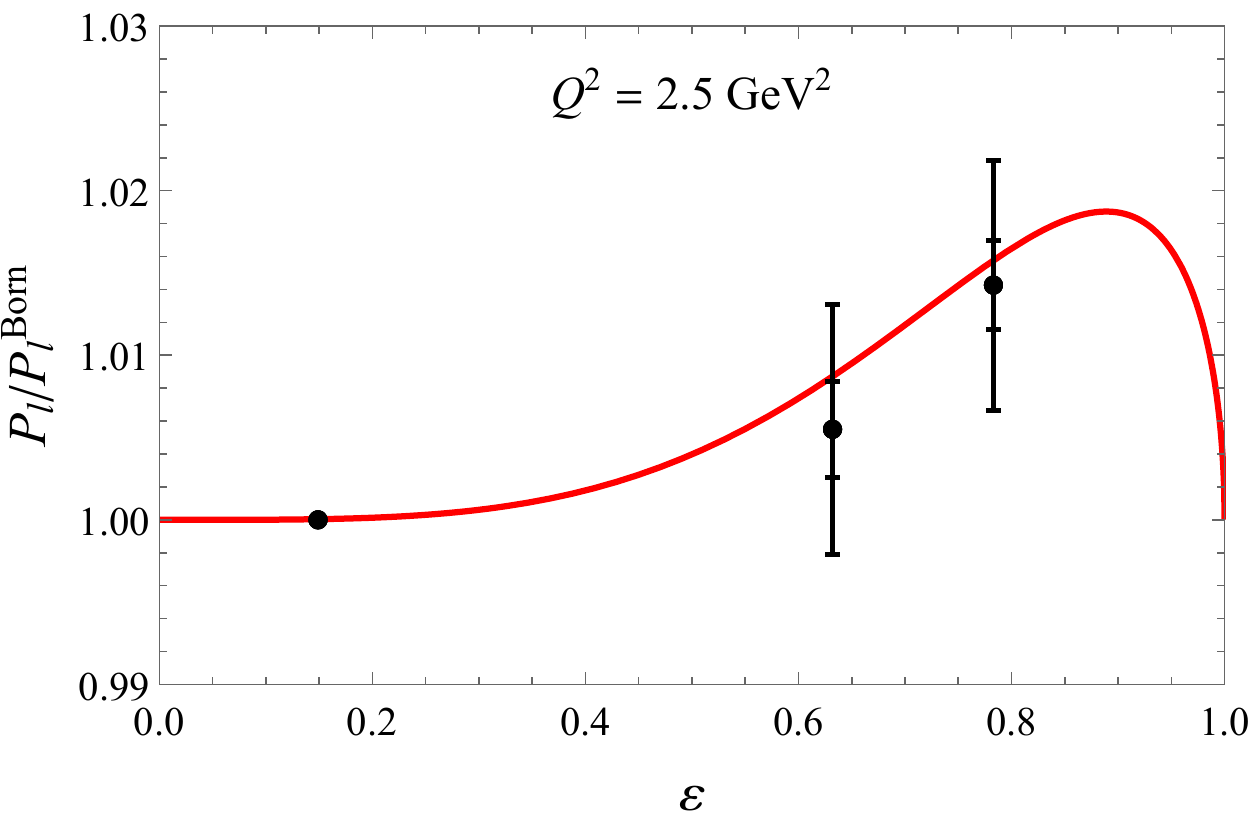}
\caption[]{Polarization transfer observables for elastic $ep$ scattering at $Q^2 = 2.5$~GeV$^2$. Upper panel: $- \mu_p \sqrt{\tau (1 + \varepsilon)/(2 \varepsilon)} P_t / P_l$, lower panel: $P_l / P_l^\text{Born}$. The JLab/Hall C data are from the GEp-$2 \gamma$ experiment~\cite{GEp2gamma:2010gvp}, 
using the updated analysis of \cite{Puckett:2017flj}. The red curves are the fit to the data. Left: $\varepsilon$-independent fit according to 
Ref.~\cite{Puckett:2017flj}; right: updated fit from Ref.~\cite{Guttmann:2010au} according to Eq.~(\ref{eq:plfit}). 
Note that the values for the polarization observables are shown without radiative corrections. 
The bulk of these corrections (virtual corrections on the lepton side and soft-photon emission  corrections), which factorize in terms of the Born cross section, drop out of the asymmetries, and the hard TPE is therefore expected to be the leading correction. It has been checked in Refs.~\cite{GEp2gamma:2010gvp,Puckett:2017flj} using the MASCARAD program~\cite{Afanasev:2001jd} that the standard radiative corrections yield a multiplicative correction around or less than 0.1\% on the asymmetries.}
\label{fig:PtPl}
\end{figure}

The data for $P_t / P_l$ of the dedicated JLab/Hall C GEp-$2 \gamma$ 
experiment~\cite{GEp2gamma:2010gvp,Puckett:2017flj}, as shown in the upper panel of Fig.~\ref{fig:PtPl},  
do not see any systematic TPE effect within their error bars of order 1\%.  
One can therefore effectively fit the observable on the lhs of \Eqref{Rratios}   
assuming an $\varepsilon$-independent part, which equals its OPE limit,
\begin{eqnarray}
- \sqrt{ \frac{ \tau (1 +\varepsilon)}{2\varepsilon}}  \frac{P_t}{P_l} \simeq 						
	R_{EM}(Q^2),
\end{eqnarray}
and extract the $Q^2$-dependence of $R_{EM}$, cf.\ \Eqref{REM}, from this observable, using the parameterization given in Ref.~\cite{Puckett:2017flj} (Global fit II).

For the longitudinal polarization observable $P_l/ P_l^\text{Born}$, shown in the lower panel of Fig.~\ref{fig:PtPl}, the deviation from unity is 6.2 times the statistical uncertainty and 2.2 times the total uncertainty~\cite{Puckett:2017flj}. The TPE correction effect has been parameterized as~\cite{Guttmann:2010au}
\begin{eqnarray}
P_l / P_l^\text{Born} &=& 1 + a_l \, \varepsilon^4 (1 - \varepsilon)^{1/2}.
\label{eq:plfit}
\end{eqnarray}
The comparison of this fit function to the data is shown in the lower panel of Fig.~\ref{fig:PtPl} for  
$a_l = 0.09$.  

Recently, an updated global analysis for $\sigma_R$, including new high-$Q^2$ data from JLab, has been given in Ref.~\cite{Christy:2021snt}.
In Ref.~\cite{Christy:2021snt}, the data for the reduced cross section have been fitted by a linear in $\varepsilon$ behavior
\begin{eqnarray}
\frac{\sigma_R(\varepsilon, Q^2)}{(\mu_p G_D)^2} = a + \varepsilon b,   
\label{eq:sigmaRfit}
 \end{eqnarray}
 with $\mu_p$ the proton magnetic moment, and $G_D(Q^2) \equiv 1/(1 + Q^2/0.71)^2$ the standard dipole. 
 As the TPE amplitudes vanish for $\varepsilon \to 1$, the proton FF $G_M$ can be extracted from Eqs.~\ref{eq:sigmaR} and \ref{eq:sigmaRfit} as
\begin{eqnarray}
 \left( \frac{G_{Mp} }{ \mu_p G_D} \right)^2 = \frac{a + b }{1 + \frac{1}{\tau} R_{EM}^2}.
 \label{eq:GMfit}
 \end{eqnarray}
The parameterization for $\sigma_R$ chosen in~\cite{Christy:2021snt} does however not have the correct $\varepsilon \to 1$ limit, in which  
the TPE amplitudes vanish.  
At a fixed value of $Q^2$, the limit $\varepsilon \to 1$ corresponds with the Regge (high-energy) limit, as 
$Q/(2\eps_1) \simeq (1 - \varepsilon)^{1/2} / \sqrt{2}$ for $\varepsilon \to 1$, with $\eps_1$ the electron lab energy.
In an expansion around $\varepsilon \to 1$, the leading TPE correction can be model independently expressed as~\cite{Tomalak:2015aoa} 
\begin{align}
& \delta_{2 \gamma}(\varepsilon \to 1, Q^2) = \alpha \pi \left(\frac{Q}{2 \eps_1}\right) + \frac{\alpha }{\pi} \left(\frac{Q^2}{M\eps_1} \right)  \ln^2 \left(\frac{Q}{2\eps_1} \right) \nn \\ &
+ c \left(\frac{Q^2}{M\eps_1} \right)  \ln \left(\frac{Q}{2\eps_1} \right) + 
 {\mathcal O}\left( \frac{Q^2}{M^2}, \frac{Q^2}{M \eps_1}, \frac{Q^2}{\eps_1^2}\right).
 \label{eq:expansion}
 \end{align}
In Eq.~\ref{eq:expansion}, the first term, which dominates for $\varepsilon \to 1$ corresponds with the McKinley-Feshbach TPE correction due to the scattering of relativistic electrons in the Coulomb field of the proton~\cite{McKinley:1948zz}, the $Q^2 \ln^2 Q$ term is due to the elastic intermediate state only, and the $Q^2 \ln Q$ term contains the effect due to inelastic intermediate states~\cite{Brown:1970te,Tomalak:2015aoa}, cf.\ discussion in Sec.~\ref{SmallMT}.

\begin{figure}[ht]
\includegraphics[width=0.45\textwidth]{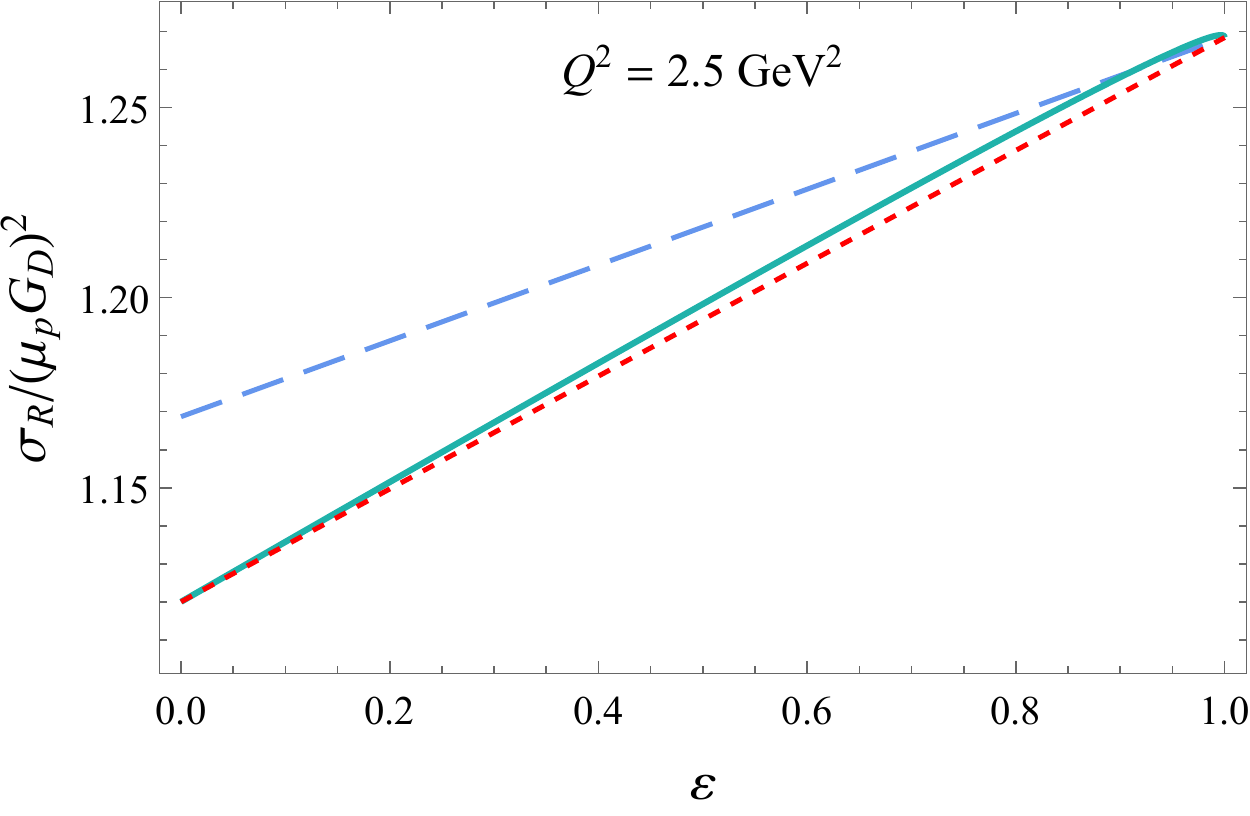}
\includegraphics[width=0.45\textwidth]{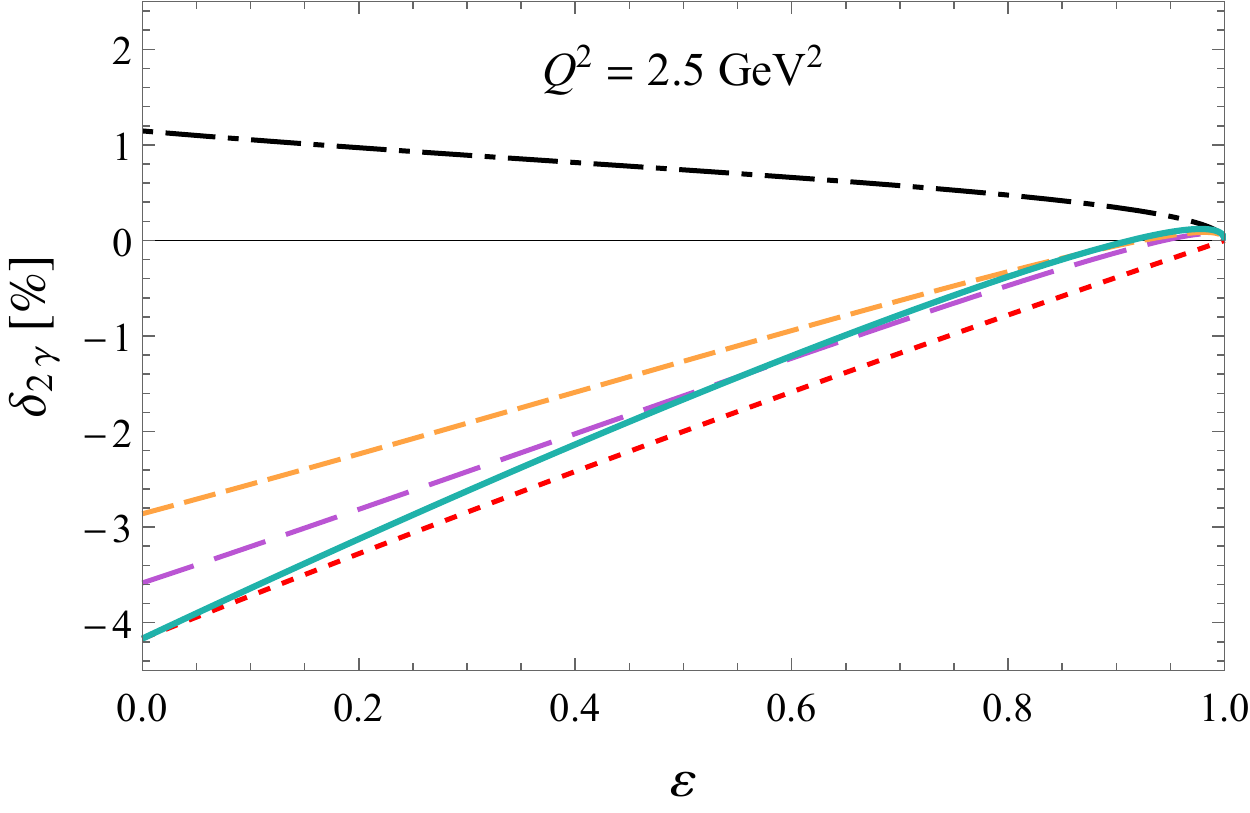}
\caption[]{Upper panel: $\varepsilon$-dependence of the reduced cross section $\sigma_R$ for $Q^2$ = 2.5 GeV$^2$, based on the linear fit of Ref.~\cite{Christy:2021snt} (red dotted curve), using the modified fit function of Eq.~\ref{eq:sigmaRfit2} (green solid curve), as well as in the OPE approximation, setting $\delta_{2 \gamma} = 0$, (blue dashed curve). Lower panel: $\varepsilon$-dependence of the TPE  correction $\delta_{2 \gamma}$ for $Q^2$ = 2.5 GeV$^2$, based on the linear fit of Ref.~\cite{Christy:2021snt} (red dotted curve), using the modified fit function of Eq.~\ref{eq:sigmaRfit2} (green solid curve), using the spline fit (purple long-dashed curve) and Pad\'e fit (yellow short-dashed curve) of Ref.~\cite{A1:2013fsc}. The black dot-dashed curve shows the McKinley-Feshbach correction, which gives the leading behavior for $\varepsilon \to 1$. }
\label{fig:sigmaR}
\end{figure}

In order to combine the empirical observation of a linearity of $\sigma_R$ over most of the $\varepsilon$ region with the correct $\varepsilon \to 1$ limit, the fit function of Eq.~\ref{eq:sigmaRfit} for $\sigma_R$ can be improved as
\begin{align}
&\frac{\sigma_R(\varepsilon, Q^2)}{(\mu_p G_D)^2} = a + \varepsilon b \nn \\
&+ (a + b) \frac{ \left(1 + \frac{\varepsilon}{\tau}  R_{EM}^2 \right) }{ \left(1 + \frac{1}{\tau}  R_{EM}^2 \right) } \frac{\alpha \pi}{\sqrt{2}} \left[(1 - \varepsilon)^{1/2} - (1 - \varepsilon)\right],   
\label{eq:sigmaRfit2}
 \end{align}
 which does not change the fit value for small $\varepsilon$
 \begin{eqnarray}
\frac{\sigma_R(\varepsilon = 0, Q^2)}{(\mu_p G_D)^2}  = a, 
\end{eqnarray}
and has the correct limit for $\varepsilon \to 1$:
\begin{eqnarray}
\frac{\sigma_R(\varepsilon \to 1, Q^2)}{(\mu_p G_D)^2} &=& (a + b) \left[ 1 + \frac{\alpha \pi}{\sqrt{2}} (1 - \varepsilon)^{1/2}  \right]. \qquad
 \end{eqnarray}
Using the fit function of Eq.~\ref{eq:sigmaRfit2}, the TPE correction $\delta_{2 \gamma}$ follows as
\begin{align}
\delta_{2 \gamma}(\varepsilon, &Q^2) = \frac{\alpha \pi}{\sqrt{2}} (1 - \varepsilon)^{1/2} \nn \\
&+ (1 - \varepsilon) \left[ 
\frac{\left( - b + \frac{a}{\tau} R_{EM}^2 \right) }{(a + b) \left( 1 + \frac{\varepsilon}{\tau} R_{EM}^2 \right)} - \frac{\alpha \pi}{\sqrt{2}}\right] .
\label{eq:delta2gafin}
\end{align}
In Fig.~\ref{fig:sigmaR} (upper panel), we compare, for $Q^2$ = 2.5 GeV$^2$, the fit functions for $\sigma_R$ of Ref.~\cite{Christy:2021snt} (red dotted curve) with the modified fit function of Eq.~\ref{eq:sigmaRfit2} (green solid curve), as well as $\sigma_R$ in the OPE approximation, setting $\delta_{2 \gamma} = 0$, (blue dashed curve).
The resulting $\varepsilon$-dependence of the TPE correction $\delta_{2 \gamma}$ for $Q^2$ = 2.5 GeV$^2$, based on the linear fit of Ref.~\cite{Christy:2021snt}, using the improved fit function of Eq.~\ref{eq:sigmaRfit2}, using the spline and Pad\'e fits of Ref.~\cite{A1:2013fsc}, are compared in Fig.~\ref{fig:sigmaR} (lower panel). For comparison, the McKinley-Feshbach correction, which gives the leading behavior for $\varepsilon \to 1$, is also shown (black dot-dashed curve).  

\begin{figure}[ht]
\includegraphics[width=0.45\textwidth]{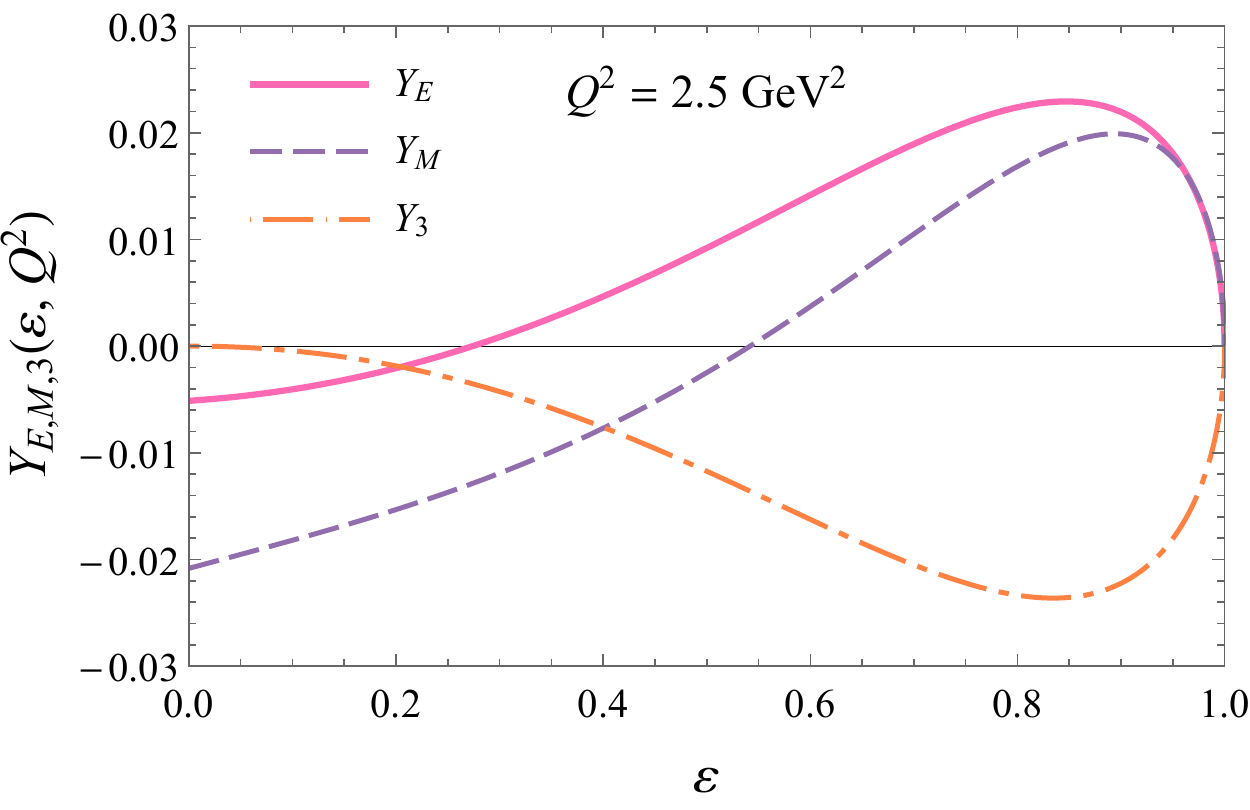}
\caption[]{$\varepsilon$ dependence for the TPE amplitudes for $Q^2 = 2.5$~GeV$^2$, based on the polarization data of 
Ref.~\cite{Puckett:2017flj}, shown in Fig.~\ref{fig:PtPl}, and the cross section fit of Ref.~\cite{Christy:2021snt}, improved using the fit function of Eq.~\ref{eq:sigmaRfit2}, given by the red curves in Fig.~\ref{fig:sigmaR}.}
\label{fig:YME3}
\end{figure}

Based on the measured $\varepsilon$-dependence 
of the three elastic $e^-p$ observables ($\sigma_R,\, P_l$, and $P_t$ at $Q^2= 2.5$~GeV$^2$ as shown in Figs.~\ref{fig:PtPl} and \ref{fig:sigmaR}), an updated empirical extraction 
of the three TPE amplitudes, 
$Y_M, Y_E$, and $Y_3$, at this $Q^2$ value was obtained, see Fig.~\ref{fig:YME3}. 

The updated empirical extraction of the TPE correction $\delta_{2\gamma}$ for general values of $Q^2$ and $\epsilon$, is compared to data from the OLYMPUS, VEPP-3 and CLAS experiments in Figs.~\ref{fig.OLYMPUS}, \ref{fig.VEPP} and \ref{fig.CLAS} (dot-dashed pink curves), respectively. Note that the empirical extraction displays a small oscillation in the low-$Q$ region, cf.\ $\varepsilon=0.45, 0.88$ panels in Fig.~\ref{fig.CLAS}. This, however, is not a true feature but a remnant that propagated from the underlying fit of $R_{EM}(Q^2)$ that was used~\cite{Puckett:2017flj}, and is enhanced due to two opposite sign contributions in the $(1 - \varepsilon)$ term in Eq.~\ref{eq:delta2gafin}.

\subsection{Dedicated experiments and comparison to theory}\label{Sec:Experiments}

\subsubsection{Single-spin asymmetries}\seclab{BeamAsymmetry}

As stated above in Sec.~\ref{SSAintro}, parity-conserving SSA for elastic $ep$ scattering are zero in the first Born approximation. Therefore measurements of either target or beam SSA may provide useful information on TPE amplitudes. While beam asymmetries arising from TPE are of the order $10^{-5}$ for electrons \cite{Afanasev:2004pu,Gorchtein:2004ac, pasquini2004resonance} and $10^{-3}$ for muons \cite{Koshchii:2019b}, the target SSA are expected at percent level \cite{Afanasev:2005prd}. The beam asymmetries were measured with high accuracy at MIT \cite{Wells:2001bn}, MAMI \cite{maas2005measurement} and Jefferson Lab \cite{armstrong2007transverse,Androic:2012bn,Abrahamyan:2012bn,PREX:2021uwt} using experimental setups designed to study parity-violating electron-helicity asymmetries. Measurements of target SSA were performed at Jefferson Lab on a polarized $^3$He target \cite{Zhang:2015kna}; results for $Q^2>$ 1 GeV$^2$ were in agreement with GPD-based calculations \cite{Afanasev:2005prd} for a neutron. 

The measurements of SSA in elastic electron scattering provided unambiguous evidence of TPE effects, while probing contributions arising from an absorptive part of the hadronic Compton amplitude entering the TPE mechanism.

\subsubsection{Cross sections and charge asymmetries}

The TPE correction, defined in Eq.~\ref{delta2gTPE}, can be accessed experimentally by measuring the ratio of cross sections from elastic $\ell ^+p$ vs. $\ell^- p$ scattering. Corrections that depend on an odd power of the lepton charge do not cancel in this ratio. This includes the leading TPE correction, $\delta_{2\gamma}$. It also includes, however, the interference of lepton and proton bremsstrahlung radiation,  $\delta_{b}$, which is comparable to $\delta_{2\gamma}$. Their sum gives us the ``odd'' part of the radiative corrections, $\delta_{\text{odd}}=\delta_{b}+\delta_{2\gamma}$.
By contrast, the ``even'' part of the corrections, $\delta_{\text{even}}$, include logarithmically enchanced terms, $\log (Q^2/m_{\ell})$, coming from lepton bremsstrahlung, vacuum polarization and vertex corrections. As a result, the even part is relatively large, implying $|\delta_{\text{odd}}| \ll |\delta_{\text{even}}|$.

The above mentioned cross section ratio can be written in the following way \cite{Afanasev:2017gsk}:
\begin{align}
    R_{\pm}^\text{exp.}&=\frac{\sigma(\ell ^+p)}{\sigma(\ell^-p)} = \frac{1+\delta_{\text{even}}-\delta_{\text{odd}}}{1+\delta_{\text{even}}+\delta_{\text{odd}}}\nn\\& \approx 1-\frac{2 \,\delta_{\text{odd}}}{1+\delta_{\text{even}}}.
\end{align}
The TPE correction can be extracted from 
\begin{equation}
\eqlab{R2gammadef}
       R_{2\gamma} \approx 1-2 \,\delta_{2\gamma},
\end{equation}
which is the 
measured ratio $R_{\pm}^\text{exp.}$ corrected for  $\delta_{b}$ and $\delta_{\text{even}}$.

Experimental possibilities are limited, as such a measurements requires both electron and positron beams of high quality and at relevant beam energies of a few GeV. In recent years, three experiments have published results, cf.\ Figs.~\ref{fig.OLYMPUS}-\ref{fig.CLAS}: an experiment \cite{Rachek:2014fam} at the VEPP-3 storage ring measured the cross section ratio with beam energies of 1.0 and 1.6 GeV with a non-magnetic spectrometer covering angles between 15$^\circ$ and 105$^\circ$. Since the experiment lacks a relative luminosity determination of sufficient precision, the data is published relative to the most forward point set to a ratio of one. However, this point is sufficiently far away from $\ep=1$ that a proper comparison with curves requires a shift of the data to match the predictions at this normalization point. CLAS at Jefferson Lab \cite{CLAS:2013mza,CLAS:2014xso} generated a wide-energy-spread electron and positron beam by converting the CEBAF electron beam to a photon beam via a converter target, subsequent pair-production from that photon beam, and by cleaning and recombining the electron/positron beam via a chicane magnet system. Using the detector setup of the CLAS collaboration in Hall B, the so produced wide-energy beam-proton scattering events are detected. By the curvature of the scattered lepton in the magnetic field, the charge of the particle can be determined.  Because of the wide-energy spread of the beam, the resulting data is sorted in large bins in $\ep$ and $Q^2$. The collaboration published data with several (non-independent) bin selections. The OLYMPUS experiment \cite{Henderson:2016dea} at the DORIS ring at DESY, Hamburg, measured using a monoenergetic beam with an energy of around 2 GeV using an open-cell internal hydrogen target. Scattered leptons and protons were detected with a spectrometer upgraded from the former BLAST detector. OLYMPUS reached the highest $Q^2$ of the three experiments of slightly more than 2~$\GeV^2$.

\subsubsection{Comparison theory vs.\ experiment} 

\begin{figure}[ht] 
\includegraphics[width=\columnwidth]{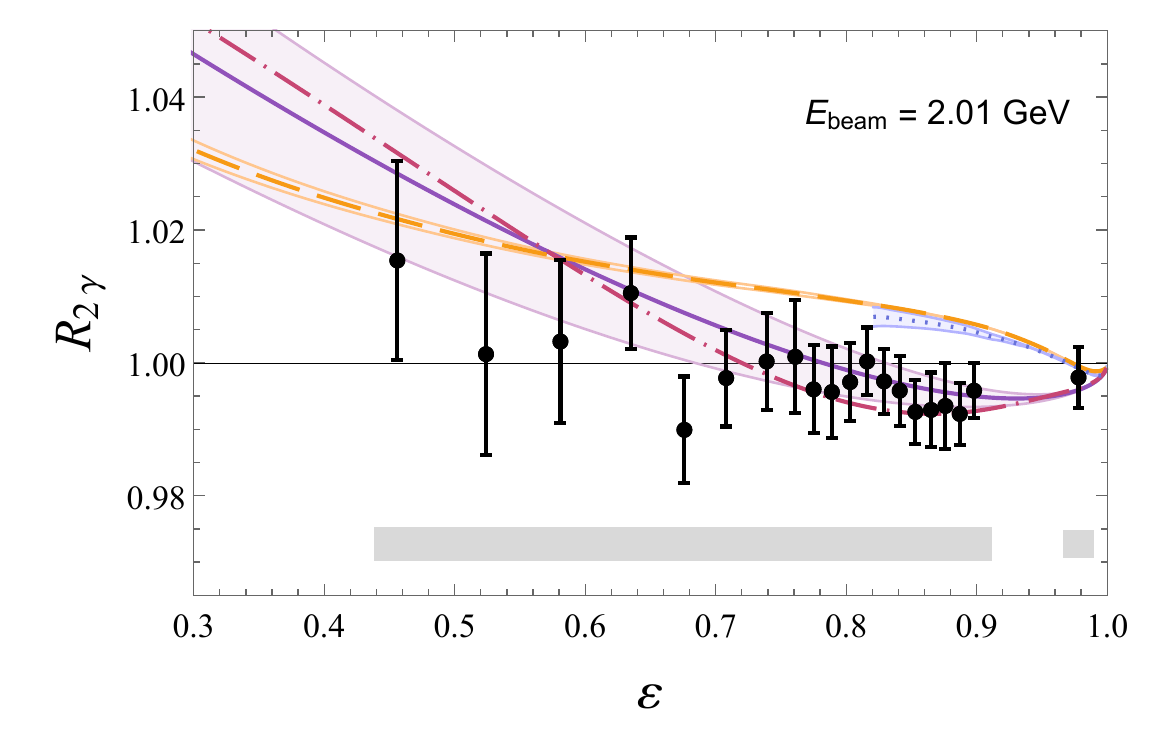} 
\caption{Comparison of predictions for $R_{2\gamma}$, \Eqref{R2gammadef}, from $ep$ scattering with beam energy $E_\text{beam}=2.01$ GeV to results from the OLYMPUS experiment (black points; statistical and uncorrelated systematic uncertainties added in quadrature; gray errorband for correlated systematic uncertainties) \cite{Henderson:2016dea}. Theoretical predictions: sum of nucleon plus spin-parity $1/2^\pm$ and $3/2^\pm$ resonances by Ahmed et al.~(dashed orange) \cite{Ahmed:2020uso}, nucleon plus $\pi N$ intermediate state by Tomalak et al.~(dotted blue) \cite{Tomalak:2017shs}. Phenomenological predictions: Bernauer et al.~(solid violet line with error band) \cite{A1:2013fsc}, Vanderhaeghen et al.~(dot-dashed pink) \cite{Guttmann:2010au}.}
\label{fig.OLYMPUS}
\end{figure}

 The OLYMPUS data (Fig.~\ref{fig.OLYMPUS}) at smaller $Q^2$ / larger $\ep$ are below unity. This is only replicated by the (mostly) phenomenological models, while theoretical predictions produce a ratio below 1 only much closer to $\ep=1$. Overall, the phenomenological predictions compare well to the data, but both predict a somewhat larger effect at larger $Q^2$\,/\,smaller $\ep$, albeit within the uncertainty of the data. Note that, calculations prior to and motivating the experimental proposal  \cite{Guttmann:2010au} predicted a large and visible deviation of the ratio from unity already around $Q^2\simeq$ 2 GeV$^2$.

\begin{figure}[ht]
\includegraphics[width=\columnwidth]{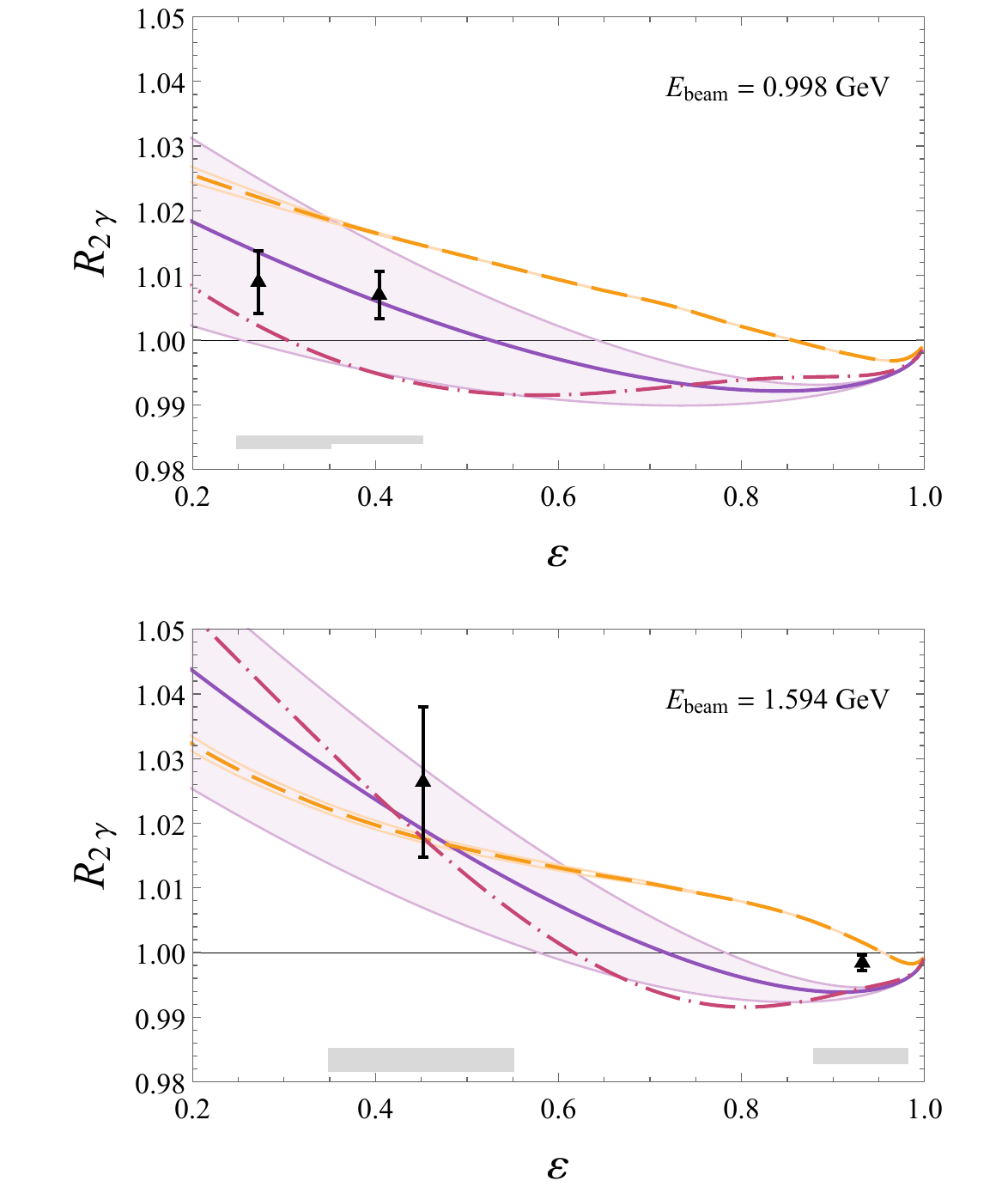}
\caption{Comparison of predictions for $R_{2\gamma}$, \Eqref{R2gammadef}, from $ep$ scattering with beam energies $E_\text{beam}=0.998$ GeV and $1.594$ GeV to results from the VEPP-3 experiment (black triangles) \cite{Rachek:2014fam} Note that the points are normalized according to the fitting model by Bernauer et al. \cite{A1:2013fsc} at the points $\varepsilon =0.931, \,\, Q^2 =0.128$ $\text{GeV}^2$ (top panel) and $\varepsilon =0.98, \,\, Q^2 =0.097$  $\text{GeV}^2$ (bottom panel). For further details see legend in Fig.~\ref{fig.OLYMPUS}.}
\label{fig.VEPP}
\end{figure}

\begin{figure}[t!]
\includegraphics[width=\columnwidth]{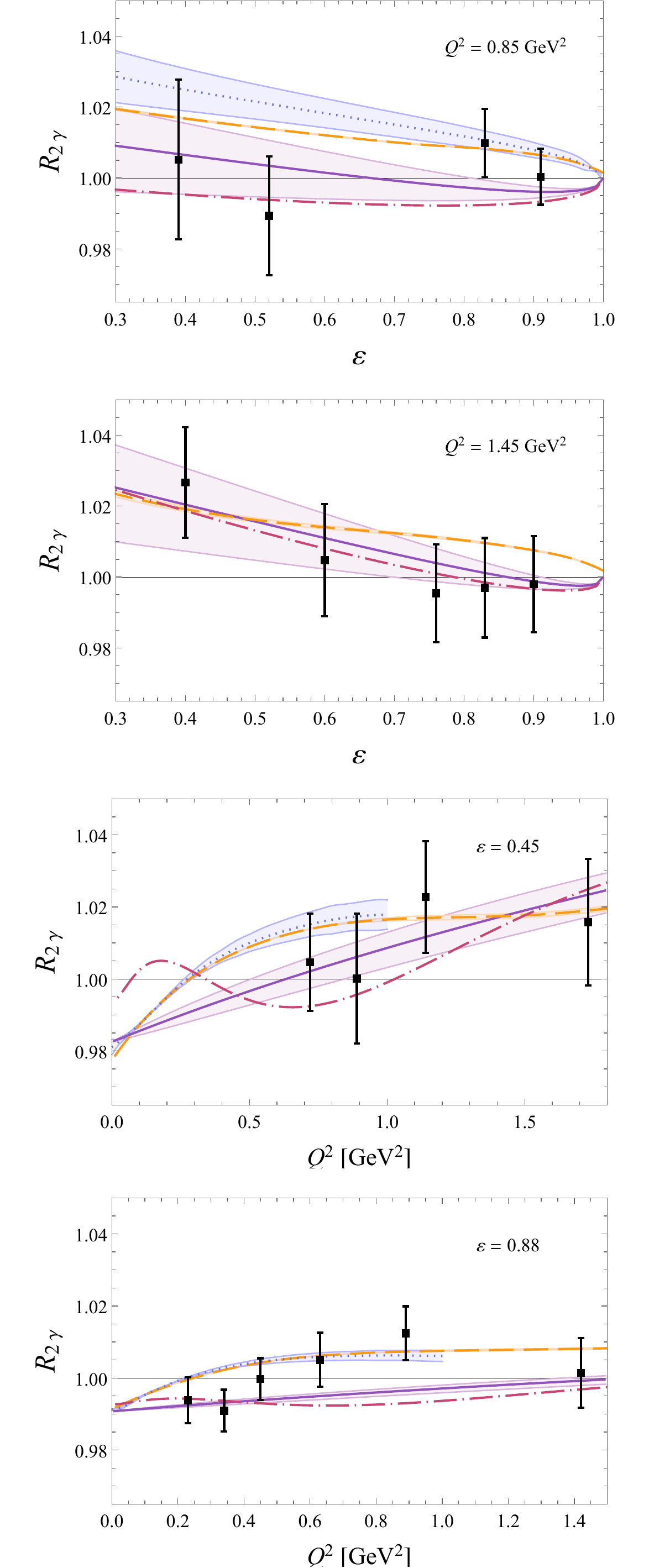}
\caption{Comparison of predictions for $R_{2\gamma}$, \Eqref{R2gammadef}, from $ep$ scattering with momentum transfer $Q^2=0.85$ GeV,  $1.45$ GeV, and photon polarization parameter $\varepsilon=0.45$ and  $0.88$ to results from the CLAS experiment (black squares) \cite{Rimal:2016toz}. For further details see legend in Fig.~\ref{fig.OLYMPUS}.} 
\label{fig.CLAS}
\end{figure}

The VEPP-3 (Fig.~\ref{fig.VEPP}) and CLAS data (Fig.~\ref{fig.CLAS}) also prefer the phenomenological extractions; however, the CLAS data are mostly compatible with Refs.~\cite{Ahmed:2020uso} and \cite{Tomalak:2017shs} as well. The CLAS $\ep=0.88$ data set and the VEPP-3 $E_\text{beam}=1.594$ GeV data set show a ratio below 1 for small $Q^2$ / large $\varepsilon$, similar to the behavior seen in the OLYMPUS result.

A global comparison of the data from the three experiments and the calculation from  \cite{Kuraev:2007dn} was done in Ref. \cite{Bytev:2019rdc}. 
The largest contribution to the charge asymmetry is due to the soft photon emission. A specific procedure was suggested to limit the effect of the different corrections applied to the data. The results do not favor an enhanced contribution of TPE in the considered kinematical range.

\subsection{Outlook} \seclab{TPEOutlook}

The OLYMPUS data gives a hint that at higher $Q^2$, TPE might not explain all of the discrepancy. Theoretical descriptions, which do not show a particular good agreement in the measured $Q^2$ range, are not at all tested at these larger $Q^2$. Precision measurements of the FFs at these large $Q^2$ however depend on an accurate description of TPE. This makes experimental determinations of TPE at these kinematics highly desirable. Unfortunately, experimental activities are limited by the availability of suitable beams.

The TPEX experiment \cite{Alarcon:2023kuv} seeks to measure TPE at DESY using beams of 2 and 3 \GeV~and up to a $Q^2$ of 4.6 $(\GeV/c)^2$. This measurement would require a new extracted-beam line at the DESY test beam facility. Current plans for future construction at DESY make it unlikely that this experiment can be realized.

Jefferson Lab is planning to construct a positron source for CEBAF. Current timelines put the availability of beam beyond 2030, if funding can be secured. Such a capability would allow for a comprehensive TPE program, with possible measurements in Halls A,~B and C \cite{Arrington:2021kdp,Bernauer:2021vbn,Cline:2021hkr,Grauvogel:2021btg}.

\begin{figure}[ht]
\includegraphics[width=0.45\textwidth]{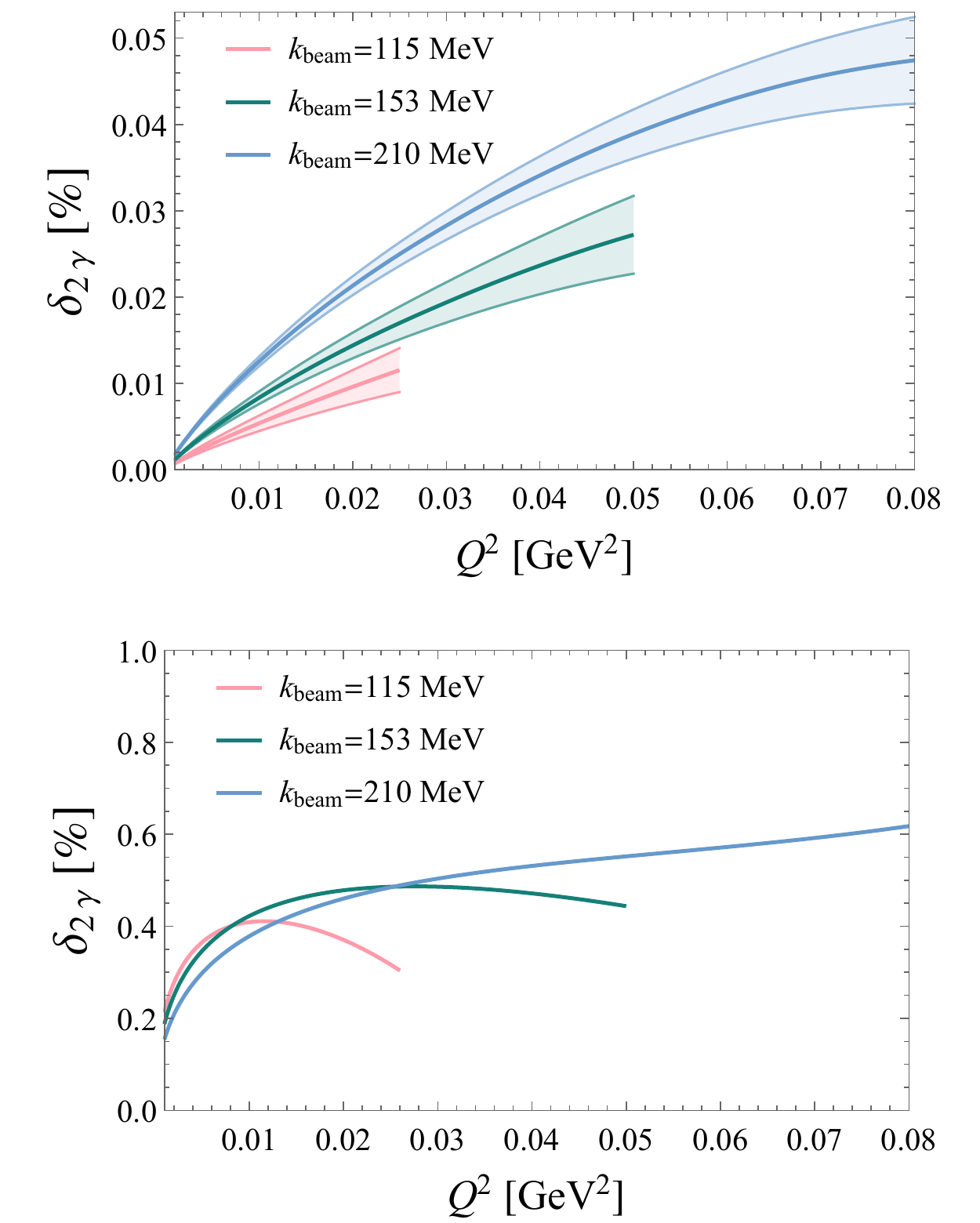}
\caption{Predictions for the inelastic (top panel) and total (bottom panel) TPE correction to $\mu^- p$ elastic scattering \cite{Tomalak:2015hva} for the three different beam momenta envisaged by the MUSE experiment \cite{MUSE:2013uhu}. While the elastic TPE dominates, the main uncertainty stems from the inelastic part. The relative uncertainty for the total TPE is thus small and not displayed here.} 
\label{fig.NearForw}
\end{figure}

At smaller $Q^2$, MUSE \cite{MUSE:2017dod,Cline:2021ehf} and AMBER \cite{Adams:2018pwt} will take data for $\mu p$ scattering with both beam charges. From this data, TPE can be determined at small $Q^2$ close to the validity of the Feshbach limit, and in the case of AMBER, at $\ep$ extremely close to 1. Theoretical predictions for the inelastic and total TPE in the kinematic region envisaged by the MUSE experiment are shown in Fig.~\ref{fig.NearForw}.

\section{Implementations and Generators for Lepton-Proton Scattering}\label{sec:Implementations}

Eventually the radiative corrections discussed above need to be made available to the experiments.
For less precise experiments it is often sufficient to do this in terms of fiducial (differential) cross sections that include a (often simplistic) model of the kinematic and geometric acceptances of the detector.
Doing this is very simple from a theoretical viewpoint as fiducial observables can easily be calculate by integrating over phase space with a measurement function $S$ as defined, e.g., in Eq.~\ref{mcLO}.
Even for very complicated measurement functions, this is a reasonably simple procedure.

However, for precision experiments, such as the current and next-generation $\ell p$ scattering experiments, fiducial cross section are insufficient; instead, corrections need to be included in the entire analysis pipeline in the form of an event generator.
Using only elastic LO events in the experimental simulation would introduce a large systematic uncertainty that would be impossible to reduce with only fiducial calculations.
Hence, the detector should be modelled using at least an NLO event generator.
To account for these effects properly, however, one really needs an NNLO generator that is ideally matched to some kind of parton shower to further capture leading logarithms beyond NNLO.
In other words, the experimental simulation needs to use the best theoretical calculation available. Similarly, it is important that the applied radiative corrections are well documented, so that future re-analyses can update the corrections with more advanced calculations.

\subsection{Overview of experimental requirements}

Experiments vary greatly in their requirements for kinematic range, supported event topology, and precision. Typically, the precision goals are informed by the size of other uncertainties, mainly the statistical precision, but also other unrelated systematic uncertainties, for example from geometry. But this can be misleading. While statistical uncertainty and often also other systematic sources have no or small correlation with kinematics, i.e.\ can be assumed to statistically average over many measurements, this is not true for radiative corrections, which have the potential to be highly correlated not only between data points of the same experiment, but even between multiple experiments if they use the same or similar formalism. 

Further, experiments might measure different event structures. In the case of $\ell p$ scattering, for example, the experiment could detect outgoing lepton, proton, or both, affecting the effectively seen corrections. Additionally, experiments can vary widely in the energy resolution, possibly requiring to describe the radiative tails accurately over rather large photon momenta ranges.  

\begin{table*}[t]
    \centering
    \caption{Characteristic parameters of published and upcoming proton form factor measurements}
      \scalebox{0.9}{
    \begin{tabular}{c|c|c|c|c|c|c}
         &\RotText{Measured particle}&\RotText{$E_\text{beam}$ [GeV]}&\RotText{$\theta$ [degrees]}&\RotText{$Q^2$ [GeV/$c^2$]}&\RotText{Effect of radius on form factor}&\RotText{Fractional contribution of $G_E$ to cross section}\\
         \hline
         GMp-12 \cite{Christy:2021snt}&$e^\prime$&2.222 -- 10.587&24.25 -- 53.5&1.577 -- 15.76&$>$100\%&0.2 -- 11.9\%\\

         MAMI FF \cite{A1:2010nsl,A1:2013fsc}&$e^\prime$&0.180 -- 0.855&15.5 -- 135&0.0033 -- 0.98&2 -- 59\%&0.23 -- 99.4\%\\
         MAMI High-$Q^2$ \cite{jmuellerthesis}&$e^\prime$&0.720 -- 1.500&15 -- 120&0.35 -- 1.95&$>$100\%&1.3 -- 92.5\%\\
         MAMI ISR \cite{Mihovilovic:2016rkr,Mihovilovic:2019jiz}&$e^\prime$&0.195, 0.330&15.21&0.001 -- 0.004&0.6 -- 2.4\%&$>$98\%\\
         MAMI Jet Target \cite{A1:2022wzx}&$e^\prime$&0.315&15 -- 40&0.007 -- 0.043&6 -- 39\%&71.2 -- 98.5\%\\
         PRad \cite{Xiong:2019umf}&$e^\prime$&1.1, 2.2&0.7 -- 6.5&0.00022 -- 0.058&0.13 -- 35\%&87.6 -- 100\%\\
        \hline
        AMBER \cite{Adams:2018pwt}&$\mu^\prime+p$&60, 100&N/A&0.001 -- 0.04&0.6 -- 24\%&91 -- 100\%\\
        MESA \cite{magixwebpage}&$e^\prime$&0.02 -- 0.105&15 -- 165&0.000027 -- 0.035
&0.016-21\%&91 -- 100\%\\
        MUSE \cite{MUSE:2017dod,Cline:2021ehf}&$e^\prime, \mu^\prime, \pi^\prime$&0.115 -- 0.21&20 -- 100&0.0016 -- 0.082
&0.96 -- 4.9\%&58 -- 99.6\%\\
        PRAD-II \cite{PRad:2020oor}&$e^\prime$&0.7, 1.4, 2.1&0.7 -- 6.5&0.00007 -- 0.056
&0.042 -- 34\%&88.6 -- 100\%\\
        ULQ2 \cite{Suda:2022hsm}&$e^\prime$&0.01 -- 0.06&0.7-6.5&0.000027 -- 0.012
&0.016 -- 7.2\%&56 -- 100\%\\

    \end{tabular}}

    \label{tab:expprop}
\end{table*}

Table \ref{tab:expprop} gives an overview over contemporary experiments \cite{Christy:2021snt,A1:2010nsl,A1:2013fsc,jmuellerthesis,Mihovilovic:2016rkr,Mihovilovic:2019jiz,A1:2022wzx,Xiong:2019umf,Adams:2018pwt,magixwebpage,MUSE:2017dod,Cline:2021ehf,PRad:2020oor,Suda:2022hsm}. We list beam energies, angle and $Q^2$ ranges, as well as detected particles. We also calculate an estimate of the effect of the proton radii on $G_E$ and $G_M$ as well as the fractional contribution of $G_E$ to the cross section, as an estimate for requirements on systematic uncertainties.  Of the planned experiments, only AMBER will measure the scattered lepton in coincidence with the recoiling proton. AMBER, PRad and PRAD-II both have forward kinematics with $\varepsilon$ very close to 1, while the other experiments tend to measure at larger angle ranges, $\varepsilon<1$. AMBER and MUSE will measure with leptons of both charges. The lower energy experiments require proper treatment of the lepton mass, and in the case of ULQ2, MUSE and likely MESA, a treatment without ultra-relativistic approximations.

Note that in most experiments at small $Q^2$, the absolute normalization of radiative corrections is less of an issue---it is generally left floating and determined by an extrapolating fit to $Q^2=0$. Typically experimental normalization uncertainties are significantly larger than radiative correction uncertainties, so that such a fit is unavoidable. On the other hand, experiments aimed to extract the radii typically measure at very small $Q^2$ to reduce systematical errors from the fit extrapolation to $Q^2=0$. Consequently, radiative corrections need to control relative, $Q^2$-dependent, errors to a very high degree: The proton radius puzzle is a 4\% difference in extracted radii, for recent reviews see Refs.~\cite{Antognini:2022xoo,Gao:2021sml,Peset:2021iul,Karr:2020wgh}. A meaningful measurement would extract the form factor slope to better than 0.5-1\% or so. A radius measurement at small $Q^2$ where the radius/slope effect is, say, 5\%, would then need to control $Q^2$-dependent systematic uncertainties over this $Q^2$ range to better than 0.05\%.  

For Rosenbluth separations at somewhat extreme kinematics, one of the form factor contributions to the measured cross section can be highly suppressed compared to the other one: $G_M$ is suppressed at small $Q^2$, while $G_E$ is at large $Q^2$. The uncertainty of radiative corrections on the cross section---in this case mainly the $\ep$-dependence at constant $Q^2$---then gets levered up by potentially large factors in the determination of the suppressed form factor. 

A further consideration is the requirement for unweighted events, or variance-reduced weights. This strongly depends on the overall speed of the remaining MC chain. For example, the Mainz A1 MC simulation code does not employ particle path tracing through magnetic fields and can simulate thousands if not millions of events per second, enabling to use fully weighted events, possibly even those with negative weights. In contrast, OLYMPUS' simulation chain included path tracing, full digitization and tracking for every simulated event, resulting in speeds of few events per second. This requires either unweighted events or at least measures to reduce the variance of the event weights.

\subsection{Overview of theoretical tools}

For now we will focus on tailor-made event generators for $\ell$$p$ experiments even though much effort has also been devoted to the development of generators at high energy (see \cite{Campbell:2022qmc} for a recent and comprehensive review).
However, we will revisit a potential synergy in Sec.~\ref{sec:mc:synergy}.
For $\ell$$p$ scattering, the most notable tools are {\sc ELRADGEN}~\cite{Afanasev:2003ic,Akushevich:2011zy}, {\sc ESEPP}~\cite{Gramolin:2014pva}, and {\sc Simul++}~\cite{Vanderhaeghen:2000ws,A1:2013fsc,Mihovilovic:2016rkr}.
Another tool in this context is {\sc McMule}~\cite{Banerjee:2020rww,Ulrich:2020frs} (as part of this topical collection, a simple TPE model is compare to the effect of NNLO corrections using McMule~\cite{McMule})  which currently cannot generate events.
However, work is underway to address this in a way that is convenient for experiments.
Looking slightly beyond $\ell p$ scattering, we have
the code by Epstein and Milner (EM)~\cite{Epstein:2016lpm} for Bhabha and M\o{}ller scattering;
{\sc MERADGEN}~\cite{Afanasev:2006xs} for M\o{}ller scattering;
{\sc MESMER}~\cite{CarloniCalame:2019mbo,CarloniCalame:2020yoz,Budassi:2021twh,Budassi:2022kqs} for muon-electron scattering;
and {\sc BabaYaga} for Bhabha scattering and photon pair production~\cite{CarloniCalame:2000pz,CarloniCalame:2001ny,CarloniCalame:2003yt,Balossini:2006wc,Balossini:2008xr}.
While these generators are not directly applicable for $\ell p$ scattering experiments, they could either be adapted into (trivially in the case of {\sc MESMER}) or help inspire new $\ell p$ generators.
Hence, we will still include them in our comparison as shown in Tab.~\ref{tab:mc}.

When we compare MC codes, we consider the order in perturbation theory they implement, whether they use resummation.
Another important point is whether they use any further approximations such as as an ultra-relativistic approximation (which treats the electron as nearly massless $E\gg m$) or the soft-photon approximation (which assumes $E,\sqrt{s} \gg E_\gamma$, also known as eikonal or peaking approximation).
We stress that the order in perturbation theory is always relative to the listed process, usually $ep\to ep$.
This means that an NLO calculation of $ep\to ep$ includes an LO calculation of $ep\to ep\gamma$ as a subset (cf. Sec.~\ref{Sec:RadCorrAdrian}).
If one wishes higher accuracy, i.e. NLO, for the latter process, more complicated calculations are required which, if properly implemented, would allow the implementation of an NNLO code.

As discussed in Sec.~\ref{Sec:RadCorrAdrian} there are two commonly used methods: slicing (cf. Eq.~\ref{slicing}) and subtraction (cf. Eq.~\ref{subtraction}).
Slicing tends to be more commonly used in NLO QED calculation as it is simpler to implement and use to generate events.

While most MC generators discussed here are fixed order, either at NLO or NNLO, some also include a resummation.
We will discuss the different strategies relevant here in Sec.~\ref{sec:mc:resum}.
Since the ability to produce unweighted events is very important from an experimental point of view, we will discuss different strategies in Sec.~\ref{sec:mc:unweight}.
Finally, while Table~\ref{tab:mc} focuses on purpose-built tools for low-energy experiments, one should not forget about the generalist tools that were developed for the LHC and other high-energy experiments.
In Sec.~\ref{sec:mc:synergy} we will discuss some potential synergies.

\begin{table*}[t]
    \centering
    \newcommand{\pol}[1]{{#1}}
     \caption{
        An overview of MC codes that are relevant for $ep$ scattering experiments.
        (N)NLO codes that use slicing are indicated with {\it w/ sl.} while codes that use subtraction are indicated with {\it w/ su.}
    }
    \scalebox{0.9}{
    \begin{tabular}{l|p{2.4cm}|l|l|p{3.0cm}|p{3.0cm}}
        \bf Name                     & \bf Process                             & \bf Order  & \bf Resummation & \bf Approximation       & \bf Unweighting \\\hline

        \sc ELRADGENv2               & polarised\qquad $\pol{e}\pol{p} \to ep$ & NLO w/ sl. & YFS exp.   & leptonic                     & flat weights for rad. events \\
        \sc ESEPP                    & $ep\to ep$                              & NLO w/ sl. & none       & $E\gg m$ for virtual, \cite{Maximon:2000hm} for TPE & Foam~\cite{Jadach:2005ex} \\
        \sc Simul++                  & $ep\to ep$                              & NLO w/ sl. & exp.       & none                         & reduced-variance weights \\
        {\sc OLYMPUS} & $ep\to ep$                              & NLO (w/ sl. opt.) & none/exp.       & single photon kinematics                         & multiple modes \\
        \sc McMule                   & $ep\to ep $              & NNLO w/ su.& planned    & point-like proton & planned \\
        & \& others & & & beyond NLO &\\
        \hline

        EM                           & $ee\to ee$                              & NLO w/ sl. & none       & none                         & Foam~\cite{Jadach:2005ex} \\
        \sc MERADGEN                 & polarised\qquad $\pol{e}\pol{e}\to ee$  & NLO w/ sl. & none       & none                         & flat weights for rad. events \\
        \sc BabaYaga                 & $e^+e^- \to e^+e^-$                     & NLO w/ sl. & matched PS & none                         & only pos. + hit-and-miss \\
        \hline

        \sc MESMER                   & $e\mu\to e\mu$                          & NNLO w/ sl.& planned    &  electronic at NNLO          & mostly pos. + hit-and-miss \\

    \end{tabular}}
   
    \label{tab:mc}
\end{table*}

\subsubsection{Parton shower and resummation}
\label{sec:mc:resum}

When calculating higher-order corrections, be it at NLO or NNLO, one often finds that the corrections are enhanced by the presence of large logarithms $L$ between disparate scales such as $\ln(m_e^2/s)$, $\ln(m_e^2/m_p^2)$, or $\ln(Q^2/m_p^2)$.
These logarithms somewhat spoil the perturbative convergence since we usually get at least one such logarithm per order in $\alpha$.
This means that NLO, and sometimes even NNLO, is insufficient.
One way out of this problem is just to compute even higher orders in $\alpha$.
However, this quickly becomes infeasible, with partial N$^3$LO corrections just about feasible.
Luckily, there is rarely a need to compute the full corrections since typically only the logarithmically enhanced matter.
These terms follow a very predictable structure and can often be calculated to all orders in $\alpha$.
This process is called resummation and it introduces its own counting.
Accounting for all terms of the form $\alpha^n L^n$ is referred to as leading logarithm (LL), the second ``tower'' $\alpha^n L^{n-1}$ is called next-to-leading logarithm (NLL) and so on.

When combining a calculation at a given level of resummation, say LL, with a fixed order calculation at e.g. NLO, one must take care to avoid double-counting.
Both calculations include the same $\mathcal{O}(\alpha L)$ term, albeit in different forms.
Ensuring that this does not spoil the result requires the two calculations to be matched at the given order.

In QED we have two main sources of large logarithms: those related to the smallness of the electron mass (usually called collinear logarithms) and those related to restricting real radiation, either through explicit cuts on the phase space integration or by considering the endpoints of distributions (usually called soft logarithms).

Logarithms can either be resummed using a numerical algorithm such as a parton shower (see below) or analytically.
A parton shower resummation contains the full LL but may contain partial results of the NLL as well, depending on method and implementation.
The development of an exact NLL parton showers is an area of very active research at the high-energy frontier.
Analytic resummation can be performed well beyond LL with state-of-the-art calculations reaching next-to-next-to-next-to-leading logarithm (N$^3$LL).
However, these calculations are observable dependent and cannot easily be used to generate events.

\paragraph{YFS exponentiation}
The simplest way to resum soft logarithms is Yennie-Frautschi-Suura (YFS) exponentiation, a procedure that is based on the universality of soft singularities in QED~\cite{Yennie:1961ad}.
In its simplest form, this accounts for large logarithms near the endpoints of differential distributions e.g. by exponentiating the singular behaviour of the distributions.
This approach has been considered long ago for LEP~\cite{Kuraev:1985hb, Skrzypek:1992vk, Cacciari:1992pz} at LL.
However, the same strategy has recently been used to calculate soft logarithms up to next-to-next-to-leading logarithm (NNLL) in the electron spectrum of muon decay.

\paragraph{YFS shower}
This form of YFS exponentiation can be used to obtain very precise results for single distributions.
However, without modification, it cannot be used to generate events.
By using the YFS approximation on both real and virtual corrections, we can construct a locally finite prescription to generate an event with an arbitrary number of real but soft photons~\cite{Schonherr:2011vna}
This shower reproduces the soft LL for any observable as the resummation happens for each phase space point.
Further, matching a YFS shower to a fixed-order calculation is relatively straightforward since one can relatively easily incorporate the fixed-order calculation in the YFS shower~\cite{Schonherr:2011vna}.

\paragraph{DGLAP-based parton showers}
An alternative resummation approach is based on the structure function approach. 
Given an electron with a fixed momentum $p$, the structure function $D(x, Q^2)$ is the probability density of observing an electron with momentum $xp$ and virtuality $Q^2$.
Convoluting a structure functions for each external electron with the fixed-order cross section then resums the collinear logarithms to LL accuracy.
The structure functions themselves are governed by the Dokshitzer-Gribov-Lipatov-Altarelli-Parisi (DGLAP) evolution equation~\cite{Gribov:1972ri,Altarelli:1977zs,Dokshitzer:1977sg}
\begin{align}
    Q^2 \frac{\partial}{\partial Q^2} D(x,Q^2) = \frac{\alpha}{2\pi} \int_x^1 \frac{\dd y}{y} P_{ee}(y) D\Big(\frac xy, Q^2\Big)\,,
\end{align}
where $P_{ee}(x)$ is the (regularised) kernel that is known up to two-loop accuracy.
A parton shower is an algorithm of solving the DGLAP equation numerically using MC methods~\cite{Ellis:1996mzs}.
This allows us not only to obtain an exact numerically solution for the structure functions but also to generate approximate photon momenta.
The latter can be improved by using the YFS theorem~\cite{CarloniCalame:2001ny}.

Work is still ongoing to reach NLL accuracy with a general parton shower.
However, analytic solution of the DGLAP equation at NLL are already possible in QED~\cite{Bertone:2019hks}.

\subsubsection{Unweighting procedures}
\label{sec:mc:unweight}
So far, we have only discussed the generation of weighted events.
This means that we generate a set of random momenta $\{p_i^\mu\}$ and use them to calculate an event weight $w_i$ that is comprised of the matrix element (squared) and a phase space mapping that accounts for the way we have generate the momenta.
We then use this event to perform some sort of histogramming, potentially involving a detector simulation.
The event weight itself may vary across many orders of magnitude and even be negative.
This increases the number of events that need to be generated to obtain a reasonable uncertainty on the final histogram since negative events will lead to (potentially severe) cancellations.

If the histogramming step is comparatively fast, one would normally stop here and just increase the number of events generated.
However, a realistic detector simulation can be extremely complicated and require hours of CPU time which is expensive even compared to the matrix element evaluation that usually takes less than a second.
Hence, our goal needs to be the reduction of the number of events that need to be passed to the detector simulation.
The easiest way to do this, is to generate events with uniform weight even if this requires the evaluation of many more events on the theory side.
Even in the absence of negative weights, this is a non-trivial task:
given a probability distribution function (PDF) that we in general only know numerically, we want to generate variables that follow this distribution.

In simple cases, it is possible to generate the event directly without weight; the energy of an emitted photon can be sampled as
\begin{align}
    E = E_\text{min} \Big(\frac{E_\text{max}}{E_{\text{min}}}\Big)^\#,
\end{align}
where $0<\#<1$ represents a uniformly distributed random number and $E_\text{min}$ some cut-off parameter.
This is possible because soft photons are distributed following the eikonal approximation.
However, this fails as soon as one considers higher-order corrections as the distribution becomes much more complicated.

A more complicated method is called \emph{hit-and-miss}:
we generate a set of weighted events with weights $w_i$ and find the maximum weight $W=\max |w_i|$.
A given event is rejected with probability $p=1-|w_i|/W$, otherwise it is accepted with weight ${\rm sign}(w_i)\times W$.
At this stage we have significantly fewer events than we started with but their weights are all $\pm W$.

This strategy can be somewhat improved upon by sampling the original distribution using an adaptive MC sampler rather than naively using uniformly distributed random numbers.
The simplest such sampler is VEGAS~\cite{Lepage:1980dq} which will sample regions with large weights more often than those of low weight, improving efficiency.
It does this by splitting each axis of the integration into (by default) 50 subdivisions that are each sampled the same.
The sizes of divisions are than adjusted to sample regions of large integrand more finely.
Note that this relies on the behaviour of the integrand being reasonably well aligned with the axis of the integration.
This has been improved upon by cells instead of a grid in FOAM~\cite{Jadach:2002kn}.
More recently, machine learning techniques were investigated for even more efficient sampling methods, e.g.~\cite{Gao:2020vdv,Bothmann:2020ywa,Gao:2020zvv}.

Much effort has been dedicated to improve unweighting procedures as hit-and-miss is very inefficient, especially in the presence of many events with negative weights $w_i<0$.
This can for example be done by optimising the event generation, i.e. ensure that fewer event with negative weights are generated in the first place by modifying the parton shower matching procedure~\cite{Danziger:2021xvr,Frederix:2020trv}.
Alternatively, one can carefully remove the events with negative weight without affecting physical observables~\cite{Andersen:2020sjs,Nachman:2020fff,Stienen:2020gns}.

Recently, this led to the development of cell resampling~\cite{Andersen:2021mvw} which works process independently and preserves all physical observables.
The idea here is to utilise the experiment's finite resolution that is anyway required to IR finiteness.
We now create a small sphere around every event with negative weight, ensuring that the sphere is smaller than the experimental resolution and cannot be probed.
Assuming enough event were generated, the sum of all weights in this sphere is positive since experiments always measure a positive cross section. 
We now replace all events by their absolute value, taking care not to disturb the sum of weights.

\subsubsection{Synergy with high energy}
\label{sec:mc:synergy}

Many MC generators were developed for the LHC physics program.
While some of these are purposed-built for certain high-energy processes, other are more general.
Some of the most mature, general-purpose tools are {\sc Sherpa}~\cite{Gleisberg:2008ta,Sherpa:2019gpd}, {\sc Herwig}~\cite{Bahr:2008pv,Bellm:2015jjp}, {\sc Pythia}~\cite{Bierlich:2022pfr}, and {\sc MadGraph\_aMC@NLO}~\cite{Bertone:2022ktl}.
While these tools were designed for hadron colliders, they could be extended for low-energy elastic $e$-$p$ scattering.
In the following, we briefly summarise the four main codes and what they could offer without major modifications.

While {\sc Sherpa} usually generates its own NLO matrix elements, it can be fed the required expressions in through a shared library.
This means that once provided with all matrix elements (including the TPE) as some kind of callable routine, it would be able to generate events at NLO accuracy out of the box.
Further, it provides a full YFS shower~\cite{Schonherr:2008av} that can resum soft logarithm; collinear contributions are taken at fixed order.

{\sc Pythia} itself is not a program but rather a library that can be used for event generation.
As such, it is trivial to link to an external matrix element provider and it could then be used for the generation of NLO events.
While naturally focused on QCD showers and hadronisation, it does include a QED shower that includes a fully coherent multiple treatment of photon radiation beyond the YFS approximation.
Further, the resummation of initial-state collinear log is possible using a lepton parton distribution function.

{\sc MadGraph\_aMC@NLO} is mainly designed for the automatic generation of NLO matrix elements which are then used in a QCD parton shower.
The automatic generation is less helpful for our purposes but could be side-stepped with some effort.
The code further provides a QED shower for final states and, similar to {\sc Pythia}, can use a lepton parton distribution function for the inclusive resummation of ISR.
Weight reduction to unit weights $w=\pm1$ is only available when combined with the parton shower to avoid the introduction of an IR cut-off.

{\sc Herwig} is a Monte Carlo event generator with QED showering.
In normal operations for LHC physics, amplitudes are included up to NLO.
However, for more complicated processes {\sc Herwig} is already set up to obtain amplitudes from external libraries.
This feature could be used to provide the generator with matrix elements relevant for $\ell$-$p$ scattering.

Facilitated by the development of many different generators, the high-energy community was forced to develop standardised interfaces for generators and event formats.
The most commonly used ones are LHEF~\cite{Alwall:2006yp} and HepMC3~\cite{Buckley:2019xhk}.
These standards are used in conjunction with the {\sc Rivet} framework~\cite{Bierlich:2019rhm}.
{\sc Rivet} supports the development of generators by encouraging experimentalists to share their analyses in a common framework that can be applied both on data and Monte Carlo samples.
New generators (at both the high-energy and low-energy frontier) will probably interface to these standards to cut down on complexity.

\subsection{Conclusion and outlook}

The precision requirements of modern $\ell p$ scattering experiments require unprecedented theoretical calculations.
A critical part of this is the development of event generators that can be used in experimental simulations.
These requirements mean that we need to go beyond leading order for the radiative process $\ell p\to \ell p\gamma$. This is also roughly equivalent to going beyond NLO for $\ell p\to \ell p$.
Currently, no event generator with these capabilities exist even though some codes (notably {\sc McMule} and {\sc MESMER}) could be adapted.

Unfortunately, the increased precision comes with a massive increase of the complexity of the distributions that need to be sampled.
Luckily this is a known problem;
the LHC community has worked on issues that are almost identical for decades.
While not all ideas are immediately transferable to our process, many are.
Additionally to the calculation of one-loop matrix elements (which is fully automatised for all standard theories), this includes methods to deal with IR singularities, negative weights, and numerical resummation using a parton shower.

Using these tried and tested techniques will allow us to construct event generators with (and beyond) NNLO accuracy for $\ell p\to \ell p$ in the coming years.

\section{Higher-order corrections in QED inelastic scattering}\label{sec:ho}
The Rosenbluth formula \cite{Rosenbluth:1950yq} as well as the Akhiezer-Rekalo  formalism \cite{Akhiezer:1968ek,Akhiezer:1974em}, respectively, for  unpolarized and polarized elastic $ep$ scattering hold only in the OPE formalism where two real FFs, functions of $Q^2$ only, completely describe the reaction. In the case of TPE, instead of two real functions of $Q^2$ one has to deal with three amplitudes, neglecting the electron mass. Three additional charge-odd amplitudes, that are proportional to the electron mass, are of the same order, generally of complex nature and also depending on two kinematical variables. The latter are relevant for observables such as transverse polarization.

Model-independent statements hold also in the annihilation region \cite{Gakh:2005hh,Gakh:2005wa}. Assuming time and parity invariance and crossing symmetry, the reactions $e^+e^-\leftrightarrow p\bar p$ are described by the same electromagnetic FFs. However, in the time-like region of transferred momenta FFs are of complex nature, even in the case of OPE.  Due to the large transferred momenta that are involved above the physical threshold one would therefore expect that TPE effects are enhanced. 

In the time-like region, the angular distribution of one of the produced particles allows one to extract the moduli of the FFs. In principle such measurement is simpler than a Rosenbluth separation, as it requires only one setting of the collider equipped with  a $4\pi$ detector. The $\varepsilon(\sim \cot^2\theta$)-linearity of the Rosenbluth plot translates by crossing symmetry into a $\cos^2\theta$ dependence of the annihilation cross section. Then, a forward-backward asymmetry arises in the presence of charge-odd contributions, as hard TPE. The sum (difference) of the cross section at corresponding angles would cancel (enhance) the TPE contribution, taking advantage of its charge-odd nature.

The study of both annihilation reactions $e^+e^-\leftrightarrow p\bar p$, related by time reversal is very instructive in this respect, as radiative corrections are in general different. In $e^+e^-$  colliders, huge effort has been done to achieve a per-mille precision, based on the lepton structure function method \cite{Baier:1980kx}. This is made necessary by the use of the ``initial state radiation'' (ISR) process \cite{Czyz:2004ua}, that allows to scan a large region of transferred momenta, even in a fixed energy collision. 

Radiative correction for the reaction $\bar p p\to e^+e^- $, induced by antiproton beams  have been particularly studied \cite{Bystritskiy:2019iki} and are being implemented in a dedicated MC event generator (cf. Sec.~\ref{sec:Implementations}),  in the frame  of the time-like FF program planned at FAIR \cite{PANDA:2016fbp}. The interest of the antiproton beam  is also related to the possibility to measure the muon-antimuon final state \cite{PANDA:2020jkm}. 

The search of charge-odd contributions to the annihilation cross section has been looked for in the BaBar data, see Ref.~\cite{Tomasi-Gustafsson:2007vpn}, with the conclusion that they vanish in the limit of 2\%, which is the order of magnitude of the interference between initial and final state radiation emission. The presence of TPE in the future data from PANDA at FAIR has been simulated in Ref. \cite{PANDA:2016fbp} pointing out that a contribution of $\ge 5\% $ would be detectable.

The QED radiative corrections to both deep-inelastic scattering and $e^+e^-$ annihilation at large virtualities
$Q^2$ and cms energies $s \gg m_e^2$ are very important for precision measurements 
since they can be 
differentially rather large in parts of the phase space. Furthermore, they strongly depend on the different 
possible measurements of the kinematic variables of the respective processes. Unlike the case in massless QCD,
the emerging logarithms $L = \ln(Q^2/m_e^2)$ or $L =\ln(s/m_e^2)$ are observable {\it phase-space} logarithms and 
cannot be resummed away as in the strictly massless case. In particular, they can be measured as has been done
in many experimental analyses in deep-inelastic scattering and $e^+e^-$ annihilation. In addition, the fermion 
mass corrections cannot be dealt with as purely massless corrections from the beginning, cf. 
e.g.~\cite{Blumlein:2020jrf}. They are, furthermore, not just given by splitting functions and massless Wilson 
coefficients, but one has to compute massive operator matrix elements (OMEs) instead of the splitting functions. 
Two 
renormalization group equations rule the corrections in the limit $Q^2 \gg m_e^2$ or $s \gg m_e^2$. \footnote{More 
generally $m_e$ can also be replaced by $m_\ell, \ell = e, \mu, \tau$} These quantities are ruled by two 
renormalization group equations, cf.~\cite{Blumlein:2000wh,Blumlein:2011mi,MUTA}, containing also the charged 
lepton masses. They read\footnote{In addition the gauge parameter $\xi$ needs to be renormalized, which we have 
suppressed here.}

\begin{align}
\label{eq:R1}
& 
\left[\left(\mu \frac{\partial}{\partial \mu} + \beta(g) \frac{\partial}{\partial g}\right) \delta_{al} 
- \gamma_m(g) m_e \frac{\partial}{\partial m_e}
+ 
\gamma_{al}(N,g) \right]\nn \\
&\times\Gamma_{li}\left(N, \frac{\mu^2}{m_e^2},g(\mu^2)\right) = 0,
\\
\label{eq:R2}
&\Biggl [\left(\mu \frac{\partial}{\partial \mu} + \beta(g) \frac{\partial}{\partial g}\right) \delta_{al} 
\delta_{kb}
- \gamma_m(g) m_e \frac{\partial}{\partial m_e}\nn \\
&- \gamma_{al}(N,g) \delta_{kb}
- \gamma_{kb}(N,g) \delta_{la} \Biggr ]\nn \\
&\times\tilde{\sigma}_{lk}\left(N, \frac{s'}{m_e^2},g(\mu^2)\right) = 0.
,\end{align}

Here $\mu$ denotes the factorization and renormalization scale, $g$ the QED coupling constant, $\beta$ the
QED $\beta$-function, $\gamma_m(g)$ the mass anomalous dimension,
$\gamma_{ij}$ the QED anomalous dimensions, $\Gamma_{li}$ are the massive OMEs and 
$\tilde{\sigma}_{lk}$ are the massless sub-system scattering coefficients. Finally, $s'$ (or $(Q^2)'$) denote the
sub-system cms energy (or virtuality). The renormalization group equations rule all corrections down to the
constant term which does not depend on the scale logarithm. Here the renormalization group equations have been written in 
Mellin space. Since the radiatively corrected (differential) scattering cross sections are observables, the 
$\mu$-dependence needs to be canceled, which is possible by a consistent expansion in the coupling constant $g$.
Usually the consideration of the radiators $\Gamma_{li}$ is therefore not sufficient, with the exception of the
leading order terms $\mathcal{O}(\alpha^k L^k)$, with $\alpha$ the fine structure constant, since there only the Born 
sub-system cross section contributes. They are leg-dependent
as the other corrections and the relevant hard scales rescale with the collinear emission variable $z$ accordingly
as will be outlined below.

The representation of the radiative corrections using Eqs.~\ref{eq:R1} and \ref{eq:R2} allows to calculate all 
but the power corrections $(m_e^2/s)^k,~~k \in \mathbb{N}, k \geq 1$. The latter terms require the full 
phase-space integrals to be carried out and are believed to yield small contributions only in the  limit of large
scales.

In Sec.~\ref{sec:2} we will consider the QED radiative corrections for 
the deep-inelastic scattering process. Section~\ref{sec:3} deals with the initial state QED radiative corrections 
to $e^+ e^-$ annihilation and Sec.~\ref{sec:4} contains the conclusions.

\subsection{Higher-order corrections to deep-inelastic scattering}\label{sec:2}

\vspace*{1mm}
\noindent
The QED corrections to deeply-inelastic scattering have been calculated in the context of different
experiments starting with those at SLAC \cite{Mo:1968cg}, CERN, with BCDMS and EMC 
\cite{Bardin:1980ii}, and with much more work for HERA, cf.~Refs.~\cite{Kwiatkowski:1990es,Arbuzov:1995id}.
These papers cover the $\mathcal{O}(\alpha)$ corrections and we will mostly discuss higher-order corrections in the 
following. Furthermore, most of the calculations refer to unpolarized-lepton unpolarized-nucleon 
scattering. Later on we will
also discuss the QED radiative corrections to polarized nucleon scattering. Early leading logarithmic  $\mathcal{O}(\alpha)$ 
approaches (LLA) are shown in Refs.~\cite{Mo:1968cg,DeRujula:1979grv,Consoli:1980nm}, see also Refs.~\cite{Kuraev:1985hb,
Kuraev:1988xn,Fadin:1989dw}. The $\mathcal{O}(\alpha)$ electroweak 
corrections, also for $\nu N$ scattering, were calculated in Refs.~\cite{Marciano:1980pb,Sarantakos:1982bp,
LlewellynSmith:1981tgo,Wheater:1982yk,Paschos:1981zx,Wirbel:1982cw,Liede:1983hw,Bardin:1986bc,Diener:2003ss,
Arbuzov:2004zr,Diener:2005me,Park:2009ft,NuTeV:2001whx} and applied to data. This has been important for the 
measurement of $\sin^2\theta_W$ from deep-inelastic data.

Complete neutral and charged current $\mathcal{O}(\alpha)$ corrections to $ep$ scattering were calculated in Refs.~\cite{Bohm:1986na,Bohm:1987cg,Spiesberger:1990fa}, calculating the bremsstrahlung contributions using MC 
integration. The neutral and charged current $\mathcal{O}(\alpha)$ corrections to deep-inelastic $ep$ scattering were
calculated analytically for the differential cross sections $\dd^2 \sigma^{\rm DIS}/\dd x/\dd Q^2$ in 
Refs.~\cite{Bardin:1987rz,Bardin:1988by,Bardin:1989vz}. The $\mathcal{O}(\alpha)$ leading log corrections to the neutral and 
charged current process in leptonic variables were computed in Ref.~\cite{Blumlein:1989gk}, correcting earlier work in Ref.~\cite{Kripfganz:1987yu}.
The LLA results agreed very well with the complete results in the neutral current case. For only leptonic 
contributions these corrections were calculated in Refs.~\cite{Beenakker:1989km,Montagna:1991ku} also. The radiative 
corrections are quite different for different choices of variables to define the differential cross sections, 
cf.~Ref.~\cite{Blumlein:2012bf}. The LLA corrections in the case of jet (or hadronic) variables were calculated 
in Ref.~\cite{Blumlein:1990wz}. In Tab.~\ref{TAB:KIN} we display the $z$ rescaling for some of the kinematic 
variables.

\begin{table*}
\centering

\caption{\label{TAB:KIN} Scaling behavior of various sets of kinematic variables for leptonic initial and final state
radiation, cf. \cite{Arbuzov:1995id,Blumlein:2012bf}. The shifted variables $\hat{Q^2}$, $\hat{y}$, the threshold $z_0$ and the Jacobian ${\cal J}$ depend on the choice of the
external kinematic variables.}
\scalebox{0.9}{
\begin{tabular}{l|c|c|c|c}

\multicolumn{5}{c}{Initial State Radiation $\hat{s} = z s$} \\
\hline
\multicolumn{1}{c}{Kinematics} &
\multicolumn{1}{|c}{$\hat{Q}^2$} &
\multicolumn{1}{|c}{$\hat{y}$} &
\multicolumn{1}{|c}{$\hat{z_0}$} &
\multicolumn{1}{|c}{${\cal J}(x,y,z)$} \\
\hline
leptonic variables
& $z Q_l^2$
& $\frac{\ds z + y_l -1}{\ds z}$
& $\frac{\ds 1- y_l }{\ds 1 - x_l y_l}$
& $\frac{\ds y_l}{\ds z + y_l -1}$
\\
mixed    variables
& $z Q_l^2$
& $\frac{\ds y_{JB}}{\ds z}$
& $y_{JB}$
& 1
\\
hadronic variables 
& $Q_h^2$
& $\frac{\ds y_h}{\ds z}$
& $y_h$
& $\frac{1}{z}$ 
\\
\hline
\multicolumn{5}{c}{Final State Radiation $\hat{s} = s$} \\
\hline
leptonic variables
& $\frac{\ds Q_l^2}{\ds z}$
& $\frac{\ds z+y_l-1}{\ds z}$
& $1 - y_l(1-x_l)$
& $\frac{\ds y_l}{\ds z(z+y_l-1)}$
\\    
mixed    variables
& $\frac{\ds Q_l^2}{\ds z}$
& $y_{JB}$
& $x_m$
& $\frac{1}{z}$
\\
\end{tabular}
}
\end{table*}

Here ${\cal J}(x,y,z)$ denotes the respective Jacobian and $x =  Q^2/(s y), y = p.q/p.l$, with 
$p$ and $l$ the nucleon and lepton 4-momentum and $q^2$ the virtuality of the process with $Q^2 = - q^2$.
$z_0$ denotes the hard bremsstrahlung threshold.
The rescaling relations for other sets of variables, such as Jaquet-Blondel variables,
the double-angle method, the $\theta y$ and  $e \Sigma$ methods, see Refs.~\cite{Arbuzov:1995id,Blumlein:2012bf}.

Let us illustrate the radiator method in lowest order LLA. One obtains

\begin{align}\frac{\dd^2 \sigma^{\rm ini(fin)}}{\dd x \dd y}& =
\frac{\alpha}{4 \pi}
L_e
\int_0^1 \dd z
 P_{ee}^{(0)}(z)  
\Biggl[
\theta(z - z_0) \nn \\
&\times{\cal J}(x,y,z)
\frac{d^2 \sigma^{0}}{\dd x \dd y}_{x=\hat{x},
y=\hat{y},s=\hat{s}}   
-  \frac{\dd^2 \sigma^{0}}{\dd x \dd y} \Biggr].
\end{align}

Here $L_e$ denotes the radiation logarithm $L_e = \ln(Q^2/m_e^2) - 1$ and we have accounted for 
a non-logarithmic correction in addition. $P_{ee}^{(0)}$ denotes the $\mathcal{O}(\alpha)$ QED splitting 
function for the $e e$ transition \cite{Fermi:1924tc}. The QED splitting functions and anomalous 
dimensions can be obtained from those of QCD \cite{Moch:2004pa,Blumlein:2021enk,
Vogt:2004mw,Ablinger:2017tan,Moch:2014sna,Blumlein:2021ryt} to three-loop order in the unpolarized and polarized cases by 
setting $T_F = 1, C_F = 1, C_A = 0$. The photon-photon splitting function needs more care.
The consistent construction of the different order radiative corrections including both the massive OMEs and the 
massless sub-system scattering cross sections is described in Ref.~\cite{Ablinger:2020qvo} 
in detail in Mellin space. Parts of the radiators depend also on generalized harmonic sums
\cite{Moch:2001zr,Ablinger:2013cf} while the others are given by harmonic sums 
\cite{Vermaseren:1998uu,Blumlein:1998if}.
However, their Mellin inversion to $z$-space can be written in terms of harmonic 
polylogarithms \cite{Remiddi:1999ew} at modified argument. Standard technologies, cf.~Ref.~\cite{Blumlein:2018cms}, 
also allow one to 
calculate the radiators in $z$-space.

We also would like to mention that the
transition matrix elements from $e^- \leftrightarrow e^+$ were given in Ref.~\cite{Blumlein:2011mi} to two-loop order. They 
are 
related to the different non-singlet anomalous dimensions $\gamma^{NS,\pm}$ \cite{Moch:2014sna,Blumlein:2021enk} 
starting at two-loop order.

Different radiative correction packages such as {\sc TERAD, HERACLES, HELIOS} and 
{\sc HECTOR} were designed  \cite{Bardin:1980ii,Kwiatkowski:1990es,Blumlein:1991ag,Akhundov:1994my,Arbuzov:1995id}. 
In addition 
to the $\mathcal{O}(\alpha L)$ corrections, which are large in certain parts of the phase space, also the $\mathcal{O}(\alpha^2 L^2)$ 
corrections were calculated for leptonic variables in Ref.~\cite{Kripfganz:1990vm} and for the whole set of kinematic 
variables in Ref.~\cite{Blumlein:1994ii}. Related radiators were obtained in Refs.~\cite{Nicrosini:1986sm,Berends:1987ab}
and later corrections in Ref.~\cite{Blumlein:2011mi}.

Likewise, complete $\mathcal{O}(\alpha)$ 
calculations have been performed for 
Jacquet-Blondel variables in Ref.~\cite{Akhundov:1993wvk} and for mixed variables in Ref.~\cite{Bardin:1995mf}. 

The 
$\mathcal{O}(\alpha^2 L)$ corrections for the latter case were computed in Ref.~\cite{Blumlein:2002fy}, extending 
the results of Refs.~\cite{Bardin:1995mf,Blumlein:1994ii}. All sub-leading corrections of this type are process 
dependent since they also depend on lower order hard scattering contributions. Therefore, resummations of 
non-leading radiators, cf.~e.g.~Ref.~\cite{Bertone:2019hks} and references therein, do not lead to a complete 
description. Furthermore, one 
has to deal with massive OMEs, cf.~e.g.~Ref.~\cite{Blumlein:2011mi}.

The QED corrections to the final state hadronic jet
are dealt with as inclusive, since they usually cannot be resolved as electromagnetic showers do also occur
in the hadronization process. Therefore, this part of the corrections is summed up according to the 
Kinoshita-Lee-Nauenberg theorem \cite{Kinoshita:1962ur,Lee:1964is}.

The radiators of the LLA corrections to higher order $\mathcal{O}(\alpha^k L^k)$ are process and radiation leg independent. Furthermore,
the non-singlet corrections are the same in the unpolarized and polarized case. The radiators for the $\mathcal{O}(\alpha^2
L^2)$ corrections were calculated in Refs.~\cite{Kripfganz:1990vm,Blumlein:1994ii}. The corresponding higher-order corrections up to $\mathcal{O}(\alpha^5 L^5)$ were computed in Refs.~\cite{Blumlein:2002bg,Jezabek:1991bx,
Skrzypek:1992vk,Przybycien:1992qe,Arbuzov:1999xso,Arbuzov:1999cq,Blumlein:2004bs,Blumlein:2007kx}, where the 
singlet corrections were calculated in Ref.~\cite{Blumlein:2004bs,Blumlein:2007kx}. One may also resum small-$x$ 
contributions of $\mathcal{O}((\alpha \ln^2(x))^k)$, which has been performed in Ref.~\cite{Blumlein:1996yz}.

In the case of leptonic variables already the $\mathcal{O}(\alpha)$ corrections do not only contain initial and final 
charged lepton radiation, but also the contributions due to the Compton peak, which was first dealt with in Ref.~\cite{Mo:1968cg} and later considered in Refs.~\cite{Bardin:1987rz,Beenakker:1989km,Blumlein:1989gk,Blumlein:1993ef}.

There are also hadronic initial state radiative corrections. In early studies 
\cite{Bohm:1986na,Bohm:1987cg,Bardin:1987rz,Bardin:1988by,Bardin:1989vz} some conceptional assumptions were still
made, which are incompatible with the massless parton model, being  generally assumed for the QCD corrections 
and
also needs to be applied to the QED corrections. The correct treatment has been given in 
Refs.~\cite{Kripfganz:1988bd,Blumlein:1990wz}. Later numerical studies were made in 
Refs.~\cite{Spiesberger:1994dm,Roth:2004ti}.

Finally, the radiative corrections to polarized-lepton polarized-nucleon scattering were also calculated at 
$\mathcal{O}(\alpha)$ in Ref.~\cite{Bardin:1996ch} and for the singlet corrections to $\mathcal{O}(\alpha^5 L^5)$ in Ref.~\cite{Blumlein:2004bs}, see also Refs.~\cite{Akushevich:1998dz,Akushevich:2011zy}.

\begin{figure}[ht]
\begin{center}
\includegraphics[width=\columnwidth,angle=0]{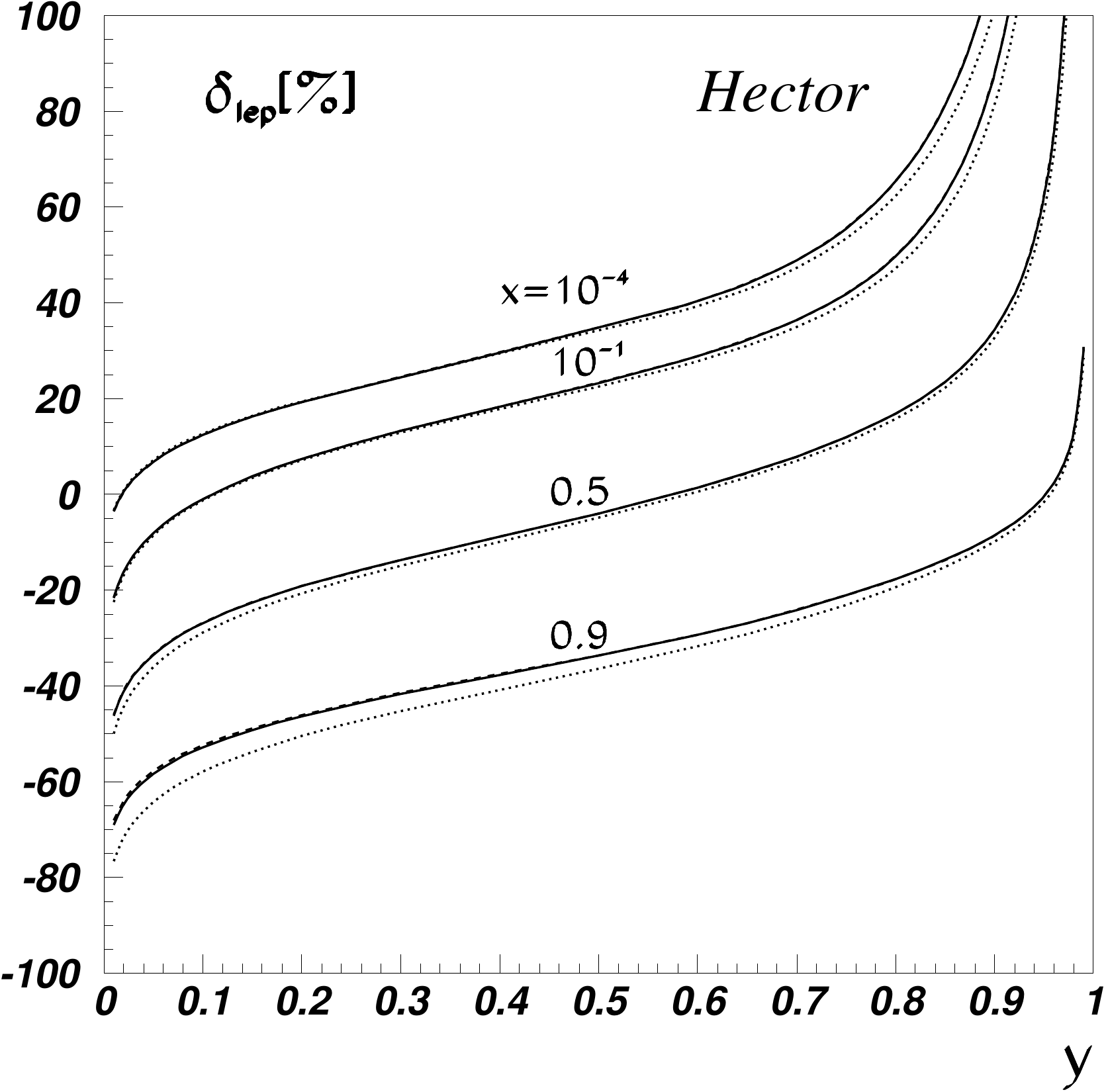}
\end{center}
\caption[]{
\label{epQED:FIG1}
Radiative corrections in leptonic variables in per cent for $E_e = 26.8$~GeV, $E_p =
820$~GeV. Dotted lines: $\mathcal{O}(\alpha)$, dashed lines: $\mathcal{O}(\alpha^2)$, solid lines: in addition soft
photon exponentiation, from \cite{Arbuzov:1995id}, \TCop (1995) by Elsevier Science.}
\end{figure}

Finally, we would like to give some numerical illustrations on the leptonic QED radiative corrections.
In Fig.~\ref{epQED:FIG1} we show the corrections in leptonic variables. It is shown that the LO
corrections deviate by some \% from the $\mathcal{O}(\alpha^2)$ corrections, which are therefore necessary. The soft 
exponentiation is adding much less. The reduction of the radiative corrections in deep-inelastic $ep$ scattering
is also possible by measuring the kinematic variables by other methods. Here we would like to mention, in 
particular, the double angle method, cf.~\cite{Blumlein:1994ii}, where the corrections are much smaller and do 
also behave rather flat as functions of the Bjorken variables $x$ and $y$.

\begin{figure}[ht]
\begin{center}
\includegraphics[width=\columnwidth,angle=0]{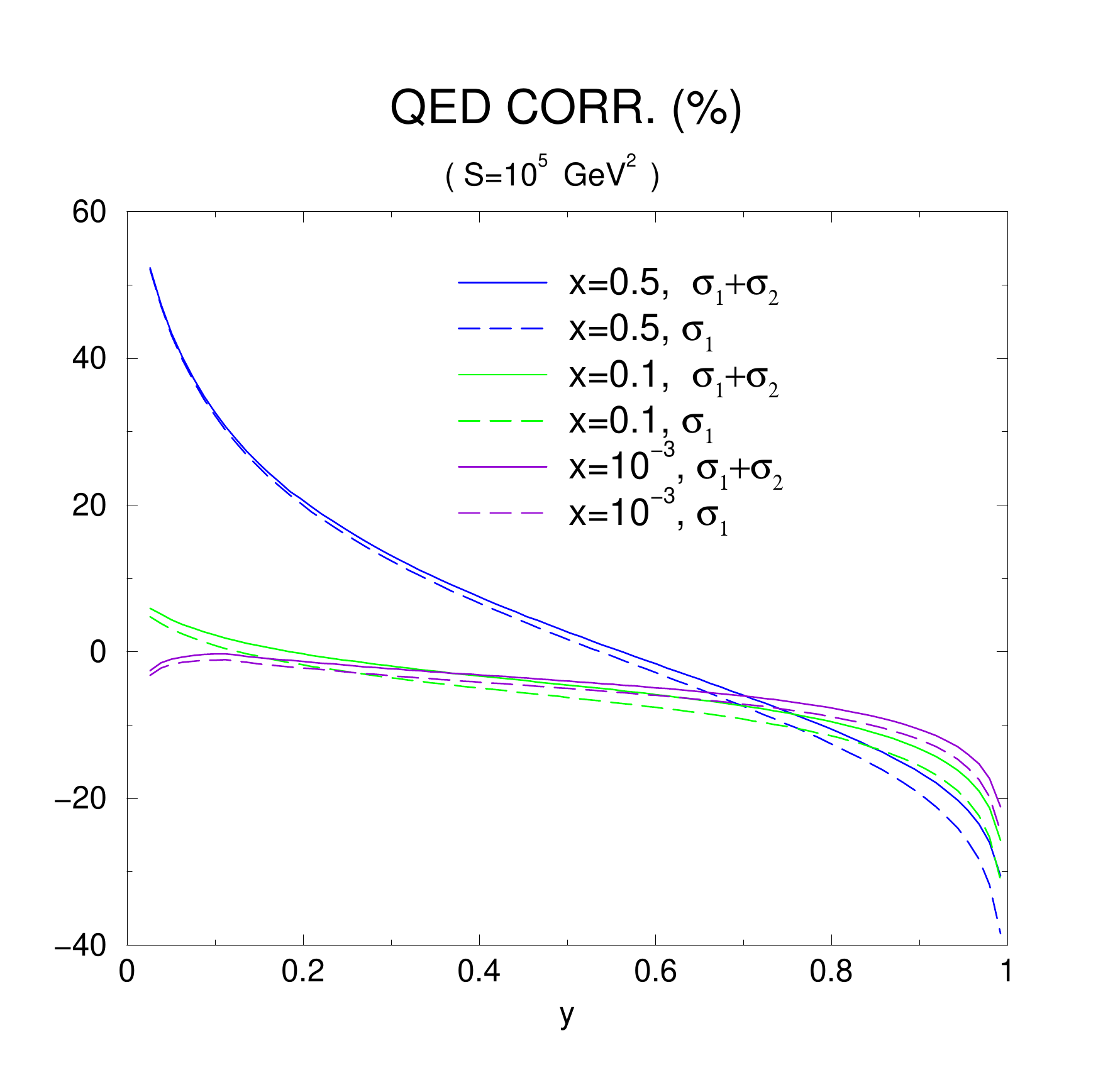}
\end{center}
\caption[]{
\label{epQED:FIG2}
Radiative corrections in mixed variables in per cent for $E_e = 26.8$~GeV, $E_p =
820$~GeV. The lines left from above to below correspond to $x = 0.5, 0.1, 0.001$ in consecutive
order. Full lines $\mathcal{O}(\alpha) + \mathcal{O}(\alpha^2 L^2) +  \mathcal{O}(\alpha^2 L)$; dashed lines: $\mathcal{O}(\alpha)$;
Figure courtesy of H. Kawamura.}
\end{figure}

In Figure~\ref{epQED:FIG2} we present numerical results in the case of mixed variables 
\cite{Bardin:1995mf,Blumlein:1994ii,Blumlein:2002fy} in $\mathcal{O}(\alpha)$, corrected by the terms of $\mathcal{O}(\alpha^2 L^2)$ 
and $\mathcal{O}(\alpha^2 L)$. The dominant behaviour is given by the $\mathcal{O}(\alpha L)$ corrections. However, several \% of the 
corrections are added by the higher-order terms included. These are required for precision measurements.

\subsection{Higher-order QED corrections to \texorpdfstring{$e^+e^-$}{TEXT} annihilation}\label{sec:3}

\vspace*{1mm}
\noindent
Unlike the case in deep-inelastic $ep$ scattering, the complete two-loop QED corrections at large cms energy 
have been calculated for $e^+ e^-$ annihilation \cite{Blumlein:2020jrf}, 
which are of central importance for precision measurements of the $Z^0$ peak, $Z^0 H^0$ production and $t\bar{t}$ 
production at threshold, cf.~\cite{Blumlein:2019pqb} and references therein.\footnote{For a recent survey see 
Ref.~\cite{Blumlein:2022mrp}. An excellent survey on the different calculations at one-loop order is given in 
Ref.~\cite{BP}.}
In earlier attempts to 
calculate these corrections or parts of it in Refs.~\cite{Berends:1987ab,Kniehl:1988id} the massless limit $m_e 
\rightarrow 0$ has been taken too early, which led to incorrect results. Using for the inclusive cross 
sections the factorization property \cite{Buza:1995ie}, one may also calculate it referring to the massless 
sub-processes \cite{Hamberg:1990np,Harlander:2002wh,Blumlein:2019srk} and the corresponding massive OMEs calculated in Ref.~\cite{Blumlein:2011mi}. Since the results of Ref.~\cite{Blumlein:2011mi} did not agree 
with those from Refs.~\cite{Berends:1987ab,Kniehl:1988id}, there was also the possibility that factorization does not 
hold in the massive case. However, it could be proven by the full massive phase-space calculation in Ref.~\cite{Blumlein:2020jrf} that this is not the case and the reasons for the differences could be given in Refs.~\cite{Blumlein:2020jrf,Blumlein:2019srk,Blumlein:2019pqb}. With the factorization relation at hand one can now 
calculate the three most leading logarithms in any order of $\alpha$. This has been done to $\mathcal{O}(\alpha^6 L^5)$ in 
Ref.~\cite{Ablinger:2020qvo}. 

\begin{table*}
    \centering
    \caption{ Shifts in the $Z$-mass and the width due to the 
different contributions to the ISR QED
    radiative corrections for a fixed width of $\Gamma_Z =  
2.4952~\GeV$  and $s$-dependent width using
    $M_Z = 91.1876~\GeV$
    \cite{ParticleDataGroup:2018ovx} and $s_0 = 4 m_\tau^2$, 
cf.~\cite{ALEPH:2005ab}; from 
\cite{Ablinger:2020qvo}, \TCop (2020) by Elsevier Science.}
\scalebox{0.9}{
    \begin{tabular}{l|r|r|r|r}
    \multicolumn{1}{c}{} &
    \multicolumn{2}{|c|}{Fixed width} &
    \multicolumn{2}{c}{$s$-dep. width} \\
    \hline
    \multicolumn{1}{c|}{Order of correction} &
    \multicolumn{1}{c|}{Peak} &
    \multicolumn{1}{c|}{Width}    &
    \multicolumn{1}{c|}{Peak} &
    \multicolumn{1}{c}{Width} \\
    \multicolumn{1}{c|}{} &
    \multicolumn{1}{c|}{(MeV)} &
    \multicolumn{1}{c|}{(MeV)}    &
    \multicolumn{1}{c|}{(MeV)} &
    \multicolumn{1}{c}{(MeV)} \\
\hline
    $\mathcal{O}(\alpha)$                    &   185.638   &   
539.408  &   181.098  &   524.978 \\
    $\mathcal{O}(\alpha^2 L^2)$                       & - 96.894   & 
-177.147  & - 95.342  & -176.235 \\
    $\mathcal{O}(\alpha^2 L)$                         &     6.982   &    
22.695  &     6.841  &    21.896 \\
    $\mathcal{O}(\alpha^2 )$                          &     0.176   & -  
2.218  &     0.174  & -  2.001 \\
    $\mathcal{O}(\alpha^3 L^3)$                       &     23.265  &    
38.560  &    22.968  &    38.081 \\
    $\mathcal{O}(\alpha^3 L^2)$                       & -   1.507  & -  
1.888  & -  1.491  & -  1.881 \\
    $\mathcal{O}(\alpha^3 L)$                        & -   0.152  &     
0.105  & -  0.151  & -  0.084 \\
    $\mathcal{O}(\alpha^4 L^4)$                       & -   1.857  &     
0.206  & -  1.858  &     0.146 \\
    $\mathcal{O}(\alpha^4 L^3)$                       &      0.131  & -  
0.071  &     0.132  & -  0.065 \\
    $\mathcal{O}(\alpha^4 L^2)$                       &      0.048  & -  
0.001  &     0.048  &     0.001 \\
    $\mathcal{O}(\alpha^5 L^5)$                       &      0.142  & -  
0.218  &     0.144  & -  0.212 \\
    $\mathcal{O}(\alpha^5 L^4)$                       & -   0.000  &     
0.020  & -  0.001  &     0.020 \\
    $\mathcal{O}(\alpha^5 L^3)$                       & -   0.008  &     
0.009  & -  0.008  &     0.008 \\
    $\mathcal{O}(\alpha^6 L^6)$                       & -   0.007  &     
0.027  & -  0.007  &     0.027 \\
    $\mathcal{O}(\alpha^6 L^5)$                       & -   0.001  &     
0.000  & -  0.001  &     0.000 \\
    \end{tabular}}
    \label{TAB1}
    \end{table*}

Besides the two-loop massive OME $A_{ee}^{(2)}$ \cite{Blumlein:2011mi} one also needs the two-loop massive OME 
$A_{\gamma e}^{(2)}$ \cite{Ablinger:2020qvo}

    \begin{align}
A_{\gamma e} &= a^{\sf \overline{MS}}
\Biggl[-\frac{1}{2} P_{\gamma e}^{(0)} L_1 
+ \Gamma_{\gamma e}^{(0)}\Biggr] \nn\\
& + {a^{\sf \overline{MS}}}^2
\Biggl[
\frac{P_{\gamma e}^{(0)}}{8} (  
P_{e e}^{(0)} +  
P_{\gamma \gamma}^{(0)} + 2 \beta_0) L_1^2
\nonumber\\ &
+ \frac{1}{2} ( P_{\gamma e}^{(1)} 
+ \Gamma_{e e}^{(0)}  P_{\gamma e}^{(0)} 
+ \Gamma_{\gamma e}^{(0)}  P_{\gamma \gamma}^{(0)}  
\nn \\
&+ 2 \beta_0 \Gamma_{\gamma e}^{(0)} ) L_1   
+ \hat{\Gamma}_{\gamma e}^{(1)}  
+ \bar{\Gamma}_{e e}^{(0)} P_{\gamma e}^{(0)}  
\nonumber\\ &
+ \bar{\Gamma}_{\gamma e}^{(0)} P_{\gamma \gamma}^{(0)}  
+ 2 \beta_0 \bar{P}_{\gamma e}^{(0)} 
\Biggr],
\end{align}

\noindent
with $L_1 = \ln(m^2/\mu^2)$, $\hat{\Gamma}_{ij} =
\Gamma_{ij}(N_F+1) - \Gamma_{ij}(N_F)$ and $\bar{f}$ are the 
$\mathcal{O}(\epsilon)$ parts of the functions $f$, where $\epsilon = 2 - D/2$ denotes 
the dimensional parameter. 

In Tab.~\ref{TAB1} we show the improvement of the peak position and the size of the width of the 
$Z^0$ boson by including higher and higher order QED initial state corrections, until we reach the 
level of $\pm 10$~keV as the theoretical error. We consider both the cases of an $s$-independent 
as well of the $s$-dependent width $\Gamma_Z$.

\begin{figure}[ht]
  \centering
  \hskip-0.8cm
  \includegraphics[width=\columnwidth]{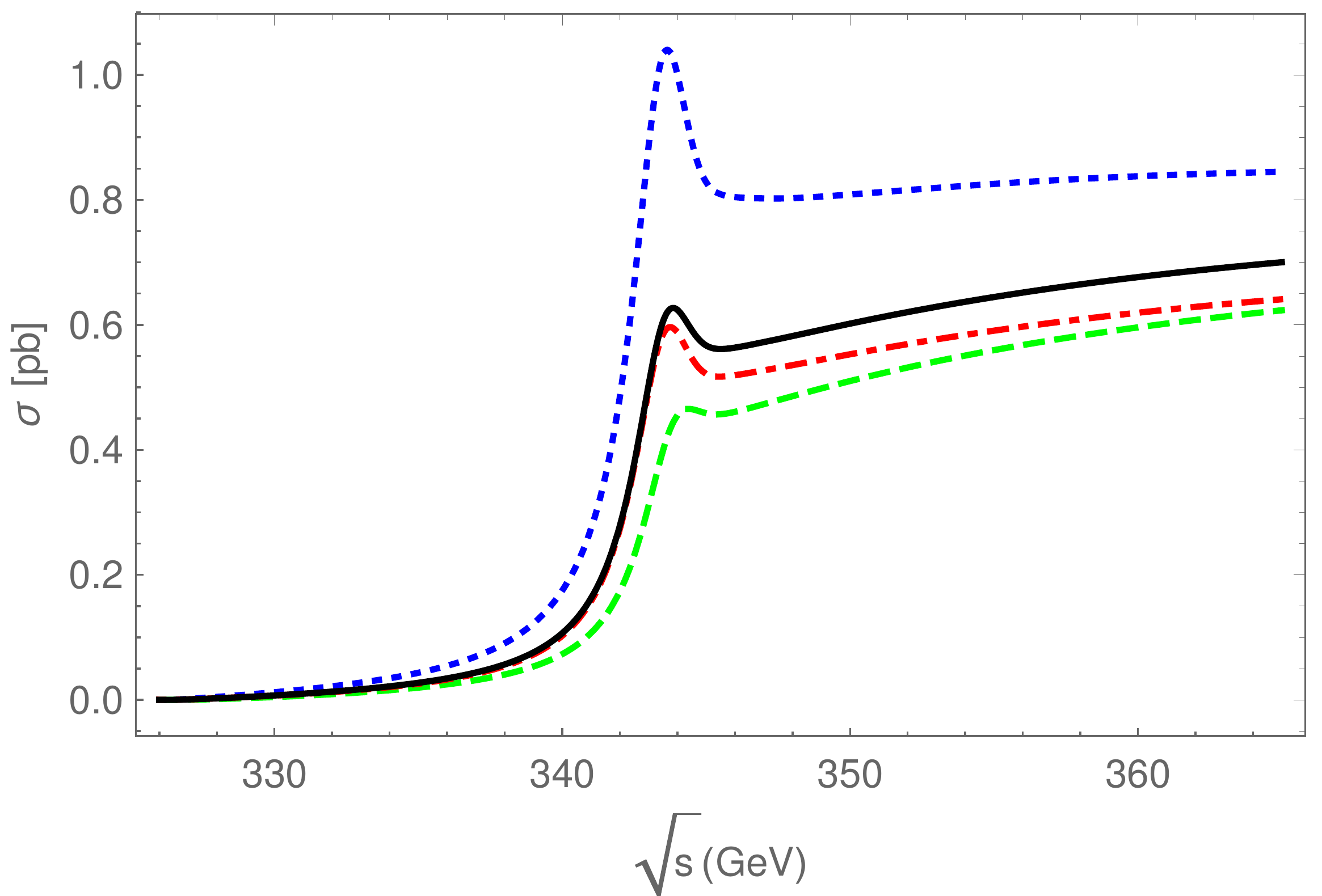}
  \caption{ The QED  ISR corrections to $e^+e^- \rightarrow 
t\overline{t}$ ($s$-channel photon exchange) in the threshold
  region for a PS-mass of $m_t = 172~\GeV$.
Dotted line $\mathcal{O}(\alpha^{0})$;
Dashed line $\mathcal{O}(\alpha)$;
Dash-dotted line $\mathcal{O}(\alpha^{2})$;
Full line $\mathcal{O}(\alpha^{2})$ + soft resummation; from 
Ref.~\cite{Blumlein:2019pqb}, \TCop (2019) by Elsevier Science.
\label{fig:TT1}}
\end{figure}

The QED initial state corrections are also very important for the measurement of inclusive 
$t\bar{t}$ production in the threshold region, which will be used for a precision measurement of 
the top quark mass in the future. Evidently higher than the $\mathcal{O}(\alpha^2)$ corrections are required
to obtain a sufficient description, cf.~Fig.~\ref{fig:TT1}.

For a future precision measurement of the fine structure constant $\alpha$ at high 
energies \cite{Janot:2015gjr} one needs also higher-order corrections to the forward-backward asymmetry. The 
corresponding LO corrections to $\mathcal{O}(\alpha^6 L^6)$ have been calculated in Ref.~\cite{Blumlein:2021jdl}
extending earlier results at $\mathcal{O}(\alpha^2 L^2)$ \cite{Beenakker:1989km}. The 
$\mathcal{O}(\alpha)$ corrections were computed in Ref.~\cite{Bohm:1989pb}.

The LLA initial state radiative corrected  forward-backward asymmetry is 
given by

\begin{eqnarray}
A_{\rm FB}(s) &=& \frac{1}{\sigma_T(s)} \int_{z_0}^1 \dd z 
\frac{4z}{(1+z)^2} 
\nonumber\\ && \times
\tilde{H}_{e}^{\rm LLA}(z) \sigma_{\rm FB}^{(0)}(zs),
\end{eqnarray}
with $\sigma_{\rm FB}^{(0)} = \sigma_{\rm F}^{(0)} - \sigma_{\rm B}^{(0)}$ and 
$\sigma_T(s)$ the radiatively corrected total cross section. The radiator is given 
by

\begin{eqnarray}
\tilde{H}_e^{\rm LLA}(z) =
\left[ H_{e}^{\rm LLA}(z)
+ H_{\rm FB}^{\rm LLA}(z)  \right],
\end{eqnarray}
where the parameter $z_0$ plays the role of an energy cut and $z = 
s'/s$. Here $H_{e}^{\rm LLA}$ is angular independent, and
denotes the leading logarithmic contributions of the usual radiators, while 
$H_{\rm FB}^{\rm LLA}$, which is 
angular dependent and obtained by the following integral

\begin{align}        H_{\rm FB}^{\rm LLA}(z) &= \int\limits_{0}^{1} {\rm d} x_1 
\int\limits_{0}^{1} {\rm d} x_2 \Biggl [ \left(
\frac{(1+z)^2}{(x_1+x_2)^2} -
        1 \right) \nn\\
        &\times\Gamma_{ee}^{\rm LLA}(x_2) \Gamma_{ee}^{\rm LLA}(x_1) 
\delta(x_1 x_2 - z) \Biggr ]. \label{eq:conv}
\end{align}

Here $\Gamma_{ee}^{\rm LLA}(x_i)$ denote the {massive OMEs} and (\ref{eq:conv}) is expanded in
$(\alpha L)$ to the respective order needed. One may use the Mellin-transform

\begin{align} 
        &\mathcal{M}[H_{\rm FB}^{\rm LLA}(z)](n) = \int\limits_{0}^{1} {\rm 
d}z z^n H_{\rm FB}^{\rm LLA}(z) \nonumber\\
&=
\int\limits_{0}^{1} {\rm d}x_1
        \int\limits_{0}^{1} {\rm d}x_2 x_1^n x_2^n \Biggl [ \left( 
\frac{(1+x_1x_2)^2}{(x_1+x_2)^2} - 1 \right) \nn \\   
&\times\Gamma_{ee}^{\rm LLA}(x_2)
        \Gamma_{ee}^{\rm LLA}(x_1) \Biggr ].
\end{align}

In $z$-space the radiator depends on cyclotomic harmonic polylogarithms 
\cite{Ablinger:2011te}.
\begin{table}[ht]
   \centering
       \caption{ $A_{\rm FB}$ evaluated at $s_-=(87.9\, {\rm GeV})^2$, $M_Z^2$
                        and $s_+=(94.3\,{\rm GeV})^2$ for the cut $z>4m_\tau^2/s$ from
        Ref.~\cite{Blumlein:2021jdl}, \TCop (2021) by Elsevier Science.}
        \label{tab:AFB1}
    \begin{tabular}{r|r|r|r}
    \multicolumn{1}{c|}{ Order} &
    \multicolumn{1}{c|}{$A_{\rm FB}(s_-)$  } &
    \multicolumn{1}{c|}{$A_{\rm FB}(M_Z^2)$} &
    \multicolumn{1}{c}{$A_{\rm FB}(s_+)$} \\
    \hline
                $\mathcal{O}(\alpha^0)$         & $-0.3564803$          
& $ 0.0225199$  & $0.2052045$
                \\
                $\mathcal{O}(\alpha L^1)$       & $-0.2945381$          
& $-0.0094232$  & $0.1579347$
                \\
                $\mathcal{O}(\alpha L^0)$       & $-0.2994478$          
& $-0.0079610$  & $0.1611962$
                \\
                $\mathcal{O}(\alpha^2 L^2)$     & $-0.3088363$          
& $ 0.0014514$  & $0.1616887$
                \\
                $\mathcal{O}(\alpha^3 L^3)$     & $-0.3080578$          
& $ 0.0000198$  & $0.1627252$
                \\
                $\mathcal{O}(\alpha^4 L^4)$     & $-0.3080976$          
& $ 0.0001587$  & $0.1625835$
                \\
                $\mathcal{O}(\alpha^5 L^5)$     & $-0.3080960$          
& $ 0.0001495$  & $0.1625911$
                \\
                $\mathcal{O}(\alpha^6 L^6)$     & $-0.3080960$          
& $ 0.0001499$  & $0.1625911$
\\
        \end{tabular}
\end{table}

\renewcommand{\arraystretch}{1.0}

\noindent
In Tab.~\ref{tab:AFB1} we illustrate the improvement of the forward-backward 
asymmetry by including higher order leading-log initial state radiative 
corrections beyond the $\mathcal{O}(\alpha^2 L^2)$ corrections \cite{Beenakker:1989km}. The 
$\mathcal{O}(\alpha^6 L^6)$ corrections stabilize the off-resonance forward-backward asymmetries to 
six digits. Of course, also non-leading log corrections are needed to be 
calculated in the future.

\subsection{Conclusions}\label{sec:4}

\vspace*{1mm}
\noindent
We have summarized the status of the higher-order QED corrections to deep-inelastic $ep$ scattering and 
for the QED initial state corrections to $e^+e^-$ annihilation. In both cases the emphasis is on the non-power 
corrections, i.e.~on the logarithmic contributions down to the constant terms of $\mathcal{O}(\alpha^k L^0)$. These 
contributions can be computed using renormalization group techniques, which involve massive OMEs on one side 
and massless sub-system scattering processes on the other side. It is important in using this approach to cancel
the dependence on the factorization scale by matching the different contributions. This is best done in 
Mellin-$N$ space in a fully analytic way. The radiators are then given by Mellin convolutions of the respective 
pieces defining the OMEs and the sub-system scattering cross sections. The result can be Mellin-inverted
analytically providing the correct radiators. The requested interplay between both types of contributions does not
allow to study QED evolution only. 

In the case of future experimental studies of $e^+e^-$ annihilation it is planned to measure the width and the 
mass of the $Z^0$-boson down to the $20$~keV range. This requires initial state corrections of up to 6th order
QED corrections by keeping three logarithmic orders.

\section{Light-Hadron Decays and Reactions}\label{sec:mesons}

The incorporation of radiative corrections is also essential for processes in a kinematical regime that is different from the previous discussions.
Recently, it has been gradually recognized that in order to achieve desired precision in experiment, radiative corrections must be incorporated as a part of MC generators (see also Sec.~\ref{sec:Implementations} for more on MC concepts), and subsequently, the demand for such corrections has increased.
Such an awakening resides in the fact that, typically, the QED NLO radiative corrections compete in size with the hadronic effects.
And thus, once hadronic parameters are to be extracted from data, ignoring radiative corrections would lead to meaningless results, the examples of which we will see below.

Historically, the approach to radiative corrections varied.
In some cases, they would be left out entirely.
If an effort was done to include at least some of the relevant corrections, others that were not negligible were ignored.
In some cases, approximate approaches (soft-photon limit or leading logarithms) were used, which often led to unreliable or misleading results.
All the above-mentioned cases introduce artificial discrepancies between theory and experiment, as also discussed in \secref{RoPTcomp}.
In other words, what is then considered to be ``measured observables'' or related ``measured hadronic parameters'' may include an unsubtracted QED part, which can turn out to be quite a significant contribution.

In the following, we will discuss the current status of QED radiative corrections in some sample decays of the lightest pseudoscalar mesons and baryons and their direct application in experiment.

\subsection{Radiative corrections for \texorpdfstring{$\pi^0$}{pi0} decays}

The NLO QED radiative corrections are now worked out for all the relevant neutral-pion decay channels: the $\pi^0$ Dalitz decay $\pi^0\to e^+e^-\gamma$, the $\pi^0$ double-Dalitz decay $\pi^0\to 2(e^+e^-)$, and the $\pi^0$ rare decay $\pi^0\to e^+e^-$.
We will address these individually in the following subsections.

\subsubsection{\texorpdfstring{$\pi^0\to e^+e^-$}{pi0->ee}}

Let us start with the last-mentioned process.
Regarding the branching ratio, this channel is loop- and helicity-suppressed with respect to the radiative decay $\pi^0\to\gamma\gamma$ by eight orders of magnitude.
It has thus been speculated that the $\pi^0\to e^+e^-$ decay might be sensitive to possible effects of new physics.
And it indeed seemed to be the case when the KTeV Collaboration published the results of their analysis on the precise measurement of the branching ratio in 2007~\cite{KTeV:2006pwx}.
The direct subsequent comparison to the Standard Model (SM) prediction was then interpreted as a 3.3\,$\sigma$ discrepancy~\cite{Dorokhov:2007bd}.
Many works followed trying to explain this difference both within (via modeling of the electromagnetic pion transition form factor) and outside the SM (by introducing various models including exotic particles).
However, it is believed that the most important contribution in this regard was done by an explicit calculation of (two-loop) virtual radiative corrections~\cite{Vasko:2011pi} that brought the discrepancy down to the $2\,\sigma$ level~\cite{Husek:2015wta}: Until then, only leading-log approximation was available~\cite{Bergstrom:1982wk,Dorokhov:2008qn} that did not reproduce well accidental cancellations among various diagrams.
New light can be shed on the branching ratio determination and the related remaining $2\,\sigma$ discrepancy: A new measurement is under preparation by the NA62 Collaboration and the results should be known, conservatively estimating it, within a few months, i.e., by the end of 2023.
In this analysis, all the known related radiative corrections~\cite{Vasko:2011pi,Husek:2014tna,Husek:2015sma,Husek:2018qdx} are already included at the MC level.

Looking at the quantity that has been measured by KTeV~\cite{KTeV:2006pwx},
\begin{equation}
\frac{B\bigl(\pi^0\to e^+e^-(\gamma),x>0.95\bigl)}
{B\bigl(\pi^0\to e^+e^-\gamma,x>0.232\bigl)}
=1.685(64)(27)\times10^{-4}\,,
\label{eq:BBKTeV}
\end{equation}
with $x\equiv m_{e^+e^-}^2/M_\pi^2$, it becomes immediately obvious why there is a need for precise knowledge of radiative corrections in order to compare this value with the predictions provided by theory, which can be written as
\begin{equation}
B\bigl(\pi^0\to e^+e^-\bigr)
\approx\bigl(6.21+0.15\widetilde\chi\bigr)\times10^{-8}\,,
\end{equation}
with $\widetilde\chi\equiv2\big[\chi^{(\mathrm{r})}(770\,\mathrm{MeV})-\frac52\big]$.
Theoretical predictions and models suggest (being rather conservative) $\chi^{(\mathrm{r})}(770\,\mathrm{MeV})\sim$\;2--3 (see, e.g., Refs.~\cite{Knecht:1999gb,Dorokhov:2007bd,Husek:2015wta,Hoferichter:2021lct}), so $\widetilde\chi$ is expected to be within the interval $(-1,1)$.
The result in Eq.~\ref{eq:BBKTeV} not only depends on the Dalitz-decay branching ratio serving here as the normalization channel, but the numerator allows for final states including soft photons with their energies limited by the requirement on the minimum of the normalized electron--positron invariant mass squared $x$.
Eq.~\ref{eq:BBKTeV} thus needs to be further processed and it was done so in the KTeV analysis via a series of steps in order to provide a ``no-radiation'' value to be compared with theory.
In other words, experimental results on the branching ratios of such processes will contain effects that need to be subtracted in some way, so they can be directly compared with theoretical predictions for branching ratios with no final-state radiation.
Such a subtraction of radiative corrections can happen at different stages of the experimental analysis, but it seems that the earlier stage this happens, the better:
Once the events accounting for photons in the final state are properly generated already at the MC level, comparing the resulting spectra with data helps to improve the quality of the analysis outcome.

\subsubsection{\texorpdfstring{$\pi^0\to e^+e^-\gamma$}{pi0->eeg}}

We have already seen that in the KTeV rare-decay search described above, the Dalitz decay was used as the normalization channel.
It is thus clear that it is essential to have the complete set of NLO QED corrections available also in this case.
It then comes as a bonus that this process is also used for normalizing purposes in measurements in the kaon sector~\cite{Husek:2018qdx}.
Furthermore, investigating the Dalitz decay itself leads to information about the singly off-shell ($\pi^0\gamma\gamma^*$) electromagnetic transition form factor (TFF) in the time-like region, i.e., direct access to the TFF slope without the need for model-dependent extrapolation from the space-like region, as used, e.g., in Ref.~\cite{CELLO:1990klc}.

\begin{figure}[ht]
    \centering
    \begin{subfigure}[t]{0.48\columnwidth}
        \centering
        \includegraphics[width=\textwidth]{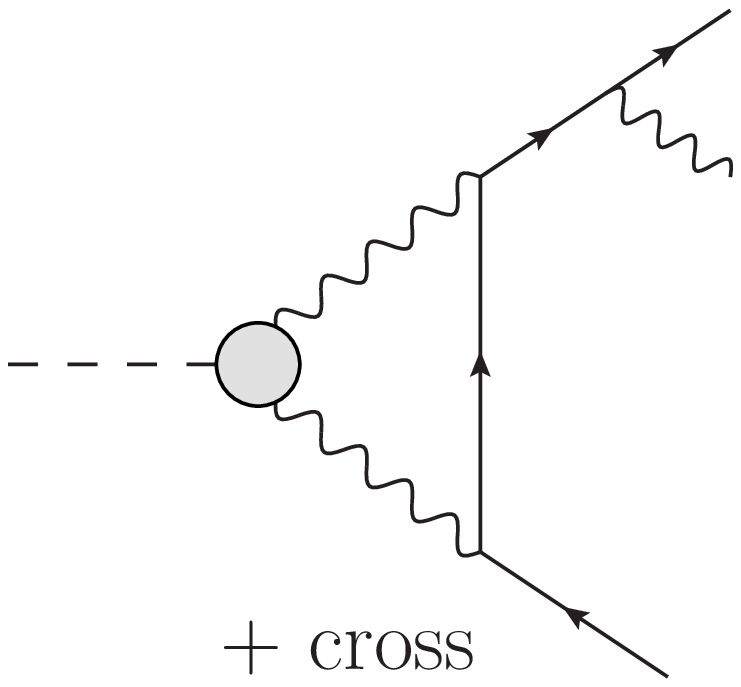}
    \end{subfigure}
    \begin{subfigure}[t]{0.48\columnwidth}
        \centering
        \includegraphics[width=\textwidth]{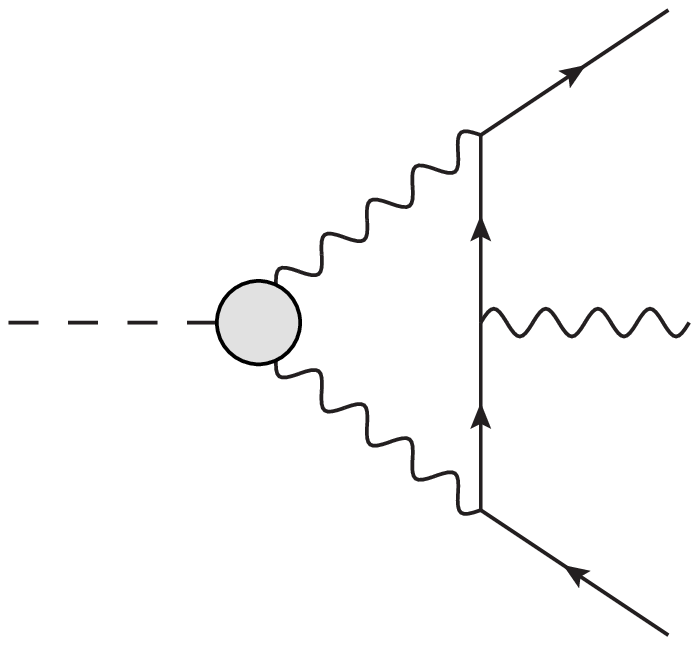}
    \end{subfigure}\\[3mm]
    \caption{One-photon-irreducible Feynman diagrams contributing to neutral-pion Dalitz-decay radiative corrections.}
    \label{fig:1gIR}
\end{figure}

The radiative corrections for $\pi^0\to e^+e^-\gamma$ were first estimated in terms of the total correction to the integral decay width~\cite{Joseph:1960zz}, and only a decade later, the first corrections to the differential decay width appeared, which used the soft-photon approximation~\cite{Lautrup:1971ew}.
What can now be considered as a classical work on the Dalitz-decay radiative corrections came a year after~\cite{Mikaelian:1972yg}.
The hard-photon corrections were included and the results were presented in terms of a table giving sample correction values throughout the Dalitz plot.
One contribution was, however, missing: the two-photon-exchange contribution, or also known as the one-photon-irreducible (1$\gamma$IR) contribution at one loop, which diagrammatically coincides with the bremsstrahlung contribution to the $\pi^0\to e^+e^-$~\cite{Husek:2014tna}; see Fig.~\ref{fig:1gIR}.
It turned out, after a vivid discussion in the literature~\cite{Tupper:1983uw,Lambin:1985sb,Tupper:1986yk}, that this contribution was not negligible, and a direct subsequent calculation would show that omitting the 1$\gamma$IR piece would introduce $\sim$\,15\% discrepancy to the measured slope~\cite{Kampf:2005tz,Husek:2015sma}.

\begin{figure}[ht]
    \centering
    \includegraphics[width=\columnwidth]{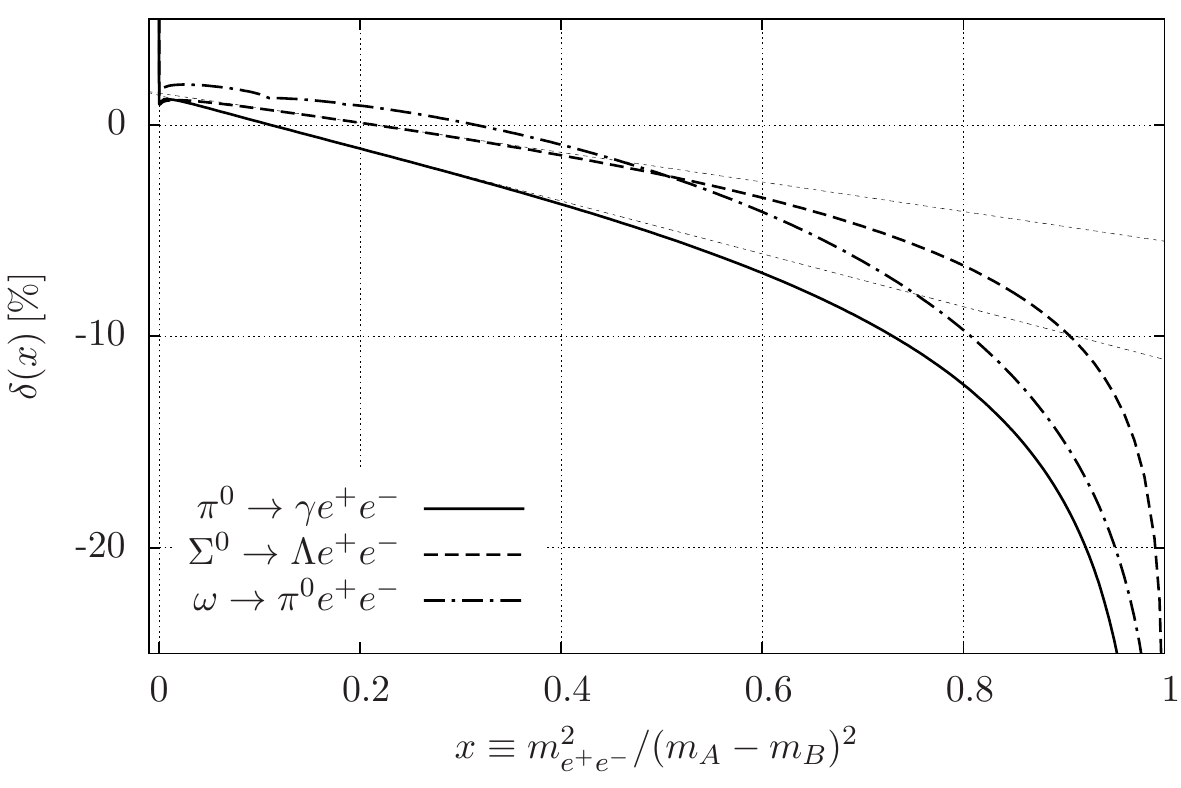}\\[4mm]
    \caption{
    Inclusive NLO QED radiative corrections for the discussed Dalitz decays $A\to Be^+e^-$.
    For the cases of $\pi^0$ and $\Sigma^0$, the slopes of the radiative corrections are denoted, from which one can infer the effects on the measured TFF slopes (see Eq.~\ref{eq:aQED}).
    }
    \label{fig:Dalitz_RC}
\end{figure}

The most precise measurement of the slope in the time-like region to date was performed by the NA62 Collaboration with the results $a_\pi=3.68(57)\%$~\cite{NA62:2016zfg}.
It is worth mentioning that the slope of the radiative corrections amounts to circa $-12.5\%$; see the low-$x$ region of Fig.~\ref{fig:Dalitz_RC}.
This corresponds to the QED correction of about $-6\%$ to the slope estimate in absolute numbers.
This in turn means that without considering the QED corrections one would inevitably obtain a negative slope of similar size as the correct result $a_\pi$ stated above.

In QED, in which hadronic effects of the $\pi^0\to\gamma^*\gamma^*$ transition are included via the electromagnetic form factor, it turns out that one can precisely determine the Dalitz-decay branching ratio, directly following from the NLO calculation of the ratio \mbox{$R=B(\pi^0\to e^+e^-\gamma(\gamma))/B(\pi^0\to\gamma\gamma)$}~\cite{Husek:2018qdx}.
For this, it is sufficient to know the NLO radiative corrections and --- from theoretical estimates confirmed by experiment --- that the TFF slope is rather small.
This then combines with the fact that the sensitivity to the only relevant (higher derivatives can be neglected since the pion mass lies well below the resonance region) hadronic parameter present ($a_\pi$) in such an equation
\begin{multline}
    R
    =\frac\alpha\pi\int\!\!\!\!\int(1+a_\pi x)^2\frac{(1-x)^3}{4x}\\
    \times\big[1+\delta(x,y)\big]\biggl(1+y^2+\frac{4m_e^2}{M_\pi^2x}\biggr)\,\mathrm{d}x\mathrm{d}y\,
\end{multline}
is suppressed because the region with high virtualities is (and the slope thus does not influence the final result much); this is represented by the $(1-x)^3$ term above.
Let us stress that such a precise determination is only possible since the complete set of NLO QED radiative corrections are now known.

Believing the above determination, it becomes surprising that the KTeV Collaboration, soon after Ref.~\cite{Husek:2018qdx} was published, presented their results on the Dalitz-decay branching ratio~\cite{KTeV:2019gud} which is 3.6\,$\sigma$ away from the mentioned QED-based result.
It is thus desirable to have another independent branching-ratio measurement.
There is ongoing activity in this direction in NA62.

\subsubsection{\texorpdfstring{$\pi^0\to 2(e^+e^-)$}{pi0->4e}}

A complete set of NLO radiative corrections in soft-photon approximation for the neutral-pion double-Dalitz decay is computed in Ref.~\cite{Kampf:2018wau}, in which the processes $P\to\bar\ell\ell\bar\ell'\ell'$ are addressed in general.
This work also considers form-factor effects and improves upon the earlier work~\cite{Barker:2002ib}.

\subsection[\texorpdfstring{$\eta^{(\prime)}$}{eta(')} Dalitz decays]{Radiative corrections for \texorpdfstring{$\eta^{(\prime)}$}{eta(')} Dalitz decays}

The knowledge of the radiative corrections for the neutral-pion Dalitz decay turns out to be very useful in different contexts, and it also serves as a starting point to the calculation of the corrections for the corresponding $\eta^{(\prime)}$ decay channels.
These come with several additional challenges that one needs to overcome in order to be able to extract valuable information on the TFFs from data.

Naive radiative corrections for $\eta\to e^+e^-\gamma$~\cite{Mikaelian:1972jn} were indeed closely inspired by the solution for its neutral-pion counterpart, and only the mass of the decaying meson would be numerically addressed.
However, one needs to take the situation a bit more seriously.
The larger rest mass of the decaying $\eta^{(\prime)}$ has further consequences.
The $\eta$-meson mass already lies above the muon-pair and pion-pair production threshold, which has some consequences for the treatment of virtual corrections.
But what brings in a real complication is the fact that $\eta'$ lies above the lowest-lying resonances $\rho$ and $\omega$.
The decays and the radiative corrections then naturally become sensitive to the widths and shapes of the resonances.
And this is a crucial difference compared to the $\pi^0$ case, where the TFF dependence was mild and could even be neglected in certain applications.
Another technical complication then resides in the fact that $\eta^{(\prime)}$ mesons carry a nontrivial strangeness content that has to be addressed as part of the TFF treatment.

\begin{figure}[ht]
    \centering
    \includegraphics[width=\columnwidth]{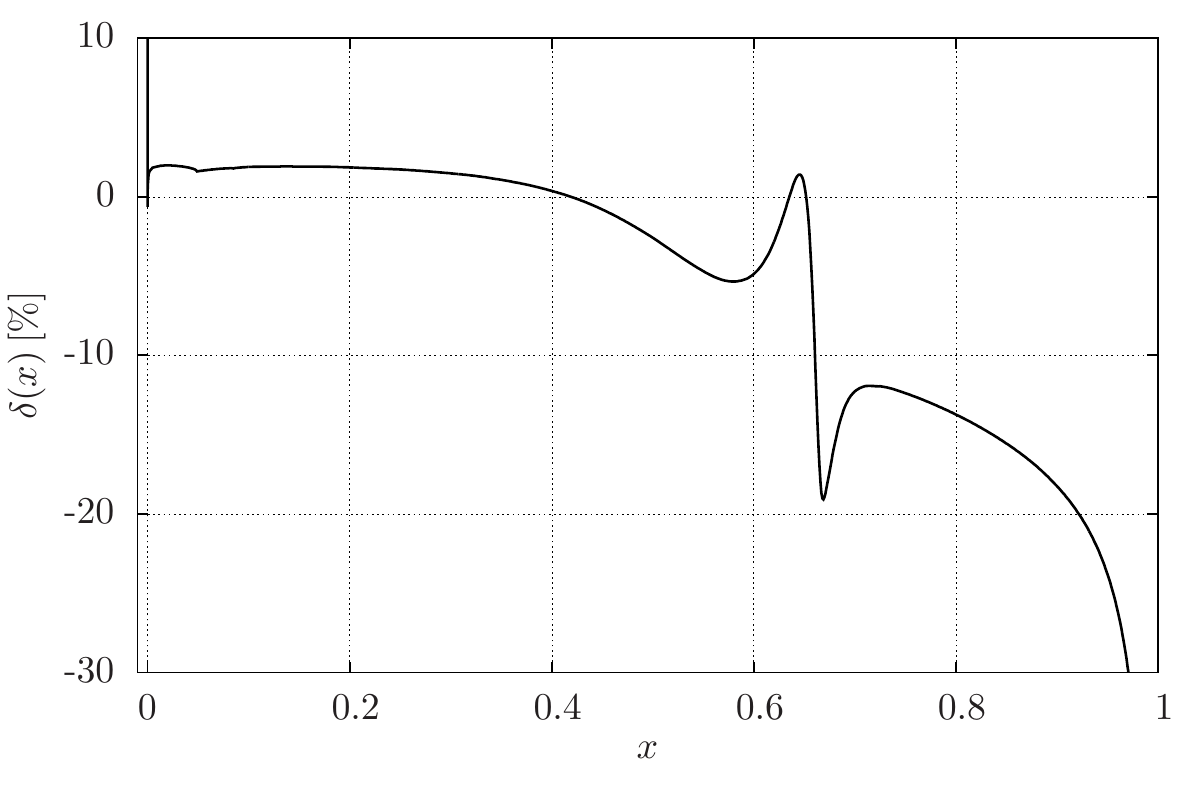}\\[2mm]
    \caption{Inclusive NLO QED radiative corrections for $\eta'\to e^+e^-\gamma$; \mbox{$x\equiv(p_{e^+}+p_{e^-})^2/M_{\eta'}^2$}.}
    \label{fig:etap_e_RC}
\end{figure}

A more advanced approach to the $\eta^{(\prime)}\to\ell^+\ell^-\gamma$ radiative corrections~\cite{Husek:2017vmo} then improves on the naive approach for $\eta\to e^+e^-\gamma$ in the following way.
Regarding the vacuum polarization contribution, the muon-loop and hadronic contributions were added, the 1$\gamma$IR contribution at one-loop level was included, and the form-factor effects (including the complete bremsstrahlung contribution) were calculated.
Moreover, no assumptions regarding the final-state lepton masses were considered, so the muon channels could be consistently treated as well.
An example of the resulting radiative corrections can be seen in Fig.~\ref{fig:etap_e_RC}.

Let us note that the developed exact treatment of the form-factor effects can also be used in the $\pi^0$ Dalitz decay.
There it introduces circa 1\% correction to the correction itself and is thus, as expected, negligible.

\subsection[\texorpdfstring{$\Sigma^0\to\Lambda e^+e^-$}{Sigma0->Lambda ee} Dalitz decay]{Radiative corrections for \texorpdfstring{$\Sigma^0\to\Lambda e^+e^-$}{Sigma0->Lambda ee} Dalitz decay}

The next technical complication in the calculation of radiative corrections to differential decay widths of Dalitz decays appears when the final-state photon is replaced by a massive particle.
The first example of such a case that we discuss here appears outside the meson sector: the Dalitz decay of the neutral Sigma baryon.

\begin{figure}[ht]
    \centering
    \begin{subfigure}[t]{0.48\columnwidth}
        \centering
        \includegraphics[width=\textwidth]{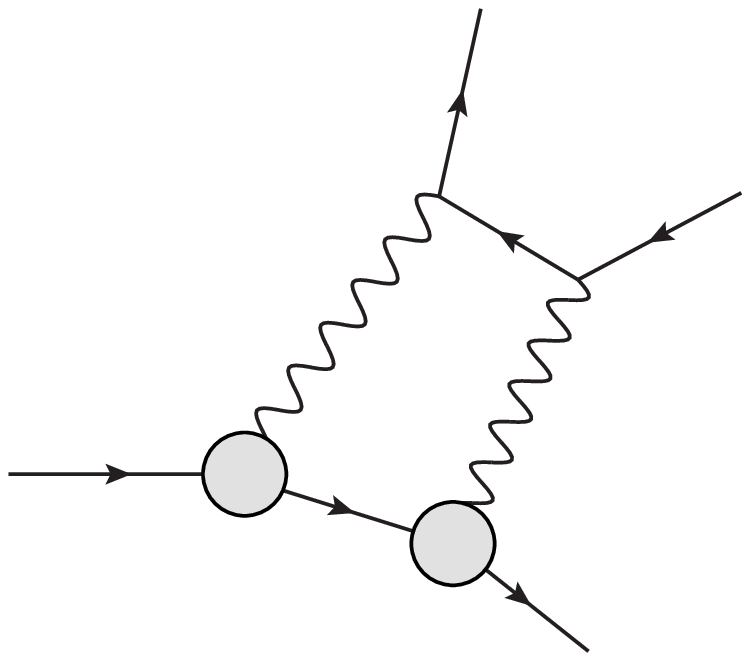}
    \end{subfigure}
    \begin{subfigure}[t]{0.48\columnwidth}
        \centering
        \includegraphics[width=\textwidth]{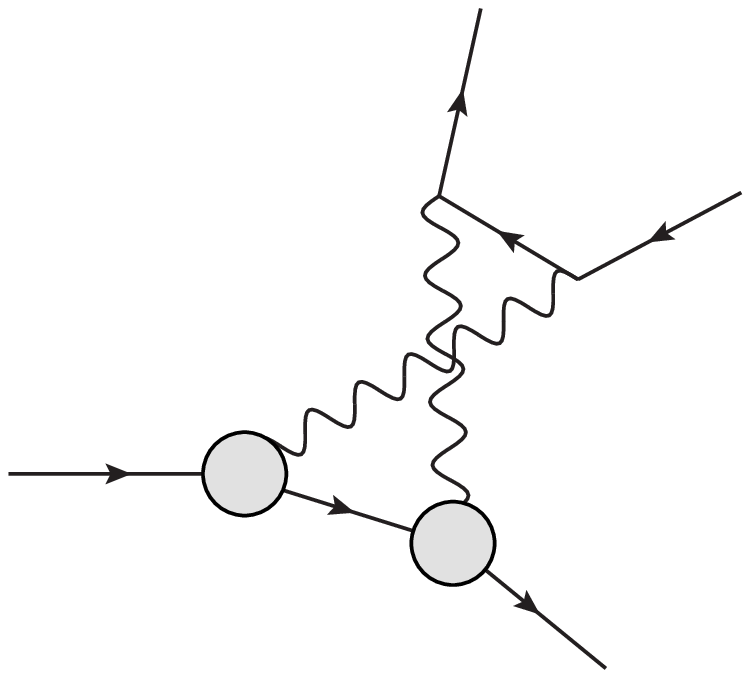}
    \end{subfigure}\\[3mm]
    \caption{1$\gamma$IR (two-photon-exchange) Feynman diagrams contributing to NLO radiative corrections for~\mbox{$\Sigma^0\to\Lambda e^+e^-$}.}
    \label{fig:1gIR_box}
\end{figure}

Radiative corrections within the soft-photon-approximation approach have been available for some time now~\cite{Sidhu:1971myo}.
The calculation of the bremsstrahlung contribution including the effects of hard-photon emission proceeds along similar lines as in the previous cases, although now the analytical integrals need to be generalized for a nonvanishing (Lambda-hyperon) mass.
It turns out that some of the integrals do not have analytical solutions anymore with such a generalization.
The counterpart of the (until now triangular) 1$\gamma$IR contribution of Fig.~\ref{fig:1gIR} are here two-photon-exchange box diagrams involving hadronic form factors; see Fig.~\ref{fig:1gIR_box}.
Although quite nontrivial to evaluate and contributing to the electron--positron asymmetry, they do not affect the one-fold differential decay width in terms of the invariant electron--positron mass squared, and thus the extraction of the $\Sigma^0\to\Lambda\gamma^*$ transition form factor.
Here lies a profound difference between the kinematical regimes of decay and scattering since, in the latter, the two-photon-exchange contribution matters, as discussed in Sec.~\ref{sec:TPE}.

Similarly to the $\pi^0$-Dalitz-decay case, the (magnetic; electric is negligible) form-factor dependence is rather mild, and one can come up with precise predictions for $\Sigma^0\to\Lambda e^+e^-$ and $\Sigma^0\to\Lambda\gamma$ branching ratios~\cite{Husek:2019wmt}.
Regarding the importance of the QED NLO radiative corrections (see Fig.~\ref{fig:Dalitz_RC}), it was found, using again the simple consideration
\begin{equation}
    a_{(+\text{QED})}-a
    \simeq\frac12\frac{\mathrm{d}\delta(x)}{\mathrm{d}x}\bigg|_{x\ll1}\,,
\label{eq:aQED}
\end{equation}
that the effect on the TFF-slope extraction is about $-3.5\%$ in absolute figures~\cite{Husek:2019wmt}.
This is quite a significant value while extracting a quantity that is estimated to be of a size $1.8(3)$\%: When ignoring the QED effects, one would thus end up with a negative magnetic TFF slope, although again of a reasonable magnitude.

\subsection[\texorpdfstring{$\omega\to\pi^0 \ell^+\ell^-$}{omega->pi0ll}]{Radiative corrections for \texorpdfstring{$\omega\to\pi^0 \ell^+\ell^-$}{omega->pi0ll}}

Once the bremsstrahlung contribution to the $\Sigma^0$ Dalitz decay is established, the results can be readily used for the $\omega\to\pi^0 \ell^+\ell^-$ decay.
Here, various extractions of the $\omega\pi V$ correlator~\cite{Dzhelyadin:1980tj,NA60:2009una,NA60:2016nad,Adlarson:2016hpp} are in tension with the simplest theoretical models, and one could wonder if applying the previously omitted QED radiative corrections (they were not available until very recently) could improve the situation.
The corrections are, as in the previous cases, negative over a large region of the Dalitz plot (see Fig.~\ref{fig:Dalitz_RC}), and one can thus expect the data points to be pushed upwards accordingly after the QED effects are subtracted.
Needless to say, these corrections are more significant in the electron channel.
A proper analysis is, of course, in place and should be part of the future experimental analyses of more precise data.

The one-photon-inclusive NLO QED radiative corrections beyond the soft-photon approximation were calculated as a bachelor project in 2021~\cite{Lindahl:bc_thesis}.

\subsection[\texorpdfstring{$K^+\to\pi^+\ell^+\ell^-$}{K+->pi+ll}]{Radiative corrections for \texorpdfstring{$K^+\to\pi^+\ell^+\ell^-$}{K+->pi+ll}}

The $K^+\to\pi^+\ell^+\ell^-$ decays allow us to access the $K^+\to\pi^+\gamma^*$ TFF.
Extracted both in the electron and muon channels, it allows us to test the lepton-flavor universality.
Most importantly, however, studying the TFF provides us with valuable information about weak transitions in the low-energy QCD sector.
Extracting the relevant related hadronic parameters from experiment then requires, as expected, the knowledge of radiative corrections in order to subtract the QED effects that are present.
Recently, following a number of previous extractions~\cite{E865:1999ker,NA482:2009pfe,NA482:2010zrc}, the TFF parameters traditionally called $a_+$ and $b_+$ were measured by the NA62 Collaboration~\cite{NA62:2022qes}; it is worth mentioning that radiative corrections were part of this analysis at the MC level~\cite{Kubis:2010mp,Husek:2022fsc}.

Regarding the radiative corrections related to the lepton part of the $K^+\to\pi^+\ell^+\ell^-$ process, there are no additional complications, and one can use the same approach as in the previous cases once the form factor is suitably expanded to account for hadronic effects in the bremsstrahlung contribution {\em beyond} the soft-photon approximation.
On the other hand, the novel nontrivial obstacle is the fact that the involved kaon and pion are charged and can thus both also radiate a bremsstrahlung photon.
Moreover, the transition $K^+\to\pi^+\gamma^*\gamma^{(*)}$ needs to be considered.
Finally, virtual corrections that complement the bremsstrahlung from the meson part are somewhat nontrivial.
Everything then should be ideally expressed in terms of a single form factor, the one that is extracted from data.
This then allows for factorizing it out or for further iterations.

\subsection{Radiative decay of \texorpdfstring{$K^+ \rightarrow e^+ \nu \gamma$}{K+ -> e+ nu g} with TREK/E36 at J-PARC}

The TREK collaboration (Time Reversal Experiment with Kaons)
was formed in 2005 to pursue a program to search for T-violation in stopped
$K^+$ decays~\cite{e06_proposal}.
When by 2010 it became clear that the
required kaon beam intensity would not be achievable in a timely manner, the
TREK collaboration still pursued a program on sensitive
beyond-the-Standard-Model physics, with experiment E36, which was proposed in
2010 using only 50 kW beam power~\cite{e36_proposal,e36_proposal-addend,e36_proposal-addend2}.

The focus of E36 was on testing lepton universality in the
SM, by a precision measurement of the ratio of two-body decay widths,
$R_K = \Gamma(K_{e2})/\Gamma(K_{\mu2})$.
The experiment was approved in 2013 and constructed at the J-PARC K1.1BR
beamline in 2014, reusing the toroidal spectrometer setup from KEK-PS
experiment E246~\cite{e246_NIM} and the CsI(Tl) barrel~\cite{ito_csi}.
A new scintillating fiber target,
a spiral fiber tracker~\cite{sft}, and
redundant PID systems (an aerogel counter array~\cite{ac},
a time-of-flight system, and
a leadglass shower calorimeter in each sector~\cite{pgc}) were implemented.
Data taking was limited to less than three months in 2015. First E36 results
have been published on radiative kaon
decays~\cite{e36_Ke2g_CsI,e36_Ke2g_GSC},
with the result for $R_K$ being expected to become available in 2023.
The E36 setup was also used for an in-situ search for dark photons, or more
generally for light neutral bosons decaying into $e^+ e^-$ pairs over a broad
mass range of 10--200\;MeV/c$^2$.
The analysis, which is still ongoing, combines a charged particle
($\mu^+$, $\pi^+$, or $e^+$) in one toroidal sector with two clusters from
$e^+ e^-$ pair detection in the segmented CsI(Tl) calorimeter.
Preliminary results from the light boson search have been
reported~\cite{e36_dp}, and the
final sensitivity is expected to be reached in 2023.

The study of radiative kaon decay allowed to extract the branching ratio for
the structure-dependent (SD) process resulting in a hard photon radiated nearly
in back-to-back kinematics with the positron. The process is accompanied by the
internal bremsstrahlung (IB), predominantly of soft photons nearly in the
direction of the positron. In the overlapping regions both processes appear,
and the IB process is calculated and simulated as a background, and
conventionally included in the two-body branching ratio for $K_{e2}$.
The latest analysis also considers internal bremsstrahlung in the
SD process $K_{e2\gamma}^\mathrm{SD}$.
The implementation of the IB radiative process in the MC simulation
follows the scheme introduced by Gatti~\cite{gatti2006}, allowing for analytic
expressions of the differential branching ratio distributions to be used in an
event generator. The experiment data were
analyzed in terms of the lepton momentum in a narrow range. The radiative
correction affecting the $K_{e2}$ and $K_{e2\gamma}$
  branching ratios were mostly due to the narrow acceptance chosen for the
  lepton momentum. In the ratio
$R_\gamma = \Gamma(K_{e2\gamma}^\mathrm{SD})/\Gamma(K_{e2})$,
the IB corrections partially cancel in the acceptance ratio.
The first publication~\cite{e36_Ke2g_CsI}
accounted for the IB acceptance effect only in the
monochromatic two-body decay $K_{e2}$. In the most recent
result~\cite{e36_Ke2g_GSC}, the IB scheme was also
implemented to correct the radiative
yield in the three-body decay $K_{e2\gamma}$.
The obtained ratio is consistent with
a recent lattice-QCD calculation~\cite{latticeQCD}, a gauged nonlocal chiral
quark model (NL$\chi$QM)~\cite{NLchiQM},
however at tension with ChPT at order $\mathcal{O}(p^4)$~\cite{bijnens1993,daphne1995}
and $\mathcal{O}(p^6)$~\cite{chen2008}, and a previous
measurement by KLOE~\cite{KLOE} at the 4-$\sigma$ level.
The latter data have been
used by NA62~\cite{NA62} to correct the universality ratio $R_K$ for the SD
background, which should be revisited.

\subsection{Radiative decay width \texorpdfstring{$\rho^- \rightarrow \pi^- \gamma$}{rho- -> pi- g} with COMPASS at CERN}

The COMPASS collaboration (COmmon Muon and Proton Apparatus for Structure and Spectroscopy) operated a multi-purpose,
high-resolution and high-rate capable spectrometer~\cite{Abbon_2007} at the CERN SPS M2 beamline with muon and hadron beams in the energy range 100--280\;GeV. A program pursued since the very beginning some 20 years ago focuses on ultra-peripheral reactions of pion and kaon beams on atomic nuclei, in the domain of momentum transfers where the electromagnetic interaction dominates --- so-called Primakoff reactions. It is intended to continue the program with a focus on kaon-induced reaction with AMBER~\cite{Adams:2018pwt}, operating at the same site as COMPASS and as its successor experiment.
\begin{figure}[ht]
    \centering
    \includegraphics[width=\columnwidth]{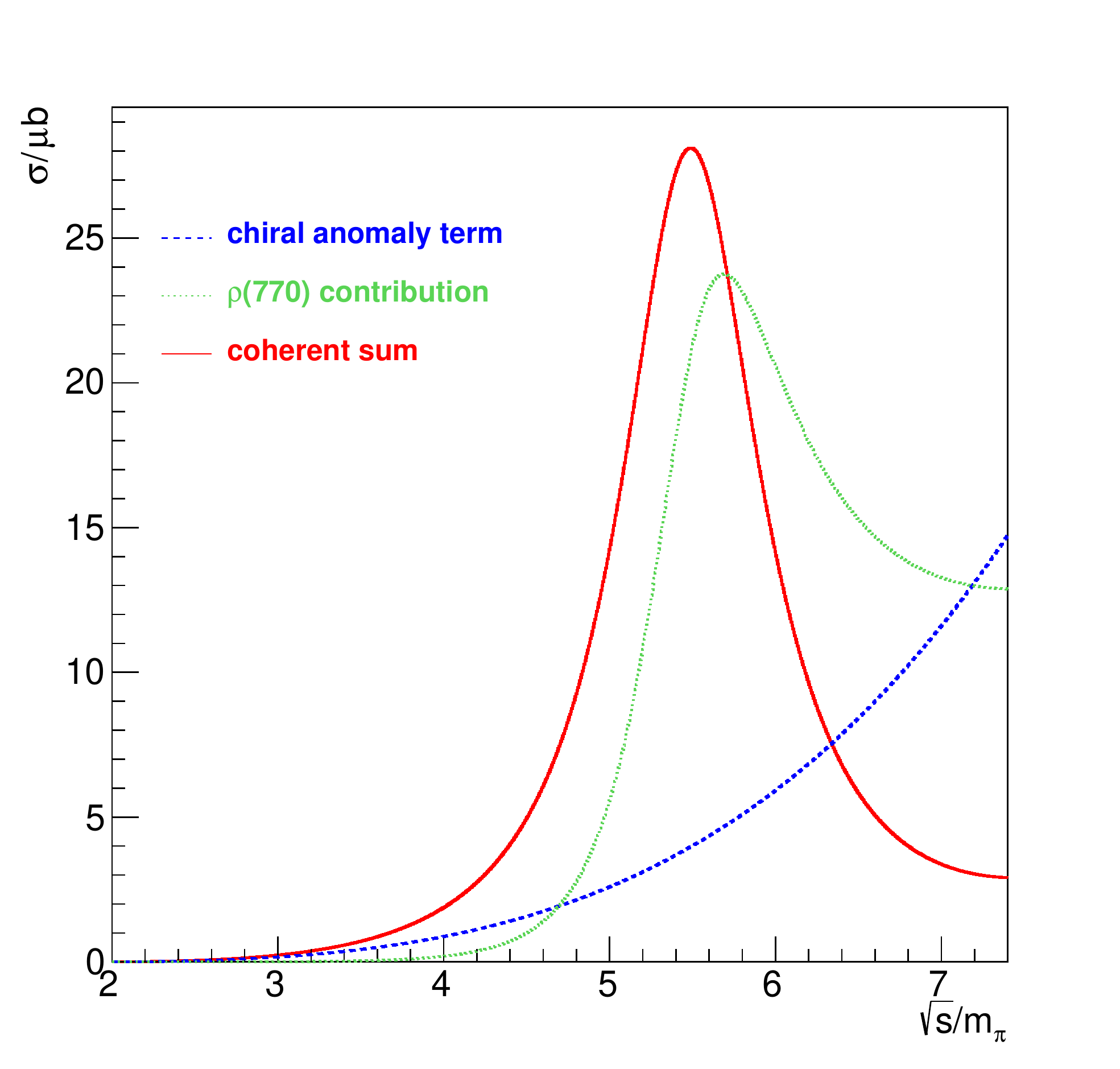}
    \caption{Contributions to the cross section for the reaction $\pi^-\gamma\rightarrow\pi^- \pi^0$ from the kinematic threshold, where the chiral anomaly dominates, to the region of the $\rho(770)$ resonance. Adapted from the respective figure in~\cite{Kaiser_2008}. The simplified model is used here to display the conceptual structure; for the data analysis to extract the chiral anomaly and the radiative width of the resonance, the dispersive framework~\cite{PhysRevD.86.116009} is used.}
    \label{fig:ChARho}
\end{figure}

One of the channels of interest is $\pi^-\gamma^{(*)}\rightarrow\pi^- \pi^0$, the photon from the exchange with the nucleus being quasi-real. According to the equivalent-photon formalism of Weizs\"acker and Williams, for a 190\;GeV pion beam on nickel, these photons cover a wide range of momenta of 0.1 to 2.0\;GeV in the rest frame of the incoming pion. Thus the full low-momentum range for the reaction becomes accessible and covers the kinematic region in which the $\rho(770)$ resonance dominates in the $s$-channel, $\pi^-\gamma^{(*)}\rightarrow\rho^-\rightarrow\pi^- \pi^0$. The cross section in the low-energy region, governed by the chiral anomaly up to the resonance region, is depicted in Fig.~\ref{fig:ChARho}. The strength of the resonance, as it appears in the reaction, depends on the radiative coupling $\rho\rightarrow\pi\gamma$, entering in inverted kinematics. The measurement of the cross section thus allows for a precision determination of the radiative width of the $\rho(770)$.

The reaction is affected by higher-order electromagnetic processes that have to be corrected for. A dominant contribution comes from a process with a $t$-channel exchange photon, coupling with the incoming photon to the outgoing $\pi ^0$ (thus $\pi^0\rightarrow\gamma\gamma$ in inverted kinematics). Corrections due to photon loops and real-photon emission have been calculated and found to be on the level of a few percent, thus are to be taken into account for a precision analysis. The formulae have been given in~\cite{PhysRevD.64.094009} and revised \cite{Kaiserpriv}. The full implementation in terms of an event generator including bremsstrahlung photons has been started.

\subsection{Summary and outlook}

In this section, we have mainly presented results regarding the so-called inclusive radiative corrections.
These are suitable to estimate the effect of QED corrections on the differential decay widths (and consequently on the extraction of TFFs or their parameters), assuming that the emission of additional photons has been ignored during the analysis.
Any additional assumptions or cuts that are made in the experiment should be reflected in the calculation, so a close experiment--theory collaboration is essential for preparing the suitable tailored approach.

As pointed out earlier, the most desirable way to implement radiative corrections is as early as at the Monte Carlo level.
Already at that stage, one might typically observe that there are events missing in the MC, and the agreement with data is not satisfactory.
This leads to larger uncertainties and extracted parameters containing unsubtracted QED parts.

The corrections in Figs.~\ref{fig:Dalitz_RC} and \ref{fig:etap_e_RC} may seem large at first.
But one needs to realize that the Dalitz decays are dominated by a one-photon exchange (i.e., by the low-$x$ region due to a pole at vanishing photon virtuality).
Integrating over the Dalitz plot then leads to the fact that the QED corrections at the dilepton production threshold will be receiving the largest weights leading to the determination of the corrections to the integrated decay width that are $\sim$\;1\% as one would expect.
These resulting values can then be independently checked by employing different methods (see, e.g., Ref.~\cite{Lautrup:1971ew}), and it was done so for $\pi^0$ and $\Sigma^0$ Dalitz decays arriving at an exact agreement.
Note that using the soft-photon approximation, the corrections are typically negative all over the Dalitz plot, and the effects of the hard-photon corrections are instrumental for obtaining the observed positive total corrections (see, e.g., Ref.~\cite{Sidhu:1971myo}).

We have stated examples of processes for which the radiative corrections have been worked out or used in the experiment.
From the mentioned examples, it is straightforward to see the importance of radiative corrections and that with improving sensitivity and higher statistics of upcoming experimental set-ups, these will become increasingly indispensable in order to extract meaningful results from data.
The list of light-hadron decays for which the exact NLO QED radiative corrections are available can thus be expected to get longer.

\section{Summary and Conclusions}\label{sec:Summary}

In this document we have outlined the state of the field for radiative corrections in $\ell p$ scattering, QED corrections to deep-inelastic scattering, and in meson decays. In the case of $\ell p$ scattering, the formalism of these corrections are laid out in detail at LO, NLO, and NNLO from a theoretical perspective. The experimental observables in unpolarized and polarized cross-section measurements are described in detail, in particular polarization transfer observables and SSAs. 

We particularly emphasize the case of the TPE, as this is an active area of work for both theoretical and experimental efforts. While the TPE is believed to be the cause of the proton form factor discrepancy, and there has been significant theoretical effort to explain the discrepancy, this has not been conclusively demonstrated in experiment. It is of paramount importance that new TPE measurements using charge asymmetric beams be performed to verify existing calculations. Simultaneously, theoretical efforts using a variety of techniques must continue to be developed to inform experimental results.

As part of the theoretical effort to treat TPE, it is incumbent upon the community to develop event generators that can be easily integrated into analysis pipelines. These generators should be ideally capable of NNLO calculations that can capture the leading logarithms beyond NNLO. Such future generators, or improvements upon existing generators, are necessary for modern precision experiments. Generators are an active area of research in the community and efforts to continue this development are a welcome contribution to the community.

Beyond the scope of $\ell p$ scattering experiments, there is more work to be done in $e^+e^-$ annihilation experiments. While the complete two-loop QED corrections have been calculated, there are future experimental measurements of the $Z^0$ mass planned which require calculations of QED initial state corrections of up to 6th order.

For mentioned light-hadron decays, detailed calculations of radiative corrections involving two-photon-exchange contributions and hard-photon corrections were briefly presented. Future high-precision, high-statistics experiments will require further developments in calculations of higher-order corrections. It is evident that future work will result in exact NLO calculations of other meson decays.

While there is still much work to be done, there has been significant progress made in the area of precision, percent-level experiments, and NNLO calculations of their associated radiative corrections. Several future experiments are planned that will test calculations of radiative corrections, and all future experiments require modern treatments of their corrections. Continued experimental and theoretical efforts in the field are crucial as we progress into the next generation of precision physics measurements.

\clearpage

\section{Acknowledgments}

We would like to thank Abilio~De~Freitas, Kay~Sch\"onwald, Hitham Hassan, Lois Flower, Marek Sch\"onherr, Jeppe Andersen, Peter Richardson, Marco Zaro, and Stefano Frixione for discussions.

E.W.C.~and J.C.B.~acknowledge the support of the National Science Foundation Grant Number PHY-2012114.

J.M.F.~acknowledges support from  the Deutsche Forschungsgemeinschaft (DFG, German Research Foundation) under Germany´s Excellence Strategy – EXC 2094 – 390783311.

F.H.~and V.S.~acknowledge support by the Swiss National Science Foundation (SNSF) through the Ambizione Grant PZ00P2\_193383.

F.H.~acknowledges support by the Deutsche Forschungsgemeinschaft (DFG) through the Emmy Noether Programme (grant 449369623) and the Research Unit FOR 5327 “Photon-photon interactions in the Standard Model and beyond --- exploiting the discovery potential from MESA to the LHC” (grant 458854507).

M.V.~acknowledges support by the Deutsche Forschungsgemeinschaft (DFG) in part through the Research Unit FOR 5327 “Photon-photon interactions in the Standard Model and beyond --- exploiting the discovery potential from MESA to the LHC” (grant 458854507), 
and in part through the Cluster of Excellence [Precision Physics, Fundamental Interactions, and Structure of Matter] (PRISMA$^+$ EXC 2118/1) within the German Excellence Strategy (Project ID 39083149). 

A.A.~acknowledges support of US National Science Foundation through Grant PHY-2111063.

A.S.~acknowledges support by the Swiss National Science foundation through the grant 207386.

M.K.~acknowledges support by Department of Energy through grants DE-SC0003884 and DE-SC0013941, and by National Science Foundation through grants PHY-1812402 and PHY-2113436.

T.H.~was supported by the Czech Science Foundation grant 23-06770S, by the Charles University grant PRIMUS 23/SCI/025, and by the Swedish Research Council grants contract numbers 2016-05996 and 2019-03779.

O.T.~is supported by the US Department of Energy through the Los Alamos National Laboratory and by LANL’s Laboratory Directed Research and Development (LDRD/PRD) program under projects 20210968PRD4 and 20210190ER. Los Alamos National Laboratory is operated by Triad National Security, LLC, for the National Nuclear Security Administration of U.S. Department of Energy (Contract No. 89233218CNA000001).

P.B.~is supported by the Natural Sciences and Engineering Research Council of Canada.

Y.U.~is supported by the UK Science and Technology Facilities Council (STFC) under grant ST/T001011/1.

Ax.~S.\ acknowledges support by the U.S.\ Department of Energy Office of Science, Office of Nuclear Physics, under contract no. DE-SC0016583.

\newpage

\bibliography{sn-bibliography,Mesons/mesons,Mesons/mesons-exp}
\end{document}